\documentclass[a4paper,fleqn,usenatbib,useAMS]{mnras}

\usepackage[T1]{fontenc}
\usepackage{ae,aecompl}

\usepackage{graphicx}	
\usepackage{amsmath}	
\usepackage{amssymb}	
\usepackage{color}
\usepackage{multicol}
\usepackage{longtable}
\usepackage{rotating}
\usepackage{pdflscape}
\usepackage{natbib}
\usepackage{enumitem}
\usepackage{appendix}
\usepackage{url}
\usepackage{float}
\usepackage{tabularx}
\usepackage{booktabs}
\usepackage[caption = false]{subfig}

\def \simless {\mathbin{\lower 3pt\hbox{$\rlap{\raise 4pt
              \hbox{$\char'074$}}\mathchar"7218$}}}
\def \simgreat {\mathbin{\lower 3pt\hbox{$\rlap{\raise 4pt
              \hbox{$\char'076$}}\mathchar"7218$}}}

\title[Stellar and dynamical masses of BCGs I]{Dynamical masses of brightest cluster galaxies I: stellar velocity anisotropy and mass-to-light ratios}

\author[Loubser et al.]{S. I. Loubser$^{1}$\thanks{E-mail:Ilani.Loubser@nwu.ac.za (SIL)}, A. Babul$^{2}$, H. Hoekstra$^{3}$, Y. M. Bah\'{e}$^{3}$, E. O'Sullivan$^{4}$, M. Donahue$^{5}$\\
$^{1}$Centre for Space Research, North-West University, Potchefstroom 2520, South Africa\\
$^{2}$Department of Physics and Astronomy, University of Victoria, Victoria, BC, V8W 2Y2, Canada\\
$^{3}$Leiden Observatory, Leiden University, PO Box 9513, 2300 RA, Leiden, The Netherlands\\
$^{4}$Harvard-Smithsonian Center for Astrophysics, 60 Garden Street, Cambridge, MA 02138, USA\\
$^{5}$Michigan State University, Physics $\&$ Astronomy Dept., East Lansing, MI 48824-2320, USA}

\date{Accepted 2020 June 09. Received 2020 May 27; in original form 2020 January 16}

\pubyear{2020}

\begin{document}
\label{firstpage}
\pagerange{\pageref{firstpage}--\pageref{lastpage}}
\maketitle

\begin{abstract}
We investigate the stellar and dynamical mass profiles in the centres of 25 brightest cluster galaxies (BCGs) at redshifts of 0.05 $\leq z \leq$ 0.30. Our spectroscopy enables us to robustly measure the Gauss-Hermite higher order velocity moments $h_{3}$ and $h_{4}$, which we compare to measurements for  massive early-type galaxies, and central group galaxies. We measure positive central values for $h_{4}$ for all the BCGs. We derive the stellar mass-to-light ratio ($\Upsilon_{\star \rm DYN}$), and velocity anisotropy ($\beta$) based on a Multi-Gaussian Expansion (MGE) and axisymmetric Jeans Anisotropic Methods (JAM, cylindrically- and spherically-aligned). We explicitly include a dark matter halo mass component, which is constrained by weak gravitational lensing measurements for these clusters. We find a strong correlation between anisotropy and velocity dispersion profile slope, with rising velocity dispersion profiles corresponding to tangential anisotropy and decreasing velocity dispersion profiles corresponding to radial anisotropy. The rising velocity dispersion profiles can also indicate a significant contribution from the intracluster light (ICL) to the total light (in projection) in the centre of the galaxy. For a small number of BCGs with rising velocity dispersion profiles, a variable stellar mass-to-light ratio can also account for the profile shape, instead of tangential anisotropy or a significant ICL contribution. We note that, for some BCGs, a variable $\beta_{z}(r)$ (from radial to tangential anisotropy) can improve the model fit to the observed kinematic profiles. The observed diversity in these properties illustrates that BCGs are not the homogeneous class of objects they are often assumed to be.  
\end{abstract}





\section{Introduction}
\label{introduction}


It is well known that most, but not all, dominant galaxies in the centres of clusters (Brightest Cluster Galaxies, BCGs) show rising velocity dispersion gradients (e.g.\ \citealt{Carter1999, Brough2007, Loubser2008, Newman2013a, Veale2017a}). In contrast, the velocity dispersion profiles of typical early-type galaxies remain flat or decrease with radius \citep{Kronawitter2000}. Historically, the rising velocity dispersion profiles have been interpreted as evidence for the existence of high mass-to-light ratio components in these galaxies \citep{Dressler1979, Carter1985}. However, as the sizes of spectroscopically observed samples of BCGs increased \citep{Loubser2008, Loubser2018}, and as the kinematics of individual galaxies were observed to larger radii in their very extended stellar haloes \citep{Murphy2014, Bender2015, Hilker2018}, the results have become more perplexing. There is a large diversity in the slopes of the velocity dispersion that we see in the centres of BCGs \citep{Loubser2018}, and at the outer radii there are often discrepancies between kinematics measured from different tracers e.g. from stars, from planetary nebulae or from globular clusters \citep{Murphy2014}. There can also be signatures of different populations in the same kinematics tracer (e.g. stars with different origin, \citealt{Hilker2018}). Since the internal kinematics directly relate to the dynamical mass profiles of the galaxies \citep{Barnes2007}, as well as to their different evolutionary paths and mass assembly mechanisms, it is important to understand this diversity in velocity dispersion profile shapes.  


Additionally, observed BCG velocity dispersion profiles are an important step towards fully resolved cluster mass profiles (e.g. \citealt{Newman2013a}), a fundamental parameter required to test galaxy formation models \citep{White1978}. Strong and weak gravitational lensing are insensitive to assumptions about the state of the matter (e.g.\ see \citealt{Hoekstra2013} for a review), and provide exquisite mass measurements that are independent of the dynamical state from 30 kpc from the cluster centre to the cluster outskirts \citep{Miralda1995, Squires1996, Hoekstra2015, Umetsu2016, Jauzac2018}. The X-ray emitting intracluster medium (ICM) bound to the cluster halo can also be a useful tracer of the dark matter content of the cluster \citep{Mahdavi2008}, however this requires assuming that the gas is in hydrostatic equilibrium. Additionally, X-ray measurements are often distorted by AGN outflows and also lack the resolution in the cluster centre required to obtain a good constraint on the slope of the total mass profile. Given the observational evidence that BCGs often lie at the bottom of the cluster potential in regular, non-interacting systems, the stellar kinematics of the BCG in such systems provides an additional measurement of the total mass at small radii ($<$ 30 kpc), since the stellar velocity dispersion probes the total gravitational potential well in which the stars are moving, and the mass contribution from gas is negligible. Specifically, measuring the BCG stellar velocity dispersion profile therefore allows us to dynamically probe the mass distribution in the central region.


Unfortunately, massive early-type galaxies exhibit a complex relationship between the gravitational potential, orbital configuration of stars, and measured line-of-sight velocity distribution ($\mathcal L_{\rm LOS}$), so that understanding the internal kinematics entails constraining both the mass and the velocity anisotropy \citep{Binney1982, Dejonghe1992, Merritt1993, Gerhard1993, Veale2017a}. Therefore, full dynamical modelling is needed, and additional information can be obtained from high-quality measurements of at least the kurtosis $h_{4}$. 


Whilst large samples of galaxies have been dynamically modelled, e.g. 28 spiral and S0 galaxies in \citet{Williams2009}, 260 early-type galaxies in ATLAS3D \citep{Krajnovic2011, Cappellari2013}, and 27 early-type galaxies from the CALIFA survey \citep{Lyubenova2016}, these surveys contain a very small number of BCGs, if any. \citet{Newman2013a} presented a detailed study of the dynamical modelling of seven cluster mass profiles using BCG velocity dispersion profiles. \citet{Bender2015} presents dynamical modelling of the BCG in Abell 2199, and \citet{Smith2017} presented dynamical modelling of the BCG in Abell 1201 using a combination of lensing and stellar dynamics. 


Here, we study a large sample of 32 BCGs, at redshifts of 0.05 $\leq z \leq$ 0.30, from the well characterised Multi-Epoch Nearby Cluster Survey (MENeaCS) and Canadian Cluster Comparison Project (CCCP) cluster samples spanning $M_{K}$ = --25.7 to --27.8 mag, with host cluster halo masses $M_{500}$ = 2.0 $\times$ $10^{14}$ to 1.5 $\times$ $10^{15}$ M$_{\sun}$ \citep{Herbonnet2019}. We derive dynamical mass models for 25 of these BCGs. Our large sample size, and the ability to constrain the dark matter halo mass component through weak lensing, and thereby limiting the number of free parameters in the dynamical models, are a considerable improvement on previous dynamical studies of BCGs. The data are described in \citet{Loubser2018} and only briefly summarised in Section \ref{data}. 

We use $H_{0}$ = 73 km s$^{-1}$ Mpc$^{-1}$, $\Omega_{\rm matter}$ = 0.27, $\Omega_{\rm vacuum}$ = 0.73 throughout, and make cosmological corrections where necessary. We refer to velocity dispersion as $\sigma$, and to $\sqrt{V^{2} + \sigma^{2}}$ as the second moment of velocity ($\nu_{\rm RMS}$), where $V$ is the rotation velocity, and we refer to the line-of-sight velocity distribution as $\mathcal L_{\rm LOS}$. The stellar mass-to-light ratio from dynamics, $\Upsilon_{\star \rm DYN}$, is determined in the rest-frame $r$-band. 

The data are described in Section \ref{data}. We present the comprehensive measurements of the higher order velocity moments in Section \ref{higherorder}, and investigate correlations with several other properties to validate our data analysis. We discuss the modelling of the stellar mass of the BCGs using the multi-Gaussian Expansion Method (MGE) in Section \ref{sec:stellarmasses}. The dynamical mass modelling, using the cylindrically-aligned Jeans Anisotropic Method (JAM), is described in Section \ref{sec:dynmasses}, including the addition of a central mass, as well as a dark matter halo mass component. The best-fitting models and parameters are discussed in Section \ref{sec:results}, including comparing the results with respect to the spherically-aligned JAM models, and incorporating the interpretation of our $h_{4}$ measurements. We finally summarise our conclusions in Section \ref{conclusions}. Various robustness tests, e.g. the effect of the mass or radius of the central (black hole) mass component, the influence of the point spread function (PSF), and the masking of foreground objects in some the images, are presented the Appendices. 

\section{Data}
\label{data}

\subsection{MENeaCS and CCCP BCG and CLoGS BGG spectra}
\label{spectra}

We use spatially-resolved long-slit spectroscopy for 14 MENeaCS BCGs and 18 CCCP BCGs, taken on the Gemini North and South telescopes with the slit aligned with the major axis of the galaxies\footnote{See Appendix \ref{SlitPA}, for a comparison between the slit position angle (PA) of the observations with the global major axis PA of the galaxies measured during the MGE fitting procedure.}. We also use \textit{r}-band imaging from the Canada-France-Hawaii Telescope (CFHT). In addition, we use weak lensing properties of the host clusters themselves \citep{Hoekstra2015, Herbonnet2019}. We also compare our measurements of $h_{3}$ and $h_{4}$ with those we make for a sub-sample of 23 Brightest Group Galaxies (BGGs) from the Complete Local-Volume Groups Survey (CLoGS) sample in the Local Universe ($D<80$ Mpc, \citealt{O'Sullivan2017}). For the $h_{3}$ and $h_{4}$ measurements of these galaxies we analyse archival spatially-resolved long-slit spectroscopy from the Hobby-Eberly Telescope (HET, see \citealt{Loubser2018}). The BCG and BGG samples are listed in Tables \ref{properties} and \ref{BGGproperties}, respectively. 

\subsection{MENeaCS and CCCP BCG imaging}
\label{images}

For both the MENeaCS and CCCP samples, we use $r{'}$ or $R$-band imaging from the CFHT, depending on the instrument used to perform the observations. Nine CCCP clusters were observed in the $R$-band with CFH12K, and the rest of the clusters were observed using the MegaCam detector and the $r{'}$ filter. The exposure times for the clusters observed with CFH12K range from 3000 to 1.4 $\times 10^{4}$ seconds for the $R$-band. The exposure times for the clusters observed with MegaCam are 4800 seconds in $r{'}$ for the CCCP and 2600 to 4800 seconds in $r{'}$ for the MENeaCS clusters. For further details on the properties of the optical images we refer the reader to \citet{Hoekstra2007} and \citet{Hoekstra2012} for the CCCP data, and \citet{Sand2012} for the MENeaCS data. 

We correct for foreground (line-of-sight) Galactic extinction of all the non-star forming BCGs by using the \citet{Schlafly2011} recalibration of the infrared-based dust map by \citet{Schlegel1998}. As discussed in \citet{Loubser2018}, the internal extinction of individual star-forming BCGs is difficult to determine, and we use average total extinction values from \citet{Crawford1999}  (see also \citealt{Mittal2015}). On average, $E(B-V)_{\rm total}$ = 0.30 for BCGs, and this total extinction is applied only to the BCGs with young stellar components.

Because our BCGs span a range of redshifts, accurately comparing them requires corrections to the observed luminosity to account for cosmological redshift and galaxy evolution. Following \citet{Hogg2002} and \citet{Houghton2012} we split this $K$ correction into two terms 
\begin{equation}
K = K_{\rm b} + K_{\rm c}
\end{equation}
where the bandpass term $K_{\rm b}$ is easily corrected in the AB magnitude system by reducing the observed brightness by (1 + $z$). However, the colour term $K_{\rm c}$ depends on the exact spectral energy distribution of the galaxies, for which we assume no colour evolution for the BCGs from redshift $z \sim 0.3$ to $z \sim 0.05$. Any colour evolution present would have a marginal affect on the stellar mass profiles. We further correct for Tolman dimming \citep{Lubin2001}, which together with $K$ gives
\begin{equation}
\mu_{\rm rest} = \mu_{\rm obs} - 7.5 \log (1+z).
\end{equation}

We describe the derivation of the stellar mass profiles from our wide-field imaging in Section \ref{sec:stellarmasses}, and the spatial extent of the stellar mass modelling as well as the masking of foreground/background features in the images in Appendix \ref{masking}. We show images of the nuclei of the BCGs in Figure \ref{MGEFig}. 

\subsection{Fundamental plane}
\label{FPlane}

\begin{figure}
\centering
   \subfloat{\includegraphics[scale=0.37]{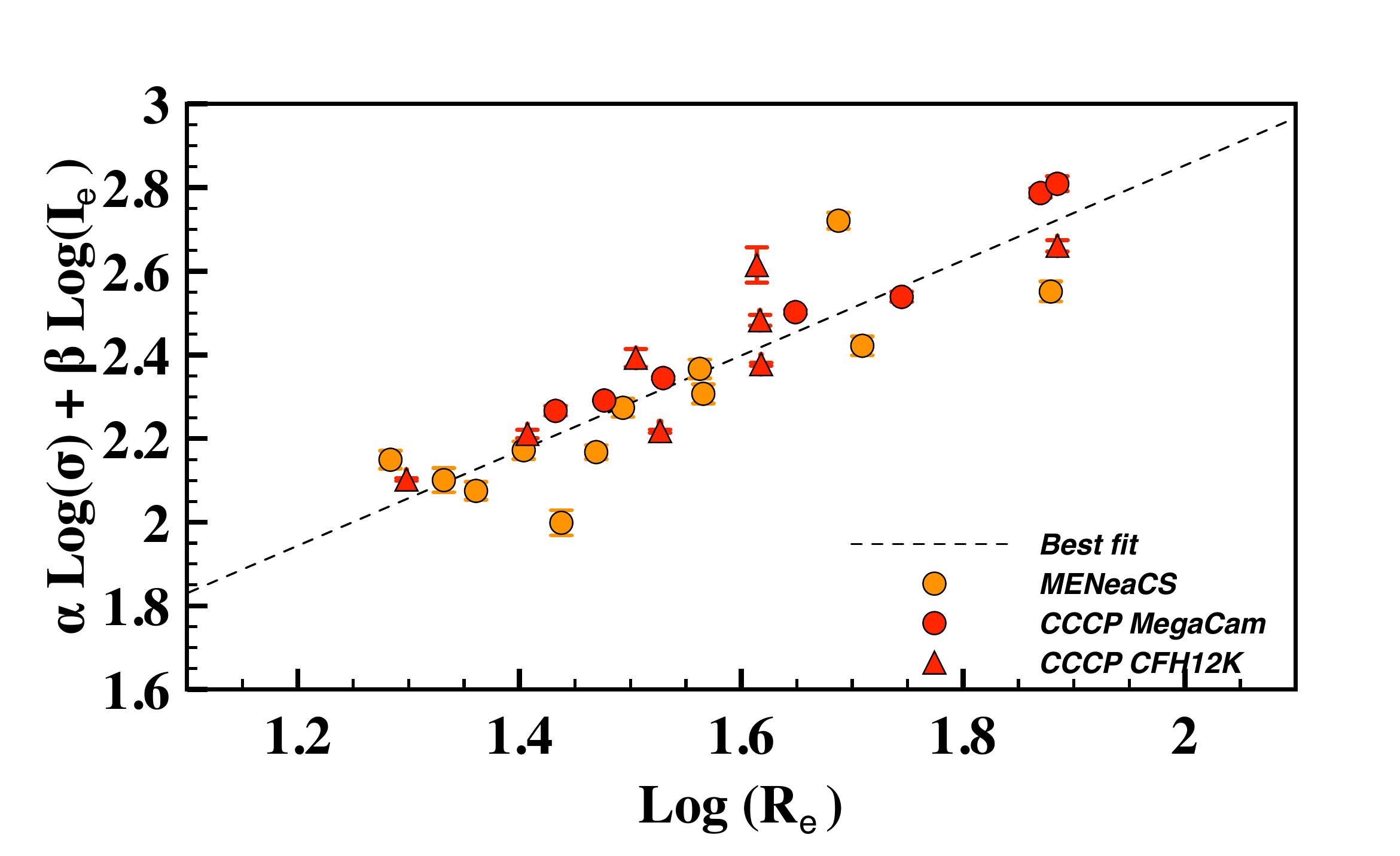}}\\
   \caption{We use our spectral and imaging measurements to plot the fundamental plane for our BCGs, using $\alpha = 1.24$ and $\beta=-0.80$ from \citet{Bildfell2008} and \citet{Bildfell_thesis}. The MENeaCS BCGs are indicated with orange symbols, and the CCCP BCGs in red (MegaCam imaging in circles and CFH12K imaging in triangles). We measure an intrinsic scatter of 0.095.}
\label{FPzoom}
\end{figure}

To illustrate the robustness of our spectral and imaging measurements, we show the fundamental plane \citep{Dressler1987,Djorgovski1987} for our data in Figure \ref{FPzoom}. We use the central velocity dispersion measurements ($\sigma_{0}$) from \citet{Loubser2018}, and we extract one-dimensional surface brightness profiles ($\mu_{r}$ in mag/arcsec$^{2}$) along the major axis and perform the corrections described above. The surface brightness is converted to a surface density ($I_{r}$ in L$_{\sun}$/pc$^{2}$), using the absolute magnitude of the Sun (in the $r$-band) from \citet{Blanton2007}. We then fit an $R^{1/4}$-law
\begin{equation}
I_{r}(R)=I_{e} e^{-7.67([R/R_{e}]^{1/4} - 1)}.
\end{equation}
For the BCGs with flat cores in the centre, we do not include the central part of the surface brightness profiles in our fits. We plot 
\begin{equation}
\log R_{e} = \alpha \log(\sigma_{0}) + \beta \log(I_{e})
\end{equation}
with $I_{e}$ in L$_{\sun}$/pc$^{2}$ and $R_{e}$ in kpc in Figure \ref{FPzoom}. We do not vary $\alpha$ or $\beta$ to minimise the residuals, but instead use $\alpha = 1.24$ and $\beta=-0.80$ as derived (for the CCCP BCGs) by \citet{Bildfell_thesis}. The $\alpha$ and $\beta$ values derived by \citet{Bildfell_thesis} are in good agreement with those by \citet{Jorgensen1996} for their sample of 226 E and S0 galaxies in 10 clusters of galaxies ($\alpha=1.24 \pm 0.07$ and $\beta=-0.82 \pm 0.02$). We measure an intrinsic scatter of 0.095, using $\mathtt{linmix\_err}$ by \citet{Kelly2007} (also see Section \ref{highermeasurements}). From this fit and plot we exclude the BCG in Abell 2055, which is known to host a BL Lac point source that is the main contributor to the light observed in the $r$-band \citep{Green2017}. We also exclude the BCGs in Abell 990, 1835, 2104 and 2390, which have problematic surface brightness profiles (because of substructure or multiple nuclei) as described in Section \ref{nuclei} and shown in Figure \ref{fig:MGE4}. 

Our measured intrinsic scatter is consistent with that of \citet{Saulder2013}, who calibrated the fundamental plane for elliptical galaxies ($z<0.2$) using 93000 galaxies from SDSS DR8, and found (for the $r$-band) intrinsic scatter between 0.0933 and 0.0956 (depending on the fitting method used), confirming the robustness of our spectral and imaging measurements.

\section{The higher order velocity moments} 
\label{higherorder}

The line-of-sight stellar velocity distribution ($\mathcal L_{\rm LOS}$) is a measure of the gravitational potential of a galaxy and thus its dynamical mass. However, there is a well-known degeneracy between mass and velocity anisotropy (e.g.\ \citealt{Binney1982, Veale2017a}) that can be alleviated somewhat by robust measurements of the Gauss-Hermite moment $h_{4}$ of the $\mathcal L_{\rm LOS}$. The $h_{4}$ moment describes the kurtosis of $\mathcal L_{\rm LOS}$, i.e.\ positive values describe a distribution more `peaked' than a Gaussian, and negative values indicate a more flat-topped distribution. The Gauss-Hermite moment $h_{3}$ describes the skewness of $\mathcal L_{\rm LOS}$, i.e.\ negative values indicate a distribution with an extended tail towards low velocities, and for positive values vice versa. For non-rotating, non-disky early-type galaxies such as the BCGs studied here, $h_{3}$ is expected to be consistent with zero.

For isothermal galaxies, the isotropic case ($\beta = 0$)\footnote{Where the stellar velocity anisotropy $\beta = 1 - (\sigma^{2}_{\phi} / \sigma^{2}_{R})$ for spherical models, or $\beta_{z} = 1 - (\sigma^{2}_{z} / \sigma^{2}_{R})$ for axisymmetric (cylindrically-aligned) models. The relation between the velocity anisotropy from spherical and cylindrical axisymmetric models is given in \citet{Cappellari2007}.} is known to correspond to flat velocity dispersion profiles and $h_{4} = 0$, whereas the radial anisotropy case corresponds to a positive $h_{4}$, and lower velocity dispersion, and decreasing velocity dispersion gradients. Lastly, the tangential anisotropy case corresponds to negative $h_{4}$, a higher velocity dispersion, and rising velocity dispersion gradients \citep{Gerhard1993, Vandermarel1993, Rix1997, Gerhard1998, Thomas2007b}. However, if there are steep gradients in the circular velocity, a rising velocity dispersion profile as well as a positive $h_{4}$ can be expected even in the isotropic case, and with additional effects on $h_{4}$ and velocity dispersion in the anisotropic cases \citep{Gerhard1993, Veale2017a}.

\subsection{Measurements of Gauss-Hermite moments $h_{3}$ and $h_{4}$}
\label{highermeasurements}

\begin{figure}
\centering
   \subfloat{\includegraphics[scale=0.3]{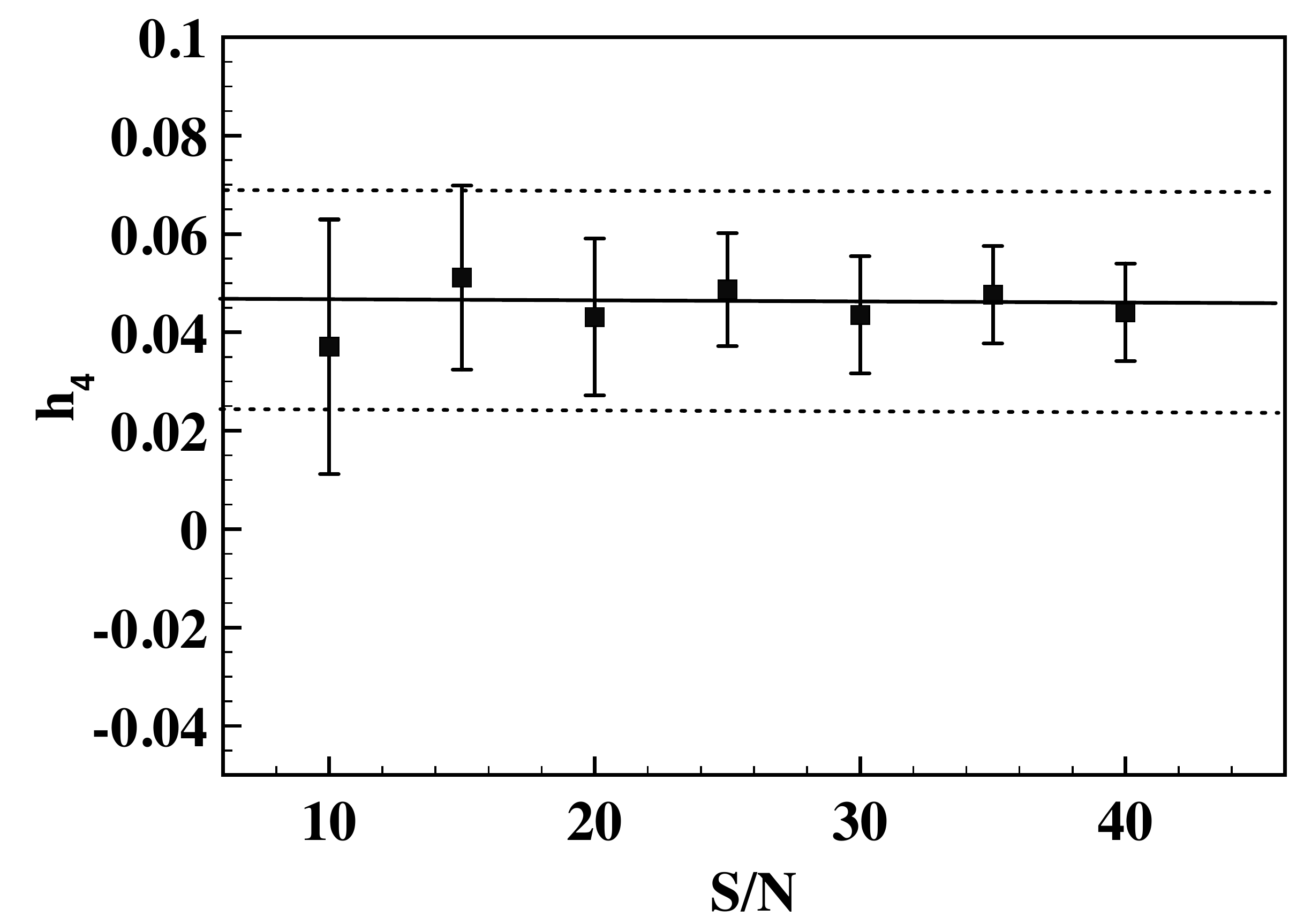}}\\
   \caption{We use the spectrum of the BCG in Abell 68 as an example, and create mock spectra of different \textit{S/N}, and repeat measurements of $h_{4}$ 50 times using Monte Carlo simulations. We plot the average measurements for the spectra with different \textit{S/N}, where the solid and dotted lines indicate our actual $h_{4}$ measurement and error bars, respectively. We conclude that we are not biased in our $h_{4}$ measurements as a result of poor \textit{S/N}.}
\label{SNbias_h4}
\end{figure}

We measure $h_{3}$ and $h_{4}$ using the inner bin, i.e. 5 kpc to either side from the BCG centre. This central aperture is the same as used in \citet{Loubser2016} (for central stellar population properties) and in \citet{Loubser2018} (for the central velocity dispersion measurements) for the BCGs. We check the sensitivity of our $h_{4}$ measurements to signal-to-noise ratio (\textit{S/N}), and we illustrate it in Figure \ref{SNbias_h4} using the BCG in Abell 68 as an example, as the spectrum is representative of the typical BCG data. We create mock spectra of different \textit{S/N}, and repeat measurements of $h_{4}$ 50 times using Monte Carlo simulations. We plot the average measurements for the spectra with different \textit{S/N}, where the solid and dotted lines indicate our $h_{4}$ measurement and 1$\sigma$-error bars, respectively. Since our central apertures all have sufficient \textit{S/N} $>$ 10, we conclude that we are not biased in our $h_{4}$ measurements as a result of poor \textit{S/N}. 

For comparison, we repeat the same central $h_{3}$ and $h_{4}$ measurements for the CLoGS BGGs, but in the central bin of 1 kpc to either side (the same central aperture as used in \citealt{Loubser2018} for the CLoGS central velocity dispersion measurements). We also directly compare our $h_{3}$ and $h_{4}$ measurements from the HET data for the CLoGS BGGs, to those measured by \citet{Vandenbosch2015} and find good agreement. Our central $h_{4}$ measurements for the four BGGs we have in common with the MASSIVE sample also agree well with their central measurements of $h_{4}$. We show the comparison in Table \ref{h4_h4} in Appendix \ref{h4}.

The $h_{3}$ and $h_{4}$ measurements for the BCGs and the BGGs are listed in Table \ref{properties} and \ref{BGGproperties}, respectively, and we plot the $h_{3}$ and $h_{4}$ measurements against the $K$-band luminosity in Figure \ref{h4_luminosity}. Since we are modelling the BCGs, we discuss their $h_{3}$ and $h_{4}$ measurements further, and only use those of the BGGs for comparison. We indicate the BCGs in MENeaCS and CCCP with young stellar components with open symbols (see \citealt{Loubser2016}). The majority of these BCGs with young stellar population components have peculiarly high measurements of $h_{4}$ ($>0.11$), in particular Abell 383, 646, 1835, 2055 and 2390. The $h_{3}$ measurements of these galaxies also deviate from zero, which together, gives a strong indication of a template mismatch \citep{Bender1994}, as they likely have some spatial variation of stellar populations in the averaged central bin. We do not use the $h_{3}$ or $h_{4}$ measurements of the six BCGs with young stellar populations (three in MENeaCS and three in CCCP) further. If these six BCGs are excluded, then $\langle h_{4} \rangle = 0.049 \pm 0.004$ and $\langle h_{3} \rangle = 0.011 \pm 0.004$ for the BCG sample. 

To quantify the correlation between $h_{3/4}$ and $M_{K}$, we assume the intrinsic (random) scatter to be normally distributed, and we use the Gibbs sampler implemented in the multivariate Gaussian mixture model routine $\mathtt{linmix\_err}$ by \citet{Kelly2007} with the default of three Gaussians. We use 5000 random draws of the sampler and take the fitted parameters as the posterior mode and the error as the 68 per cent highest posterior density credible interval. For $h_{4}$ vs $M_{K}$ (BCGs, with pivot at 26.5), we find a shallow slope = 0.0121 $\pm$ 0.0070, with an intrinsic scatter of 0.0105 $\pm$ 0.0104, and correlation coefficient 0.625 (with a zero point = 0.0512 $\pm$ 0.0045). For $h_{3}$ vs $M_{K}$ (BCGs), we find a slope consistent with zero (0.0004 $\pm$ 0.0077), with an intrinsic scatter of 0.0144 $\pm$ 0.0121, and correlation coefficient 0.026 (with a zero point = 0.0083 $\pm$ 0.0048). 

\begin{table}
\caption{Properties and central higher order moment measurements of the BCGs. A $\star$ indicates that imaging was observed in the $R$ filter, all other imaging is in the $r'$ filter. Lastly, we also list the black hole mass ($M_{\rm BH}$) used as central mass component ($M_{\rm CEN}$) in Section \ref{blackhole}.}
\label{properties}
\begin{scriptsize}
\begin{tabular}{l c r r c}
\hline
Name & $z$ &   $h_{3}$ & $h_{4}$ & $M_{\rm BH}$ \\
            &      &     &       &    (10$^{9}$ M$_{\sun}$)   \\             
\hline
\multicolumn{5}{c}{MENeaCS}\\
Abell 780 & 0.054 & 	   0.011$\pm$0.032 & 0.109$\pm$0.038   &  1.96 \\
Abell 754 & 0.054 &   --0.005$\pm$0.019	& 0.039$\pm$0.019  &  1.38\\ 
Abell 2319 & 0.056  &      --0.004$\pm$0.006  &	0.082$\pm$0.014 &  2.09\\
Abell 1991 & 0.059  & 	       0.007$\pm$0.016   &  	0.069$\pm$0.034 &  1.16 \\
Abell 1795 & 0.063 &   0.026$\pm$0.030	& 0.082$\pm$0.021 &   0.85 \\
Abell 644 & 0.070 &        0.017$\pm$0.008	& 0.055$\pm$0.017 &   2.12 \\
Abell 2029 & 0.077 &       0.023$\pm$0.029	& 0.020$\pm$0.034 &   1.29\\
Abell 1650 & 0.084  &      0.006$\pm$0.029 &	0.048$\pm$0.014 &  7.58 \\
Abell 2420 & 0.085 &      0.036$\pm$0.065	& 0.034$\pm$0.031 & 1.22 \\
Abell 2142 & 0.091 &    0.012$\pm$0.017 &	0.068$\pm$0.020 &    2.37\\
Abell 2055 & 0.102  &     --0.017$\pm$0.044	& 0.159$\pm$0.042 &  2.93\\
Abell 2050 & 0.118 &    --0.009$\pm$0.019  &	0.062$\pm$0.009 & 0.56\\
Abell 646 & 0.129  &   --0.053$\pm$0.048  &	0.109$\pm$0.061 &   1.87\\
Abell 990 & 0.144 &    --0.022$\pm$0.046	&  0.063$\pm$0.026 &  1.38\\
\hline
\multicolumn{5}{c}{CCCP}\\
Abell 2104 & 0.153 &    0.036$\pm$0.012  &	0.020$\pm$0.030 &   0.51 \\
Abell 2259 & 0.164 &    0.027$\pm$0.028  &	0.059$\pm$0.026 &  2.00\\
Abell 586 & 0.171 &   0.004$\pm$0.025   & 	0.036$\pm$0.024 &  1.17\\
MS 0906+11$^{\star}$ & 0.174 & 0.029$\pm$0.017 & 	0.046$\pm$0.005 & 1.26 \\
Abell 1689$^{\star}$ & 0.183 &    --0.008$\pm$0.032  &	0.063$\pm$0.034 &  2.06\\
MS 0440+02 & 0.187 &    --0.005$\pm$0.006 &	0.065$\pm$0.021 &   4.94\\ 
Abell 383$^{\star}$ & 0.190 &     0.076$\pm$0.090	& 0.143$\pm$0.050 &   5.87\\
Abell 963$^{\star}$ & 0.206 &        0.019$\pm$0.007  &	0.036$\pm$0.016 &  2.72\\
Abell 1763$^{\star}$ & 0.223 &   0.005$\pm$0.010 &	0.030$\pm$0.005 &  3.93\\
Abell 1942 & 0.224&   0.031$\pm$0.013 &	0.056$\pm$0.025 &  1.40\\
Abell 2261 & 0.224 &   0.026$\pm$0.017  &	0.048$\pm$0.008 & 8.00\\
Abell 2390$^{\star}$ & 0.228 &      0.076$\pm$0.029 &	 0.129$\pm$0.033 & 2.97\\
Abell 267$^{\star}$ & 0.231 &    0.020$\pm$0.025 &	0.039$\pm$0.019 &  2.11\\
Abell 1835 & 0.253  &   0.045$\pm$0.011 & 	0.143$\pm$0.019 &  1.25\\
Abell 68$^{\star}$ & 0.255 &      --0.001$\pm$0.009  &	0.044$\pm$0.022 & 2.22\\
MS 1455+22$^{\star}$ & 0.258 &    0.015$\pm$0.018 &	 0.056$\pm$0.013 &  5.83\\
Abell 611 & 0.288 &      --0.036$\pm$0.020  &  	0.045$\pm$0.019 & 1.78\\
Abell 2537 & 0.295 &    --0.003$\pm$0.007 &	0.075$\pm$0.016 &  1.88\\
\hline
\end{tabular}
\end{scriptsize}
\end{table}

\begin{table}
\caption{Central higher order moment measurements of the CLoGS BGGs.}
\label{BGGproperties}
\centering
\begin{tabular}{l r r}
\hline
Name & $h_{3}$ & $h_{4}$\\
\hline
\multicolumn{3}{c}{High density sample}\\
NGC0410 &	--0.025$\pm$0.016 &	0.051$\pm$0.015 \\
NGC0584	 & 0.037$\pm$0.053 &	0.109$\pm$0.038\\
NGC0777	& --0.012$\pm$0.026 &	0.056$\pm$0.033\\
NGC0924	 & 0.014$\pm$0.069 &	0.082$\pm$0.038\\
NGC1060	 & --0.021$\pm$0.019 & 	0.045$\pm$0.028\\
NGC1453	 & --0.030$\pm$0.034 &	 0.067$\pm$0.034\\
NGC1587	 & --0.004$\pm$0.038 &	 0.077$\pm$0.044\\
NGC2563	 & 0.055$\pm$0.056 &	0.059$\pm$0.079\\
NGC4261	 & 0.011$\pm$0.020 &	0.068$\pm$0.037\\
NGC5353	 & 0.021$\pm$0.023 &	0.093$\pm$0.027\\
NGC5846	 & --0.048$\pm$0.043 &	 0.077$\pm$0.064\\
NGC5982	 & --0.029$\pm$0.060 &	0.071$\pm$0.039\\
NGC6658	 & 0.022$\pm$0.086 &	0.115$\pm$0.087\\
NGC7619	 & 0.002$\pm$0.022 &	0.071$\pm$0.032\\
\hline
\multicolumn{3}{c}{Low density sample}\\
NGC0315	 & 0.009$\pm$0.032 & 0.057$\pm$0.037\\
NGC0524	 & 0.009$\pm$0.089 &	0.146$\pm$0.073\\
NGC1779	 & 0.009$\pm$0.094 &	0.076$\pm$0.061\\
NGC2768	 & 0.011$\pm$0.037 &	0.013$\pm$0.083\\
NGC3613 &	--0.006$\pm$0.031 &	0.025$\pm$0.057\\
NGC3665	 & 0.016$\pm$0.054 &	0.037$\pm$0.055\\
NGC5127	 & 0.006$\pm$0.070 &	0.053$\pm$0.053\\
NGC5490	 & --0.014$\pm$0.030 &	0.038$\pm$0.049\\
NGC5629	 & 0.002$\pm$0.068 &	0.019$\pm$0.050\\
\hline
\end{tabular}
\end{table}

\subsection{Discussion: Gauss-Hermite moments $h_{3}$ and $h_{4}$}
\label{h4_discussion}

\begin{figure}
\centering
   \subfloat{\includegraphics[scale=0.35]{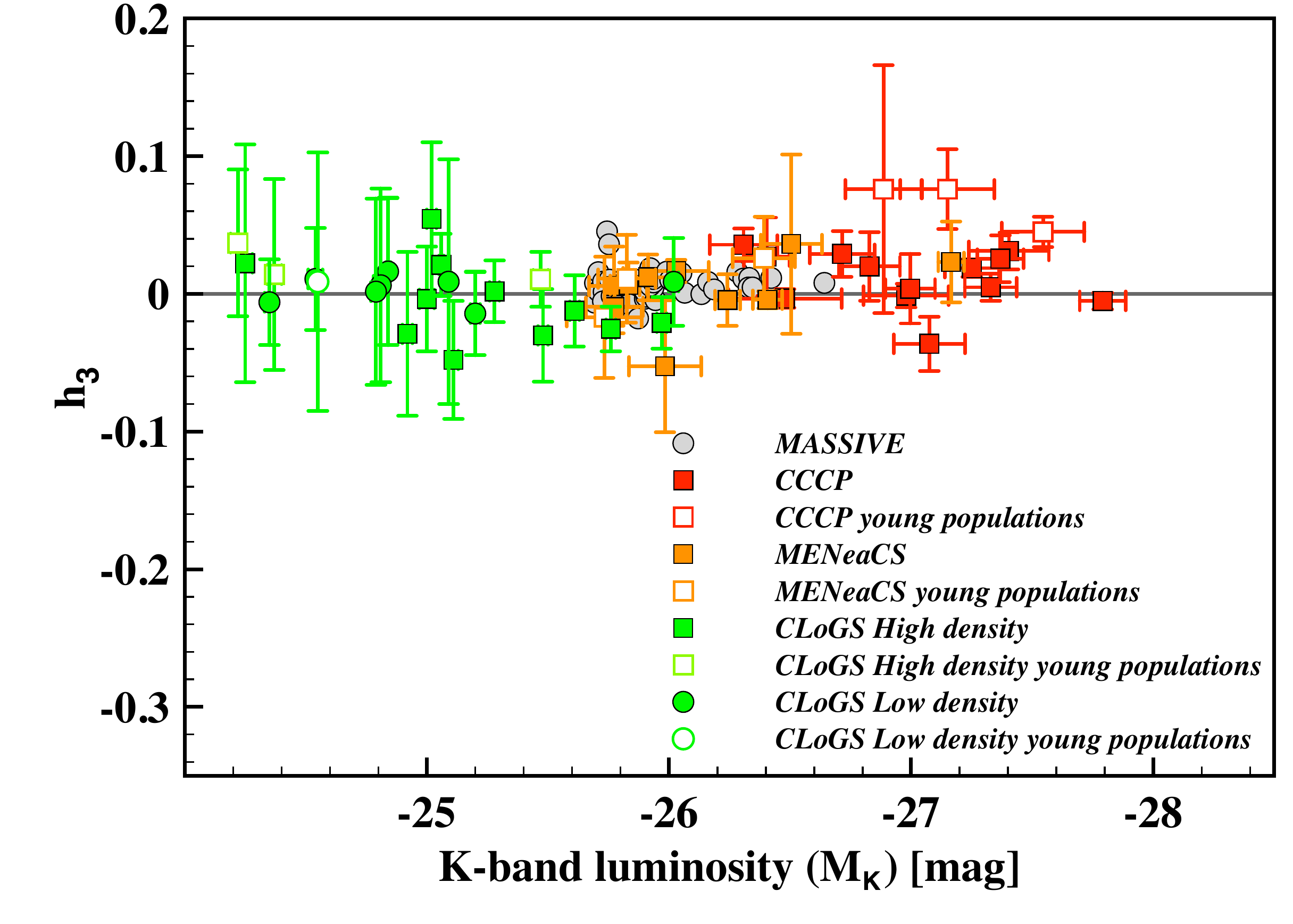}}\\
    \subfloat{\includegraphics[scale=0.35]{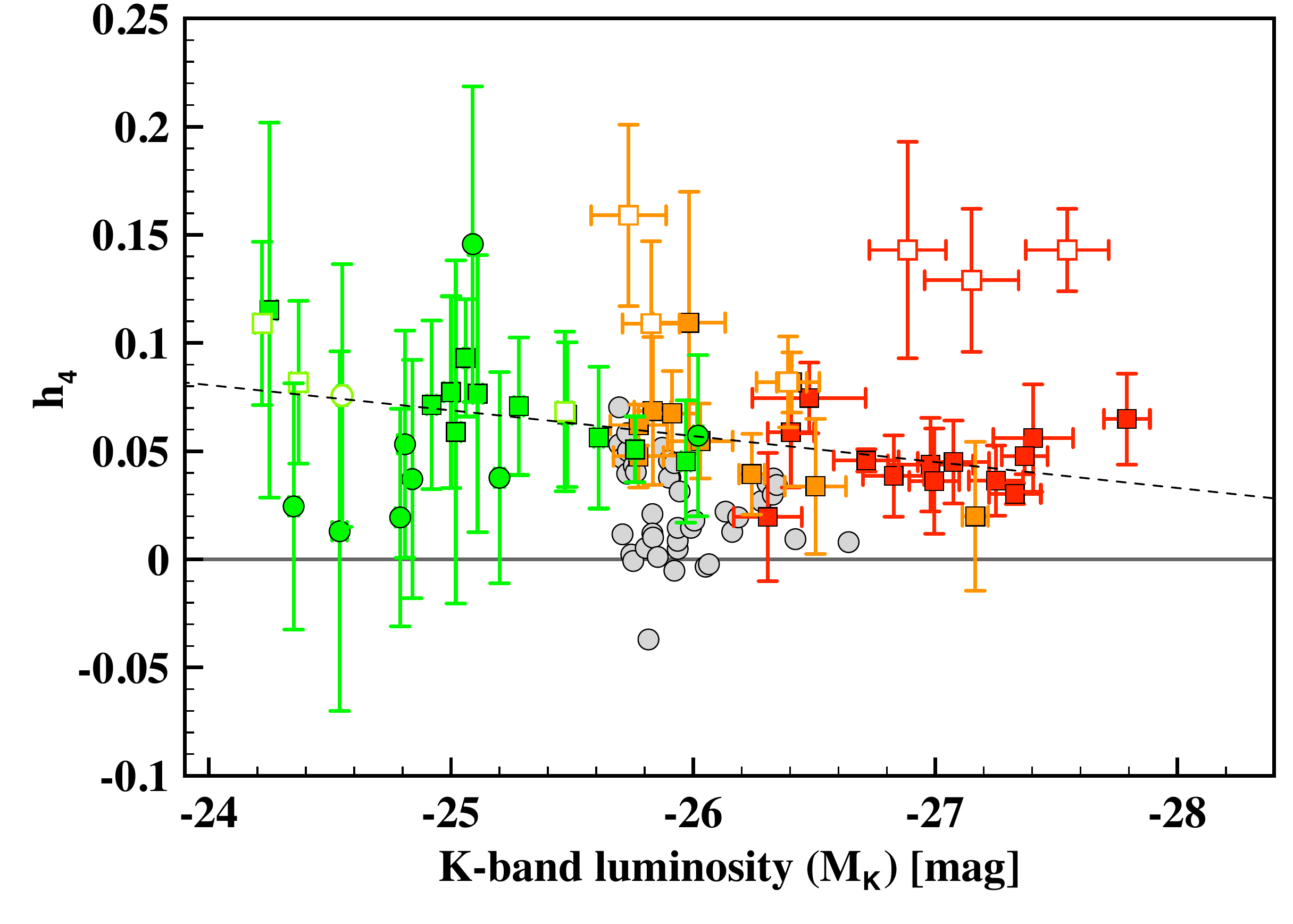}}
   \caption{Top: the central $h_{3}$ measurements vs $K$-band luminosity. Here we also plot the averaged $\langle h_{3} \rangle$ within the effective radius for MASSIVE  \citep{Veale2017a} for comparison (grey). Bottom: the central $h_{4}$ measurements vs $K$-band luminosity. We also plot the central $h_{4}$ for MASSIVE for comparison. The BCGs with young stellar components are plotted with open symbols, and likely suffer from template mismatch. The CLoGS BGGs with young components are results from preliminary analysis in preparation.}
\label{h4_luminosity}
\end{figure}

In the MASSIVE survey of elliptical galaxies closer than 108 Mpc, \citet{Veale2017a} show an anti-correlation between $h_{3}$ gradient and rotation velocity ($V$, as a function of radius) for each massive elliptical galaxy in their sample classified as a fast rotator. Here, we only use the central binned value of $h_{3}$, and as shown in \citet{Loubser2018}, we do not have any fast rotators in our BCG sample. Nevertheless, we also test for any gradients within our central measurements of $h_{3}$, as well as $h_{4}$, with repeated measurements in smaller bins. We find suggestions of weak gradients in $h_{3}$ within our central apertures, but not in $h_{4}$. 

\citet{Veale2017a} show (their figures 9 and 11) $h_{3}$ vs $M_{K}$, and $h_{4}$ vs $M_{K}$ for the MASSIVE survey, and they find an indication that the more luminous galaxies have a higher $h_{4}$. The MASSIVE survey also find their galaxies to have positive $\langle h_{4} \rangle$ (averaged over their radial range), and the most luminous galaxies have $\langle h_{4} \rangle \sim 0.05$ while less luminous galaxies have a range of values between 0 and 0.05. We show $h_{3}$ vs $M_{K}$, as well as $h_{4}$ vs $M_{K}$, for all BCGs and BGGs, and MASSIVE for comparison, in Figure \ref{h4_luminosity}. As described above, we find positive $h_{4}$ values for all of our BCGs and BGGs as shown in Figure \ref{h4_luminosity}, but for $h_{4}$ vs $M_{K}$ (for the BCGs) over a much larger magnitude range (than MASSIVE), we do not see the correlation described above, and instead find a very weak slope (--0.012 $\pm$ 0.007, indicated in Figure \ref{h4_luminosity}) in the opposite direction with the brighter BCGs having a slightly lower $h_{4}$ (and an intrinsic scatter of $h_{4}$ of 0.011). We find a slope (and zero-point) consistent with zero for $h_{3}$ vs $M_{K}$ (for the BCGs). We also investigated correlations between $h_{4}$ vs central velocity dispersion ($\sigma_{0}$), velocity dispersion slope ($\eta \pm \Delta \eta$), and $M_{500}$ (from \citealt{Herbonnet2019}), and see no correlations between these properties (not shown here).

\citet{Veale2017a} and \citet{Veale2018} find a correlation between their $h_{4}$ gradients and outer velocity dispersion gradients. \citet{Carter1999}, however, find that all three of their nearby BCGs show a positive and constant $h_{4}$ moment with radius, despite the fact that one of their BCGs, NGC6166, has a rising velocity dispersion profile. Even though they find different results, both groups interpret their results as an indication that the increase in velocity dispersion is not associated with a change in velocity anisotropy from radial to tangential orbits, but more likely to be a consequence of circular velocity gradients or an ICL component as discussed in Section \ref{ICL}. 

In their higher-order kinematics analysis of the SAMI survey, \citet{Vandesande2017} suggest a bias towards positive measurements of $h_{4}$ (their Figure 3, top right panel), by plotting $\sigma_{\rm m2} - \sigma_{\rm m4}$ against $h_{4}$, where $\sigma_{\rm m2}$ is the velocity dispersion measured assuming the stellar velocity distribution ($\mathcal L_{\rm LOS}$) is a pure Gaussian, and $\sigma_{\rm m4}$ is the velocity dispersion measured by fitting a truncated Gauss-Hermite series \citep{Vandermarel1993, Gerhard1993} to parametrise the $\mathcal L_{\rm LOS}$. They find a bias in that there is a tendency for $h_{4}$ to be positive, even where $\sigma_{\rm m2} - \sigma_{\rm m4}$ is zero\footnote{Since the Gauss-Hermite series is by construction orthogonal, a non-zero value of $h_{4}$ would not affect the best-fit value of $\sigma_{\rm m4}$, at least for a $\mathcal L_{\rm LOS}$ that is perfectly fit by a Gauss-Hermite polynomial.}. They investigate whether the positive $h_{4}$ values are the result of instrumental resolution, template mismatch, or different seeing conditions, and conclude that it is none of these factors and that there must be a physical reason for positive $h_{4}$. \citet{Veale2017a} also speculate that their positive measurements for $h_{4}$ are due to a physical origin rather than template mismatch. We have repeated the test performed in \citet{Vandesande2017}, and also conclude that we expect our $h_{4}$ measurements to be positive, and that this is due to a physical origin rather than an observational bias or a tempate mismatch in the measurements (with the exception of the star forming BCGs where their unusually high $h_{4}$ measurements are possibly due to a template mismatch). To test this possibility in more detail, we have extracted mock line-of-sight velocity distributions ($\mathcal L_{\rm LOS}$) for the BCGs in the Hydrangea cosmological hydrodynamical galaxy cluster simulations \citep{Bahe2017}. Outside of the central few kpc, where the simulations are affected by their limited resolution, these mock line-of-sight $\mathcal L_{\rm LOS}$ are best fit with truncated Gauss-Hermite series with $h_{4} \sim 0.05$, in good agreement with the values we measure for our observed BCGs. We therefore conclude that our positive $h_{4}$ measurements are physical, rather than reflecting observational biases.


\section{CCCP and MENeaCS BCGs Stellar masses}
\label{sec:stellarmasses}

We model the BCGs using a mass model which incorporates the stellar mass distribution, a central mass concentration representing a supermassive black hole, and a dark matter halo. The primary mass component of this dynamical modelling is the stellar mass description resulting from the MGE formalism (e.g.\ as used by \citet{Cappellari2006} for SAURON galaxies, by \citet{Williams2009} for S0 galaxies, and  \citet{Scott2013} for the ATLAS3D galaxies), which we describe in detail below.

\subsection{Multi-Gaussian Expansion (MGE)}
\label{MGE}

We use the MGE method \citep{Monnet1992, Emsellem1994}, as implemented by \citet{Cappellari2002}, to obtain the stellar mass distribution from the $r$-band photometry (and thereby less affected by dust obscuration). This allows the photometry to be reproduced in detail, including ellipticity variations with radius, where appropriate. While the MGE method lacks any direct physical association with intrinsic properties of the galaxies (e.g.\ cores), it reproduces the observed surface brightness photometry more accurately than simpler parametrizations. 

The MGE procedure starts by determining the galaxy's average ellipticity ($\epsilon$), position angle (PA), and coordinates of the luminosity-weighted centre ($x_{\rm cen}, y_{\rm cen}$). The galaxy image is then divided into four quadrants, and photometric profiles are measured along sectors uniformly spaced in angle from the major axis to the minor axis. Surface brightness profiles from the four quadrants are averaged together, and each is then fitted as the sum of Gaussian components. The best-fitting MGE model surface brightness is then determined iteratively by comparison with the observed surface brightness, after having been convolved with the instrumental PSF (see detailed description in \citealt{Cappellari2002}). 

The PSF-convolved MGE-outputs are further described and presented in Appendix \ref{masking}, with a green line indicating the position of the slit from which the kinematic profiles \citep{Loubser2018} are derived on images of the nuclei of the BCGs. Our MGE fits extend beyond the effective radii ($R_{e}$) for all the BCGs modelled here (on average it extends to $\sim$3.2$R_{e}$), with the exception of Abell 68 for which it extends to $\sim$0.73$R_{e}$ (where $R_{e}$ is 41 kpc). Therefore in all cases, the MGE modelling extends well beyond the available kinematics (>15 kpc). We masked some foreground/background features in the images and a discussion and examples are presented in Appendix \ref{masking}. The MGE procedure uses instrumental units, and we convert the model to physical units in Section \ref{conversion}, and using the equations given in \citet{Cappellari2002}.

We emphasise that the MGE method is not limited to axisymmetric (oblate) galaxies. \citet{VandenBosch2008} choose to use MGE fitting for stellar surface densities to use in their triaxial orbit based models instead of fully non-parametric methods. They emphasised that MGE models can reproduce a large variety of densities, which appears realistic when projected along any viewing direction, including mass models with radially varying triaxiality, multiple photometric components and discs. This reliable reproduction has since been confirmed using simulated galaxies where the ground truth is known: \citet{Li2016}, in their article assessing the JAM method using the Illustris simulation, demonstrate that the MGE formalism can deal with generalized geometries including triaxial shapes. \citet{He2019} use massive galaxy clusters ($M_{200} > 5 \times 10^{14}$ M$_{\sun}$) from the Cluster-EAGLE hydrodynamic simulation, and also apply MGE to various central cluster galaxies (the majority of which can be classified as prolate shape), and report that MGE fits most within an error of 10 per cent. The MGE method can reproduce isophotal twists, whereas most other models of surface brightness e.g. concentric ellipsoids \citep{Contopoulos1956, Stark1977, Binney1985} or non-parametric methods \citep{Magorrian1999} fail to reproduce ellipticity variations and isophotal twists. 

We have compared our numerical values of the MGE parametrisation of the surface brightness of our BCGs (presented in Appendix \ref{masking}) to the $r$-band surface brightness profiles in \citet{Bildfell2008} and \citet{Bildfell_thesis} (independent measurements, not using MGE, from the same $r$-band images), and find agreement within 0.25 mag/arcsec$^{2}$. We also compared our profiles to the $g$-band profiles measured by \citet{Kluge2020} for the three BCGs we have in common with their sample, by accounting for the $g$-$r$ colour gradients from \citet{Bildfell_thesis}, and we find no significant differences.

\subsection{Conversion to physical units}
\label{conversion}

We use the distance to the BCG, the exposure times of the images, the imaging plate scale (arcsec/pixel, 0.206 for the $R$-filter and 0.187 for the $r{'}$-filter), the zero-point of the filter (in AB magnitudes), and the extinction, to convert each MGE model to physical units using equation (1) in \citet{Cappellari2002} and the standard photometry formulas \citep{Holtzman1995}. The surface brightness $\mu$ (in mag/arcsec$^{2}$) is converted to surface brightness density $I'$ (in L$_{\sun}$/pc$^2$) by adopting the absolute magnitude of the Sun in the $r{'}$-band as 4.64 and in $R$-band as 4.61 from \citet{Blanton2007}. 

This gives the distance-independent results $I'_{j}$, the dispersion $\sigma'_{j}$ (in arcsec) along the major $x'$-axis, and the flattening $q'_{j}$ for each Gaussian ($j$). The total luminosity of each Gaussian is
\begin{equation}
L_{j} = 2\pi I'_{j} \sigma_{j}^{'2} q'_{j}
\end{equation}
where the galaxy distance is used to convert $\sigma'_{j}$ to kpc. The total MGE surface brightness $\Sigma$ is:
\begin{equation}
\Sigma(x', y') = \sum_{j=1}^{N} I_{j}' \mathrm{exp} \bigg[ - \frac{1}{2 \sigma_{j}^{'2}} \bigg(x'^{2} + \frac{y'^{2}}{q_{j}'^{2}} \bigg) \bigg]
\end{equation}
where the model is composed of $N$ Gaussian components.

Starting from these values, the deprojection from surface density to intrinsic density is performed for the axisymmetric case in JAM (Section \ref{sec:dynmasses} for the cylindrically-aligned models, and Section \ref{spherical} for the spherically-aligned models). For a given inclination $i$, the MGE surface density can be deprojected analytically \citep{Monnet1992} to obtain the intrinsic stellar mass density. Although this deprojection is non-unique, it represents a reasonable choice, which produces realistic intrinsic densities, that resemble observed galaxies when projected from any inclination \citep{Cappellari2006}.  

\subsection{Influence of the point-spread function (PSF)}
\label{stellarpsf}

The ground-based measurement of any galaxy surface brightness profile will be distorted by the inherent limitations of atmospheric and detector resolution \citep{Saglia1993, Schombert2012}. To determine the PSF we measure the surface brightness profile of an unsaturated star close to the BCG (see e.g. \citealt{Hoekstra2007}). For our imaging data, the characteristic maximum scale of the PSF (FWHM) is 1$\arcsec$ (though most of the data have a sub-arcsecond PSF). This is taken into account in the MGE analysis through a PSF convolution as described in Appendix A of \citet{Cappellari2002}. We illustrate the sensitivity of our modelling results to the PSF in Figure \ref{table_PSF} in Appendix \ref{PSF}, and we find that the effect of the PSF on the measured stellar mass profiles is minimal. 

\subsection{Multiple nuclei and substructure}
\label{nuclei}

We also use the output from our surface brightness analysis to detect possible multiple nuclei and substructure in the core, especially at the locations where the longslit was placed. Steep contours that can be seen in the MGE images are most likely stellar objects in the line-of-sight. Because of the symmetry applied (along the major and minor axes) in the MGE algorithm, these small, single objects very rarely influence the Gaussian parameterisation significantly. There are however, four BCGs (shown in Appendix \ref{masking}), where the structures along the line-of-sight make it very difficult to accurately derive the stellar mass profile. Additionally, three BCGs, Abell 586, MS0440+02 and MS0906+11 (all of them in the CCCP sub-sample) have prominent, large multiple nuclei. Multiple nuclei are impossible to fit with a single set of Gaussians as used in the MGE algorithm. These three BCGs with clear multiple nuclei, as well as the four very problematic MGE fit cases described above, are therefore excluded from the dynamical mass modelling below. 


\section{CCCP and MENeaCS BCG Dynamical masses}
\label{sec:dynmasses}

In this Section, we use the results from the MGE analysis, in combination with the observed kinematic data, to infer the mass distribution of the BCGs. The Jeans Anisotropic Method (JAM) is a generalization of the axisymmetric Jeans formalism, which can be used to model the stellar kinematics of galaxies. For the cylindrically-aligned models, we assume a constant stellar mass-to-light ratio ($\Upsilon_{\star \rm DYN}$), and a velocity ellipsoid that is aligned with cylindrical coordinates ($R$, $z$) and characterized by the anisotropy parameter $\beta_{z} = 1 - (\sigma^{2}_{z} / \sigma^{2}_{R}$). The anisotropy parameter $\beta_{z}$ describes the flattening of the velocity dispersion ellipsoid in the vertical direction, with $\beta_{z} = 0$ corresponding to isotropy, $0 < \beta_{z} < 1$ corresponding to radial anisotropy and $\beta_{z} < 0$ corresponding to tangential anisotropy. Since the intrinsic shape of BCGs can be oblate, prolate or triaxial, as discussed in more detail below, we also use the axisymmetric Jeans equations under the assumption of an anisotropic (three-integral) velocity ellipsoid aligned with the spherical polar coordinate system \citep{Cappellari2020} in Section \ref{spherical}, and discuss how it affects the results obtained in this section. 

When solving the Jeans equations, one is left with two unknown parameters, the radial profiles of mass and anisotropy, in a single equation, leading to the mass-anisotropy degeneracy (see the review by \citealt{Courteau2014}). A promising approach is to use the kurtosis, $h_{4}$ \citep{Binney1982}, and we discuss our dynamical modelling best-fitting parameters together with the interpretation from our $h_{4}$ measurements in Section \ref{additionh4}. 

To infer BCG mass profiles from the observed luminosity distribution (as parametrised from the MGE models) and the kinematic profiles, we use the axisymmetric case of the JAM (adapted for our data). Dynamical modelling studies of BCGs are rare, and some studies suggest that BCGs are more typically triaxial or prolate \citep{Fasano2010}, but this approach has been used for BCGs before (see \citealt{Smith2017}), and it is a sensible first step before attempting more general but computationally intensive orbit-based methods \citep{VandenBosch2008}. The effect of isophote twisting observed in elliptical galaxies with increasing ellipticity is often used as an indicator of a triaxial shape \citep{Kormendy1996, Emsellem2007, Krajnovic2008}. However, as discussed in \citet{Li2018}, it is not straightforward to determine whether a galaxy is oblate or not (especially for slow rotators). There are both axisymmetric oblate spheroids and triaxial ellipsoids among the most massive early-type galaxies. \citet{Krajnovic2018} present stellar velocity maps of 25 massive early-type galaxies with 14 of them the BCGs in clusters richer than the Virgo cluster. These 14 BCGs can be classified as: five prolate (i.e. rotation around the major-axis suggestive of a triaxial or close to prolate intrinsic shape), four triaxial and five oblate. 

The best models for these triaxial objects are particle-based, but the triaxial approximation also contains degeneracies so that no unique solution can be obtained \citep{Rybicki1987}. When interested in global galaxy quantities or test the results of more general models, it is still useful to construct simpler and approximate models \citep{Cappellari2008, Cappellari2020}. The inherent limitations of our long-slit data does not justify more sophisticated modelling approaches. We refer the reader to \citet{Cappellari2006} and \citet{Cappellari2008} for the details and the solutions to the Jeans equations. We also adapted JAM to include the dark matter mass component as described in Section \ref{AddDM}.

Even though the rotational velocities for this sample of BCGs are negligible, we keep our analysis general and take the velocity dispersion ($\sigma$) and velocity ($V$) profiles from \citet{Loubser2018}, and compute the second moment of velocity $\nu_{\rm RMS}$ profile for each BCG. We assume symmetry about the minor axis and average the measurements on both sides of the galaxy centre (inversely weighted by the errors on $\nu_{\rm RMS}$). The errors on the $\nu_{\rm RMS}$ measurements are obtained from an error spectrum propagated through the data reduction process, and divided by a factor $\sqrt{N}$ where $N$ is the number of pixel rows added to form each combined bin in the kinematics profile.  

We describe the construction of the dynamical mass models, and all assumptions made, below. Similarly to Section \ref{stellarpsf}, the second velocity moment is convolved with the PSF (of the spectral observations) before making comparisons with the observed quantities (see \citealt{Cappellari2008}). We again find that the effect of the PSF is minimal on the measured dynamic mass profiles. 

\subsection{The angle of inclination}

Round, non-rotating, massive ellipticals of the kind we are studying are best fitted with an inclination of 90$\degr$ \footnote{This can also stem from the fact that edge-on axisymmetric models have more degrees of freedom than face-on models (where rotation disappears), which is not necessarily taken into account when the goodness-of-fit is calculated.}. This is in agreement with the fact that the vast majority of these galaxies, with flat nuclear surface-brightness profiles, always appear nearly round on the sky \citep{Cappellari2006, Fasano2010}. They cannot all be flat systems seen nearly face-on, as the observed fraction is too high \citep{Tremblay1996}.    

We therefore assume a configuration $i = 90 \degr$, but the MGE description for the stellar mass is consistent with inclinations as low as $i = 68 \degr$ (the limit arises from the highest-ellipticity Gaussian in the fit to the projected luminosity). Allowing inclination as a free parameter for their BCG, \citet{Smith2017} find that high inclinations ($i > 80 \degr$) are somewhat favoured, and that none of the parameters of interest have significant covariance with $i$ \citep{Smith2017}. Hence, no information is lost by including inclination as a fixed parameter (at $i = 90 \degr$) in the models discussed. Detailed tests for the effects of inclination are summarized in \citet{Smith2017}. Furthermore, \citet{VanDerMarel1991} noted that the $\Upsilon_{\star \rm DYN}$ derived from fitting Jeans models is only weakly dependent on inclination, due to the fact that the increased flattening of a model at low inclination is compensated by a decrease in the observed velocities, due to projection effects.


\subsection{Adding a central (black hole) mass component}
\label{blackhole}

To eliminate unnecessary free parameters, we do not fit for the mass of a supermassive black hole (BH) in the centre of the BCG in the dynamical modelling. The spatial resolution is also not sufficient to accurately constrain its value. We set the mass of the BH equal to that predicted by the $M_{\rm BH} - \sigma$ relation. We use the relation by \citet{McConnell2011} given by
\begin{equation}
\frac{M_{\rm BH}}{10^8\ \rm M_{\sun}} = 1.9 \bigg( \frac{\sigma}{200\ \rm km\ s^{-1}} \bigg)^{5.1}
\end{equation}
and we use the central velocity dispersion, $\sigma_{0}$, from \citet{Loubser2018}. The BH mass values used in the dynamical modelling are listed in Table \ref{properties}. 

\citet{Smith2017}, who include an unresolved central mass concentration in the form of an extra Gaussian with very small scale radius in their MGE formalism, suggest that this component could also represent stellar mass not reflected in the luminosity distribution, e.g.\ due to an increasingly heavy IMF towards the galaxy centre \citep{MartinNavarro2015, VanDokkum2017}. We add a central Gaussian component representing a super-massive black hole (with mass determined from the $M_{\rm BH} - \sigma$ relation) with a radius of influence of 0.2 arcsec from the centre, and we label this mass component $M_{\rm CEN}$. Several studies pointed out that BCGs follow a steeper $M_{\rm BH} - \sigma$ relation than other massive early-type galaxies (see discussion in \citealt{Mehrgan2019}). In Table \ref{BH} in Appendix \ref{sec:BH}, we test how sensitive the best-fitting parameters are to changes in the mass or radius of the black hole mass component, and we illustrate that the resulting changes in the best-fitting parameters are negligible, and therefore independent of the relation we use. For the BCG in Abell 68, which we use as an example, a black hole 10 times more massive than the black hole mass we used will give a change of 0.02 in $\beta_{z}$ and 0.03 in $\Upsilon_{\star \rm DYN}$.  


\subsection{Adding a dark matter halo mass component}
\label{AddDM}

We explicitly include a dark matter halo as a third mass component in our models. The radial range of our long-slit kinematic data is insufficient to allow us to constrain the radial profile of the halo. We rather attempt to probe what influence the mass of the dark halo has on the kinematic measurements (<20 kpc), assuming that the haloes follow the one-parameter density profile as described below. 

We assume that the halo is spherical and characterized by the two-parameter double power-law NFW profile \citep{Navarro1996}. We then adopt the approach introduced by \citet{Rix1997} and followed by e.g.\ \citet{Napolitano2005, Williams2009} to reduce the dark halo density profile to a function of a single parameter ($M_{\rm DM}$). The NFW-profile\footnote{As shown in Appendix \ref{Delta} and summarised in Section \ref{SumDM}, we find that our dynamical modelling is robust against the dark matter distribution or the value used for the concentration parameter within the radial range used here.}:

\begin{equation}
\rho_{\rm DM}(r) = \frac{\rho_{\rm s}}{(r/r_{\rm s})(1+r/r_{\rm s})^{2}}
\end{equation}
where $r_{\rm s}$ is the scale-length, can be rewritten as a function of $M_{\rm DM}$, the total dark matter mass inside $r_{200}$
\begin{equation}
\rho_{\rm DM}(r) = \frac{M_{\rm DM}}{4 \pi A(c_{200})} \frac{1}{r(r_{\rm s}+r)^{2}}
\end{equation}
and  
\begin{equation}
A(c_{200}) = \ln (1+c_{200}) - \frac{c_{200}}{1+c_{200}}. 
\end{equation}
Thus, the concentration parameter is $c_{200}=r_{200}/r_{\rm s}$.

It is known that halo concentration correlates with virial mass \citep{Bullock2001} and at $z \sim 0$, we use the approximation for the WMAP cosmology from \citet{Maccio2008}\footnote{The choice of cosmology has a negligible influence on our results.}:
\begin{equation}
\log c_{200} = 0.917 - 0.104 \log (M_{\rm DM}/[10^{12} h^{-1} \rm M_{\sun}]).
\end{equation} 
We note that this relationship is not only cosmology dependent, but also redshift dependent. However, we find that the observational errors on $M_{200}$ and $r_{200}$ (obtained from weak lensing results by \citet{Herbonnet2019}) are by far the dominant uncertainty, and the change in this relation from $z=0.3$ to $z=0$ is neglected\footnote{Also see \citet{Prada2012} and \citet{Klypin2016} for the negligible redshift evolution between $z=0.3$ and $z=0$, and \citet{Dutton2014} for a redshift dependent relation for Planck cosmology.}. 

Since we have additional constraints on the total mass in the cluster available, we approximate $M_{\rm DM}$ and $r_{200}$ from weak lensing observations \citep{Herbonnet2019} available for 23 of the 25 MENeaCS and CCCP clusters we model here. The values for $M_{200}$ obtained from weak lensing are for the total mass. Thus, $M_{\rm DM}=\alpha M_{200}$, where $\alpha = \Omega_{\rm M} /(\Omega_{\rm M} - \Omega_{\rm b})$. Assuming the baryon fraction within $r_{200}$, $f_{\rm b, 200}$, is equal to the cosmological baryon fraction, i.e.\ $f_{\rm b, 200} = \Omega_{\rm b} / \Omega_{\rm M}$ = 0.17, $M_{200} \sim 1.2 M_{\rm DM}$. This choice of the cosmological value has a negligible influence on our results (also see Appendix \ref{Delta}).

We perform a multi-Gaussian expansion, similar to the procedure used for the stellar mass profiles, of this one-dimensional, single parameter profile in units of M$_{\sun}$/pc$^{2}$ to allow us to easily include it in the total potential for which we derive model kinematics. We generate an MGE NFW profile that extends beyond the virial radius of the cluster, and optimise the number of Gaussians used to describe the profile in order to achieve an accurate fit to the NFW profile at very small radii. We derive the best-fitting stellar mass-to-light ratio ($\Upsilon_{\star \rm DYN}$) by using a constant value to scale (weight) only the stellar mass component (in units of L$_{\sun}$/pc$^{2}$) in the total mass model which consists of the three mass components: stellar, central and dark matter halo. 


\subsection{Summary of dynamical mass models}
\label{SumDM}

To summarise our methodology described in the previous Sections: we find the stellar mass density distribution of each MGE model by assuming axisymmetry and a constant stellar mass-to-light ratio, $\Upsilon_{\star \rm DYN}$. We add a central mass component for a supermassive black hole, and a dark halo that follows a spherically symmetric NFW profile and assumes the correlation between halo concentration and halo mass. 

Since the galaxy inclination $i$ is fixed at 90$\degr$, and the dark matter mass $M_{\rm DM}$ within $r_{200}$ approximated from weak lensing results, the resulting JAM models have two free parameters: (i) the anisotropy ($\beta_{z}$), and (ii) the stellar mass-to-light ratio ($\Upsilon_{\star \rm DYN}$). The mass model is used to calculate an estimate of the observed stellar kinematics, by solving the Jeans equations. This predicted quantity is then compared to the observed kinematic profile $\nu_{\rm RMS} = \sqrt{V^{2} + \sigma^{2}}$. The two parameters $\beta_{z}$, and $\Upsilon_{\star \rm DYN}$ are then adjusted until the predicted stellar kinematics best match the observations, placing constraints on those parameters. For this we use the reduced $\chi^{2}$ statistic, where the degree of freedom is the number of fitted datapoints (between five and seven) minus two (for free parameters).

For the total mass models using all three mass components, we derive the errors on the best-fitting parameters ($\beta_{z}$, and $\Upsilon_{\star \rm DYN}$ as given in Table \ref{DynModsTable}) by incorporating the 1$\sigma$ errors on the weak lensing masses, as well as an estimated error of $\pm$10 per cent on the calculated value for the concentration parameter. We show an example of the effect of these uncertainties on the best-fitting parameters in Appendix \ref{Delta}.

It is not currently possible to simultaneously fit the observed second moment of velocity $\nu_{\rm RMS}=\sqrt{V^{2} + \sigma^{2}}$, as well as $h_{4}$, as full modelling of the stellar orbits (Schwarzschild modelling), as well as the use of Integral Field Unit (IFU) kinematic maps instead of long-slit observations, would be necessary. 


\section{Results: best-fitting models}
\label{sec:results}

\begin{figure}
\centering
\includegraphics[width=1.1\linewidth]{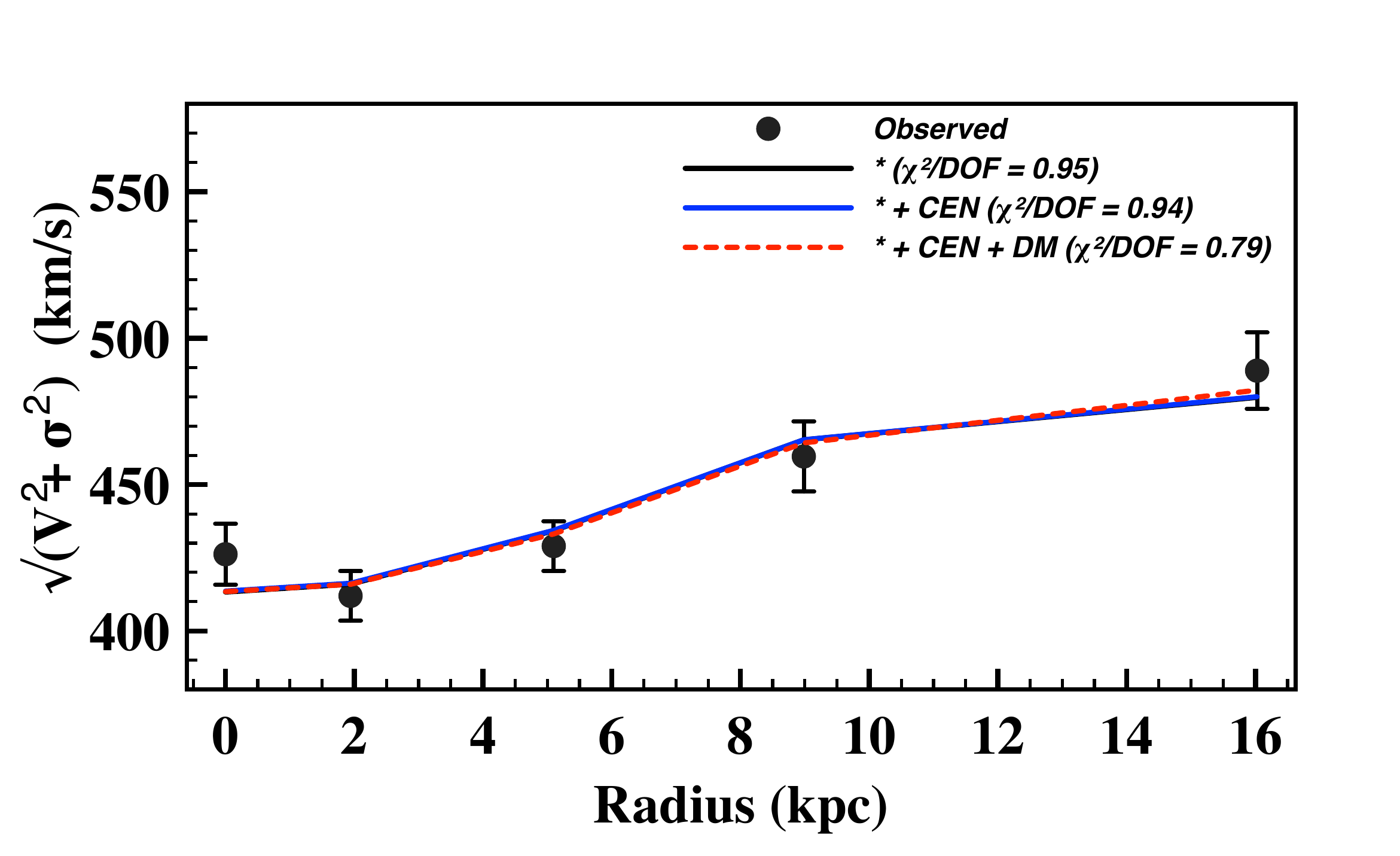} \\
\includegraphics[width=1.1\linewidth]{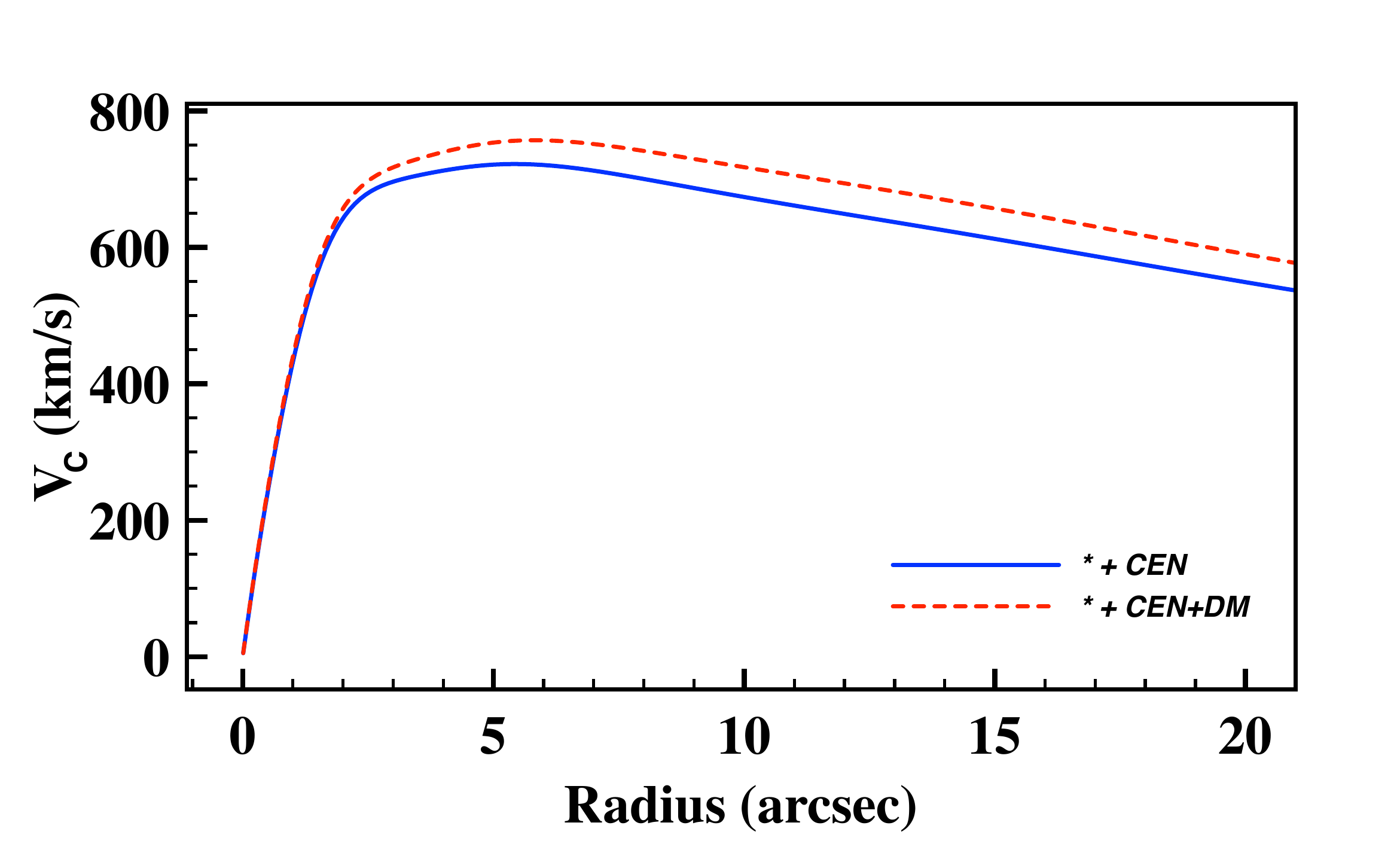}
\caption{Top: The averaged second moment of velocity ($\sqrt{V^{2} + \sigma^{2}}$) profile for Abell 2261 with three combinations of mass models shown. Bottom: Circular velocity curves for the total mass (red dashed), and for the central plus stellar mass only (blue solid).}
\label{A2261example}
\end{figure}

We find the best-fitting parameters for three different combinations of the mass models described above. We first find the best fit for just the stellar mass component (this fit is indicated with a $\star$ below), then the best fit for the stellar mass and a central mass component representing the BH ($\star$ + CEN), and lastly a fit with the stellar mass component, central mass component and dark matter halo mass component ($\star$ + CEN + DM). We plot the best-fitting dynamical mass profile for the BCG in Abell 2261 as an example in Figure \ref{A2261example}, and we tabulate the best-fitting parameters in Table \ref{DynModsTable} (for $\star$ + CEN + DM), and in Table \ref{AdditionalTable} in Appendix \ref{Additional} (for $\star$ and $\star$ + CEN for completeness). We plot the best-fitting dynamical mass profiles with the observed kinematics of all the BCGs in the Appendix \ref{DynMods}.

As mentioned in Section \ref{nuclei}, from our total sample of 32 BCGs we exclude Abell 586, 990, 1835, 2104, 2390, MS0440+02, MS0906+11 because of stellar substructure. We do the dynamical modelling of the remaining 25 BCGs, but for Abell 644 and 2319 we do not model the case with dark matter included ($\star$ + CEN + DM) since no weak lensing masses are available in \citet{Herbonnet2019}. For the interpretation of our dynamical modelling results, we removed Abell 2055, which had a best-fitting $\beta_{z} = -2.25$. This BCG is known to host a BL Lac point source that is the main contributor to the optical emission observed \citep{Green2017}. We also remove Abell 963, which also has an unusual $\beta_{z} = -1.13$. Observational indications suggest that typical massive elliptical galaxies are isotropic or radially anisotropic in their central regions (e.g. \citealt{Gerhard2001}; \citealt{Cappellari2007}), and theoretical models of galaxy formation predict that elliptical galaxies should be almost isotropic in the centre to radially biased in the outskirts \citep{Barnes1992, Hernquist1993, Nipoti2006}. As we discuss in Section \ref{OtherEs}, the exception is that of galaxies with rising velocity dispersion gradients; nevertheless strong tangential anisotropy is not expected. The stellar mass profile for this BCG is steeply declining with radius, while the observed $\nu_{\rm rms}$ profile has a very shallow slope and then increases with radius.

Our fits to the observed kinematics are restricted to within $<$20 kpc from the galaxy centre, where the stellar mass is expected to be the dominant contribution, yet there are still sufficiently many data points to constrain the inner shape of the kinematic profiles \citep{Loubser2018}. Having determined the best-fitting parameters for each galaxy, we can also compute circular velocities ($V_{\rm c}$) of the total mass and stellar plus central mass components using the numerical techniques described in \citet{Cappellari2002}. The circular velocity curves\footnote{For elliptical galaxies, where stars move in elongated non-closed orbits, the notion of circular velocity $V_{\rm c}$ has a purely formal sense of being the velocity of conventional test particles in circular orbits.} provide an intuitive measure of the mass enclosed as a function of radius, which is proportional to $V_{\rm c}^{2}$. We again show the results for Abell 2261 as an example in Figure \ref{A2261example}. 

We also use the circular velocity curves to derive the enclosed mass within a 15 kpc radius sphere, for the stellar plus central mass and total (dynamical) mass curves. The results are also listed in Table \ref{DynModsTable}, and the difference between the two values is the dark matter mass enclosed in a sphere with radius 15 kpc. On average we find that this is $8.2 \pm 2.6$ per cent of the total (dynamical mass). We also fitted $r^{1/4}$-laws to the $r$-band surface brightness profiles as described in Section \ref{FPlane}, deriving the effective radius, $R_{e}$. If we exclude  the BCG in Abell 2055, 990, 1835, 2104 and 2390, as described in Section \ref{FPlane}, then we find the average $R_{e} = 40.0 \pm 17.8$ kpc. Thus, for comparison with other samples of galaxies where the radial range is expressed as a factor of $R_{e}$, 15 kpc constitutes $\sim$ 0.38$R_{e}$, on average.

\subsection{Distribution of the best-fitting parameters}
\label{param_distrib}

We find that adding the central mass component, and the fixed dark matter halo mass component, do not give a significantly better or worse fit to the observed kinematics, than just the stellar mass component alone (see a typical example in Figure \ref{A2261example}). In general, increasing $\Upsilon_{\star \rm DYN}$ shifts the predicted $\nu_{\rm RMS}$ to higher velocities at all radii. Adding the $M_{\rm DM}$ mass component decreases $\Upsilon_{\star \rm DYN}$ on average by 8.3 $\pm$ 2.9 per cent over our kinematic range, and increases $\beta_{z}$ by on average 0.04 (Abell 2055 and Abell 963 not included). 

\begin{figure}
\centering
\includegraphics[scale=0.4]{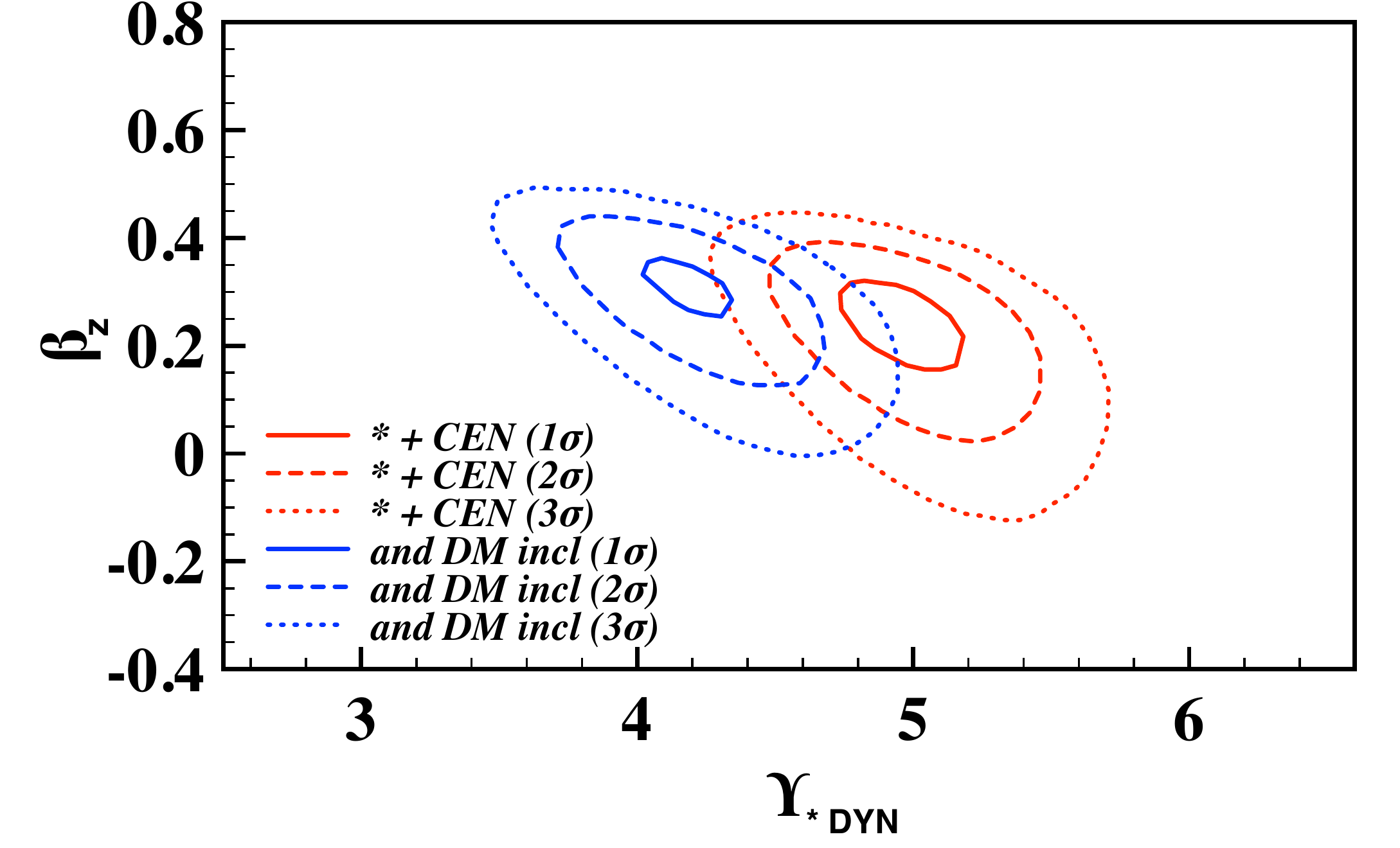}
\caption{We show the free parameters $\beta_{z}$ and $\Upsilon_{\star \rm DYN}$, with the 1$\sigma$ (68 per cent), 2$\sigma$ (95.4 per cent), 3$\sigma$ (99.73 per cent) confidence contours indicated, for the fit with just the stellar and central mass components i.e.\ assuming dark matter does not contribute to the total mass (red), and the stellar, central and dark matter components (blue) for the BCG in Abell 68 as an example. Adding the $M_{\rm DM}$ mass component decreases $\Upsilon_{\star \rm DYN}$ on average by 8.3 $\pm$ 2.9 per cent over our kinematic range, and increases $\beta_{z}$ by on average 0.04. This figure also illustrate that we can actually constrain $\beta_{z}$ and $\Upsilon_{\star \rm DYN}$ quite well separately, and that even without using $h_{4}$ we can get reasonably good constraints on the $\Upsilon_{\star \rm DYN}$.}
\label{A68SCD}
\end{figure}

We show values for the parameters $\beta_{z}$ and $\Upsilon_{\star \rm DYN}$, with the confidence levels indicated, for the fit with just the stellar and central mass components (red), and the stellar, central and dark matter components (blue, with a goodness-of-fit of $\chi^{2}/DOF$=2.99) for Abell 68 in Figure \ref{A68SCD} as an example, to illustrate the influence of the degeneracy between the two parameters. The figure also illustrate the change of the best-fitting parameters, for Abell 68, when the fixed dark matter component is included. Figure \ref{A68SCD} also illustrates that we can actually constrain $\beta_{z}$ and $\Upsilon_{\star \rm DYN}$ quite well separately, and that even without using $h_{4}$ we can get reasonably good constraints on the $\Upsilon_{\star \rm DYN}$.

\begin{figure}
\centering
\includegraphics[scale=0.52, trim = 0mm 5mm 0mm 35mm, clip]{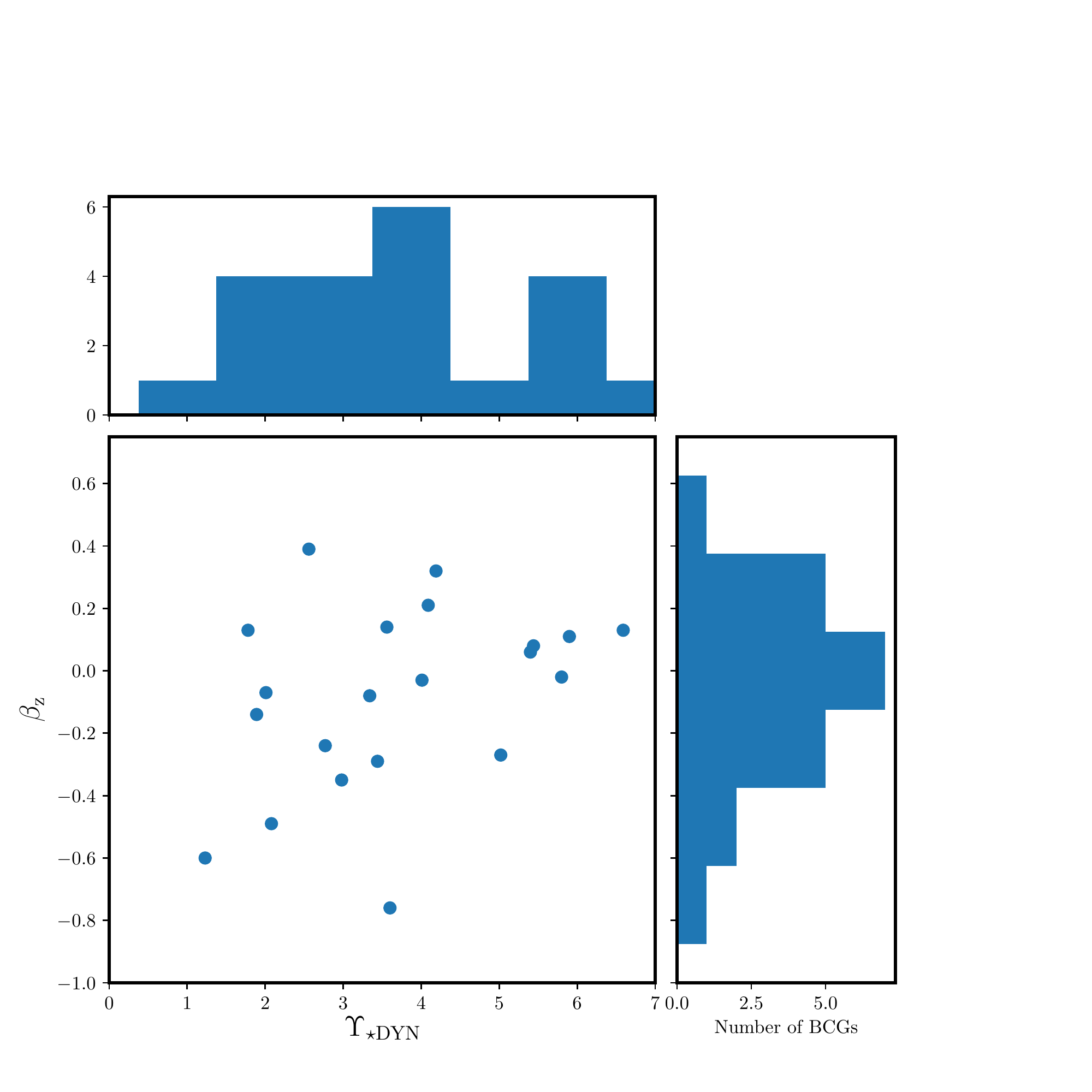}
\caption{We show the distributions of the best-fitting free parameters ($\Upsilon_{\star \rm DYN}$, $\beta_{z}$) for the mass model with all three components (* + CEN + DM). The dynamical modelling results of the BCGs in Abell 2055 and 963 are excluded. The typical error on $\beta_{z}$ is 0.02, and the typical error on $\Upsilon_{\star \rm DYN}$ is $< 0.2$ (see Table \ref{DynModsTable}). We find a wide range of best-fitting values for both parameters.}
\label{Histograms}
\end{figure}

\begin{figure}
\centering
\includegraphics[width=0.9\linewidth]{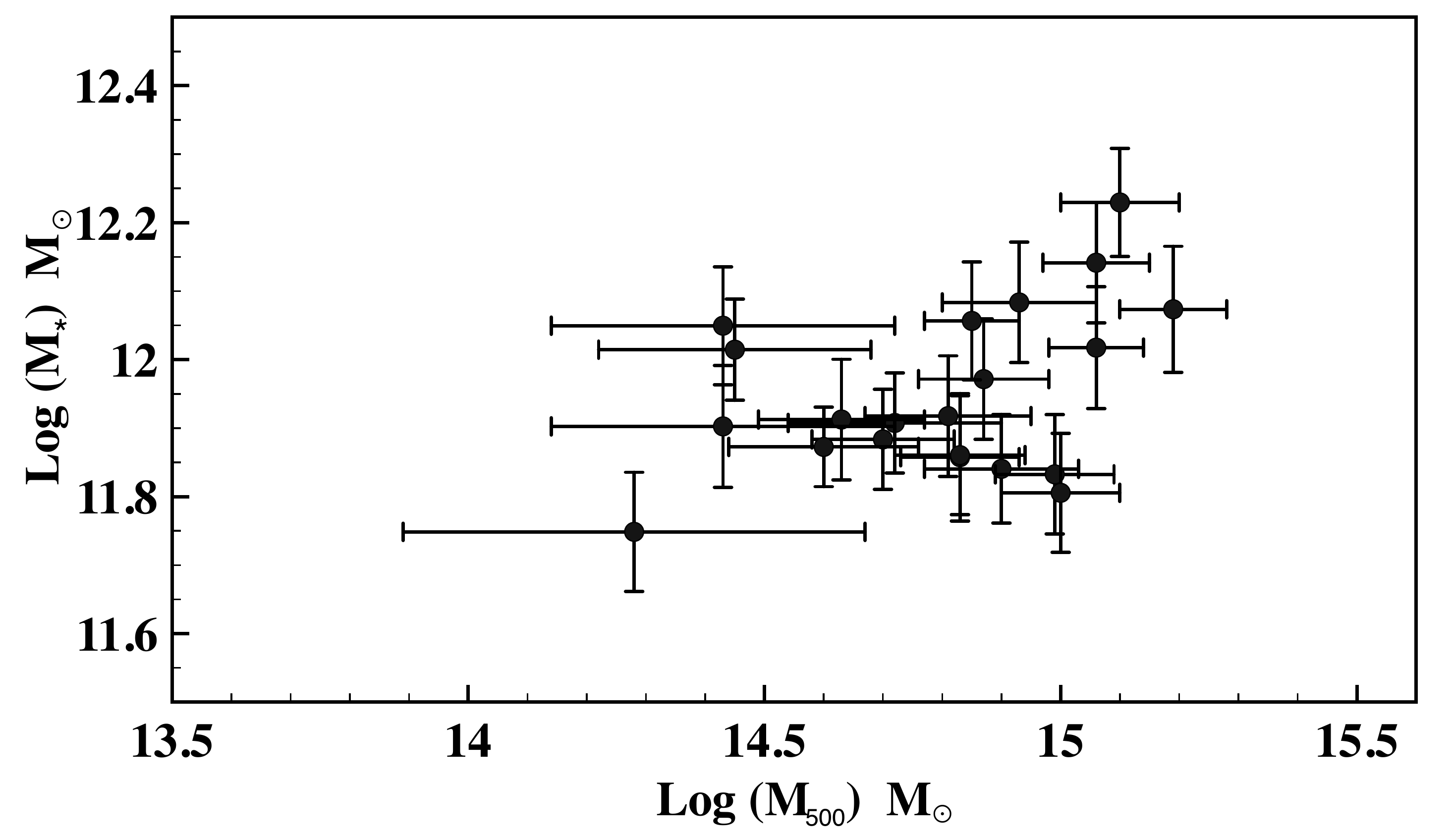} \\
\includegraphics[width=0.9\linewidth]{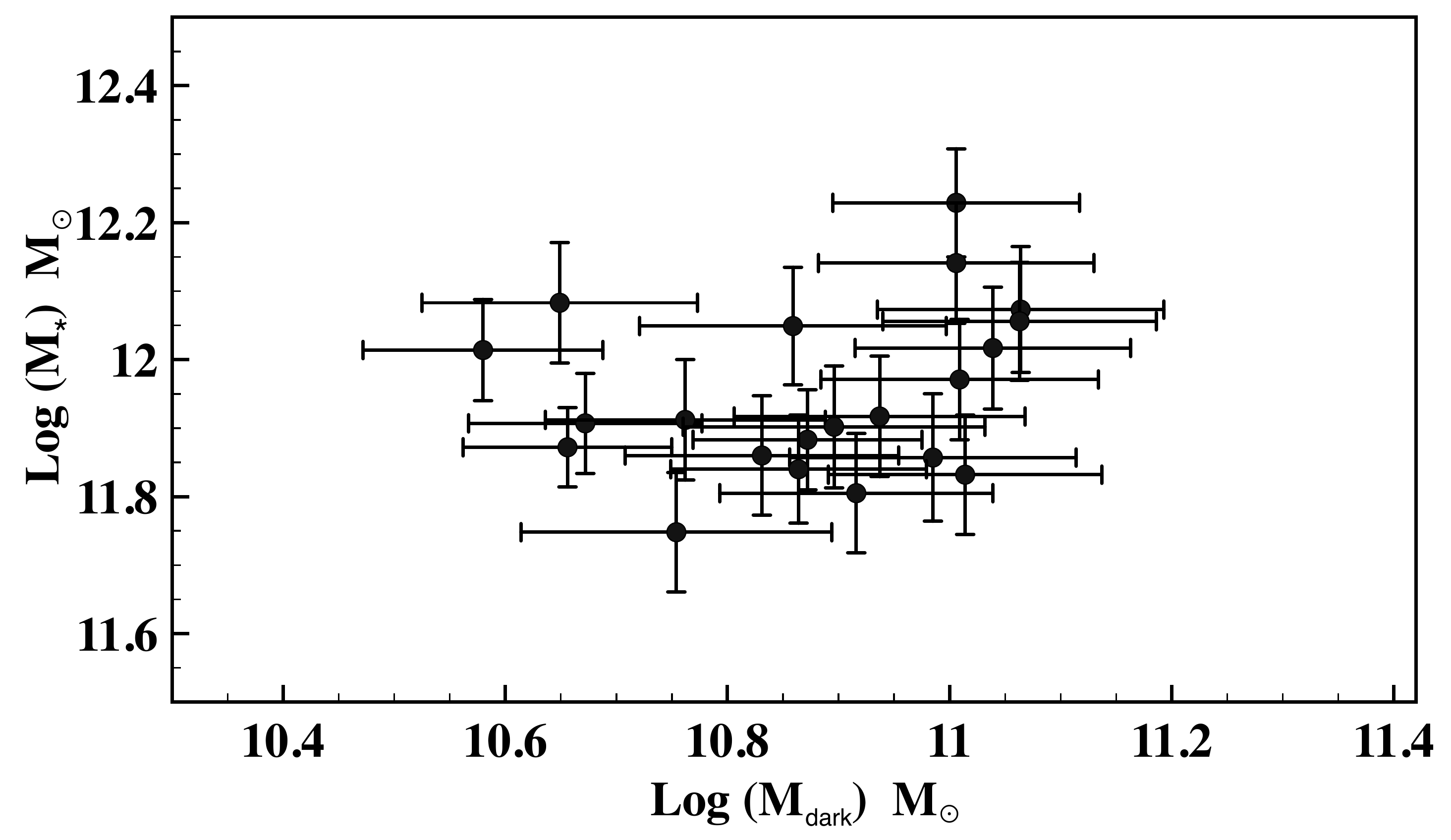} 
\caption{We show the enclosed stellar mass ($M_{\rm stellar}$) for a sphere with a radius of 15 kpc, against the halo mass at $M_{500}$ from weak lensing \citep{Herbonnet2019}, as well as against the enclosed dark matter mass ($M_{\rm dark}$) for a sphere with a radius of 15 kpc.}
\label{Mstellar_MDM}
\end{figure}

We further show the distributions of the best-fitting free parameters ($\Upsilon_{\star \rm DYN}$, $\beta_{z}$) for the mass model with all three components (* + CEN + DM) in Figure \ref{Histograms}. We also plot the enclosed stellar mass ($M_{\rm stellar}$) for a sphere with a radius of 15 kpc, against the dark matter halo mass at $M_{500}$ from \citet{Herbonnet2019} in Figure \ref{Mstellar_MDM}, as well as against the enclosed dark matter mass ($M_{\rm dark}$) for a sphere with a radius of 15 kpc. We find a weak correlation for the stellar mass against the halo mass.

\begin{table*}
\centering
\caption{Table with best-fit solution for the $\textbf{$\star$ + CEN + DM}$ scenario (the other two scenarios are presented in Table \ref{AdditionalTable} in Appendix \ref{Additional} for completeness). $\Upsilon_{\star \rm DYN}$ is constant with radius. For Abell 644 and Abell 2319, no weak lensing mass are available. See discussion in Section \ref{param_distrib} regarding the best-fitting solutions for Abell 963 and Abell 2055.}
\begin{tabular}{lrrrrrr}
\toprule
 & $\beta_{z}$  &  $\Upsilon_{\star \rm DYN}$  &  $\chi^{2}/DOF$ & $M_{\rm stellar}$ & $M_{\rm dyn}$ & $M_{\rm dark}$ \\  
 &  &  &  & $\times 10^{11}$ (M$_{\sun}$) & $\times 10^{11}$ (M$_{\sun}$) & $\times 10^{10}$ (M$_{\sun}$) \\         
\midrule
 Abell 68 & 0.32$\pm$0.02 & 4.19$\pm$0.19 & 2.99 &	 7.2$\pm$1.5 & 8.2$\pm$1.7 & 9.7$\pm$2.9 \\
 Abell 267 & 0.14$\pm$0.01 & 3.56$\pm$0.10 & 11.82 &	 8.1$\pm$1.4 & 8.5$\pm$1.5 & 4.7$\pm$1.1\\ 
 Abell 383 & --0.07$\pm$0.01 & 2.01$\pm$0.10  & 6.79 &	 11.2$\pm$2.2 & 11.9$\pm$2.9 & 7.2$\pm$2.3\\
 Abell 611 & --0.24$\pm$0.02 & 2.77$\pm$0.11 & 6.44 & 7.6$\pm$1.3 & 8.4$\pm$1.4 & 7.4$\pm$1.8\\
 Abell 646 & --0.27$\pm$0.02 & 5.02$\pm$0.10 & 0.33 & 10.3$\pm$1.8 & 10.7$\pm$1.9 & 3.8$\pm$0.9 \\
 Abell 754 & 0.39$\pm$0.02 & 2.56$\pm$0.08 & 0.96& 6.9$\pm$1.3 & 7.6$\pm$1.5 & 7.3$\pm$1.9 \\
 Abell 780 & --0.35$\pm$0.03 & 2.98$\pm$0.13 & 4.84 & 8.0$\pm$1.6 & 8.8$\pm$2.1 & 7.9$\pm$2.5\\
 Abell 963 & --1.13$\pm$0.03 & 2.12$\pm$0.06 & 10.12 & 32.9$\pm$6.7 & 34.4$\pm$7.0 & 15.5$\pm$4.5\\
 Abell 1650 & --0.02$\pm$0.02 & 5.80$\pm$0.11 & 3.53 & 7.2$\pm$1.4 & 7.9$\pm$1.6 & 6.8$\pm$1.9\\
 Abell 1689 & --0.76$\pm$0.02 & 3.60$\pm$0.09 & 7.51 & 11.8$\pm$2.5 & 13.0$\pm$2.7 & 11.6$\pm$3.4\\
 Abell 1763 & --0.14$\pm$0.02 & 1.89$\pm$0.06 & 2.47 & 12.1$\pm$2.5 & 12.9$\pm$2.6 & 4.5$\pm$1.3\\
 Abell 1795 & --0.08$\pm$0.02 & 3.34$\pm$0.09 & 4.81 & 6.8$\pm$1.4 & 7.8$\pm$1.6 & 10.3$\pm$2.9\\
 Abell 1942 & --0.60$\pm$0.02 & 1.23$\pm$0.05 & 4.08 & 9.4$\pm$1.9 & 10.4$\pm$2.1 & 10.2$\pm$2.9\\
 Abell 1991 & 0.06$\pm$0.02  & 5.40$\pm$0.27 & 6.34 & 5.6$\pm$1.1 & 6.2$\pm$1.6 & 5.7$\pm$1.8 \\
 Abell 2029 & 0.13$\pm$0.01 & 6.59$\pm$0.11 & 2.47 & 13.8$\pm$2.8 & 14.8$\pm$3.0 & 10.1$\pm$2.9\\
 Abell 2050 & --0.03$\pm$0.01 & 4.01$\pm$0.08 & 0.93 & 7.4$\pm$1.0 & 7.9$\pm$1.4 & 4.5$\pm$1.0 \\
 Abell 2055 & --2.25$\pm$0.35 & 1.27$\pm$0.05 & 2.80 & 5.0$\pm$0.9 & 5.0$\pm$2.2 & 0.1$\pm$0.1 \\
 Abell 2142 & 0.11$\pm$0.02 & 5.90$\pm$0.14 & 3.63 & 6.4$\pm$1.3 & 7.2$\pm$1.4 & 8.2$\pm$2.3 \\
 Abell 2259 & 0.21$\pm$0.02 & 4.09$\pm$0.12 & 3.20 & 8.2$\pm$1.7 & 8.7$\pm$1.8 & 5.8$\pm$1.7\\
 Abell 2261 & --0.29$\pm$0.02 & 3.44$\pm$0.06 & 0.79 & 16.9$\pm$3.1 & 18.0$\pm$3.2 & 10.1$\pm$2.6 \\
 Abell 2420 & 0.08$\pm$0.02 & 5.44$\pm$0.19 & 1.45 & 8.3$\pm$1.7 & 9.1$\pm$2.0 & 8.7$\pm$2.6\\
 Abell 2537 & --0.49$\pm$0.04 & 2.08$\pm$0.06 & 1.72 & 10.4$\pm$2.1 & 11.5$\pm$2.3 & 10.9$\pm$3.1\\
 MS1455+22 & 0.13$\pm$0.01 & 1.78$\pm$0.05 & 0.68 & 11.4$\pm$2.3 & 12.5$\pm$2.5 & 11.6$\pm$3.3 \\
\bottomrule
\end{tabular}
\label{DynModsTable}
\end{table*}

\subsubsection{BCGs with low $\Upsilon_{\star \rm DYN}$}

Figure \ref{Histograms} shows a spread in best-fitting $\Upsilon_{\star \rm DYN}$ between values $1<\Upsilon_{\star \rm DYN}<7$. We investigate all the BCGs for which the best-fitting $\Upsilon_{\star \rm DYN} < 3$ from the dynamical modelling (for the $\star$ + CEN + DM cases), since this is lower than what we expect for BCGs that are typically passively evolving. For Abell 383, 780, 2055 and MS1455+22, the BCGs have young stellar population components (see \citealt{Loubser2016} and Loubser et al., in prep), explaining the low stellar mass-to-light ratio. Similarly, Abell 611, 963 and 2537 have significant age gradients in the SSP-equivalent stellar population ages derived for the inner (0 -- 5 kpc) and outer (5 -- 15 kpc) apertures, and therefore also had more recent star formation in the centre (but not enough or recent enough to identify or constrain the younger stellar component, see discussion in \citealt{Loubser2016}). The same is true for Abell 1942, which has an SSP-equivalent stellar population age of $\sim$ 4 Gyr for both the inner and outer apertures. The two exceptions, that have no young stars and still have a lower stellar mass-to-light ratio, are Abell 754 ($\Upsilon_{\star \rm DYN} = 2.56$), and Abell 1763 ($\Upsilon_{\star \rm DYN}  = 1.89$) for which we find a (relatively) older SSP-equivalent stellar population age of $\sim 7.5$ Gyr. 

For the BCG in Abell 754 (PGC025714), our measurements of the central velocity dispersion, kinematic profile, and stellar populations agree very well with those made by \citet{Brough2007}, \citet{Spolaor2010}, and \citet{Groenewald2014} (from independent data and analysis). Our surface brightness profile derived from MGE agrees with that derived by \citet{Bildfell2008} and \citet{Bildfell_thesis} to within $\sim$0.2 mag. Abell 754 has the fourth lowest dynamical mass estimate for the central 15 kpc (see Table \ref{DynModsTable}), and a corresponding low central velocity dispersion (295 $\pm$ 14 km s$^{-1}$) compared to the average central velocity dispersion of the BCGs modelled here ($\langle \sigma_{0} \rangle = 324 \pm 3\ \rm km\ \rm s^{-1}$).

For the BCG in Abell 1763 (Leda2174167), a wide-angle tail radio galaxy, we find a very peaked surface brightness profile in our MGE analysis (see Appendix \ref{masking}), similar to what we typically find for the BCGs with young stellar components (e.g. see also Abell 383 and MS1455+22), but contrary to the surface brightness profile measured by found by \citet{Bildfell2008} and \citet{Bildfell_thesis}. We also find stellar population properties that agree with other evidence for no recent star formation by e.g.\ \citet{Crawford1999}, \citet{Hoffer2012}, and \citet{Rawle2012}. Abell 1763 does not have a low dynamical mass in the centre (see Table \ref{DynModsTable}, and a high central velocity dispersion of 362 $\pm$ 2 km s$^{-1}$), but it has one of the lowest contributions of dark matter mass in the centre (3.4 per cent) on account of its high stellar mass and brightness (M$_{K} = -27.33$ mag) compared to the average in our sample (M$_{K} = -26.52$ mag). 

As we show and discuss in Section \ref{spherical}, changing our cylindrically-aligned JAM models to spherically-aligned JAM models, results in the best-fitting $\Upsilon_{\star \rm DYN}$ parameters changing by up to $\sim$ 15 per cent (higher for BCGs with rising velocity dispersion profiles, and lower for BCGs with decreasing velocity dispersion profiles). This can possibly account for the low $\Upsilon_{\star \rm DYN}$ measured for Abell 1763, but will cause the $\Upsilon_{\star \rm DYN}$ for Abell 754 to be even lower.

\citet{Li2016} also assess the effectiveness of the (cylindrically-aligned) JAM-technique using cosmological hydrodynamic simulations from the Illustris project. They find that the enclosed total mass (within 2.5$R_{e}$, i.e. more than five times our radial range) is well constrained to within 10 per cent, but that there is a degeneracy between the stellar mass and dark matter mass components. For prolate galaxies, they determine that the JAM-recovered stellar mass is on average 18 per cent higher than the input values and the dark matter mass 22 per cent lower (and therefore an underestimation of the dark matter fraction). Interestingly, in a similar test performed using Schwarzschild modelling in \citet{Thomas2007a}, and applied to Coma galaxies in \citet{Thomas2007b}, they find the opposite. Their recovery accuracy of the total mass is three per cent for oblate galaxies and 20 per cent for prolate galaxies. In \citet{Thomas2007a}, all recovered stellar mass-to-light ratios are lower than the true values. Comparing different dynamical models (see Section \ref{spherical}) is necessary to give an estimate on any systematic over- or underestimation of $\Upsilon_{\star \rm DYN}$. 

\subsection{The correlation between stellar anisotropy ($\beta_{z}$) and velocity dispersion profiles}
\label{anisotropy}

\begin{figure}
\centering
   \includegraphics[scale=0.29]{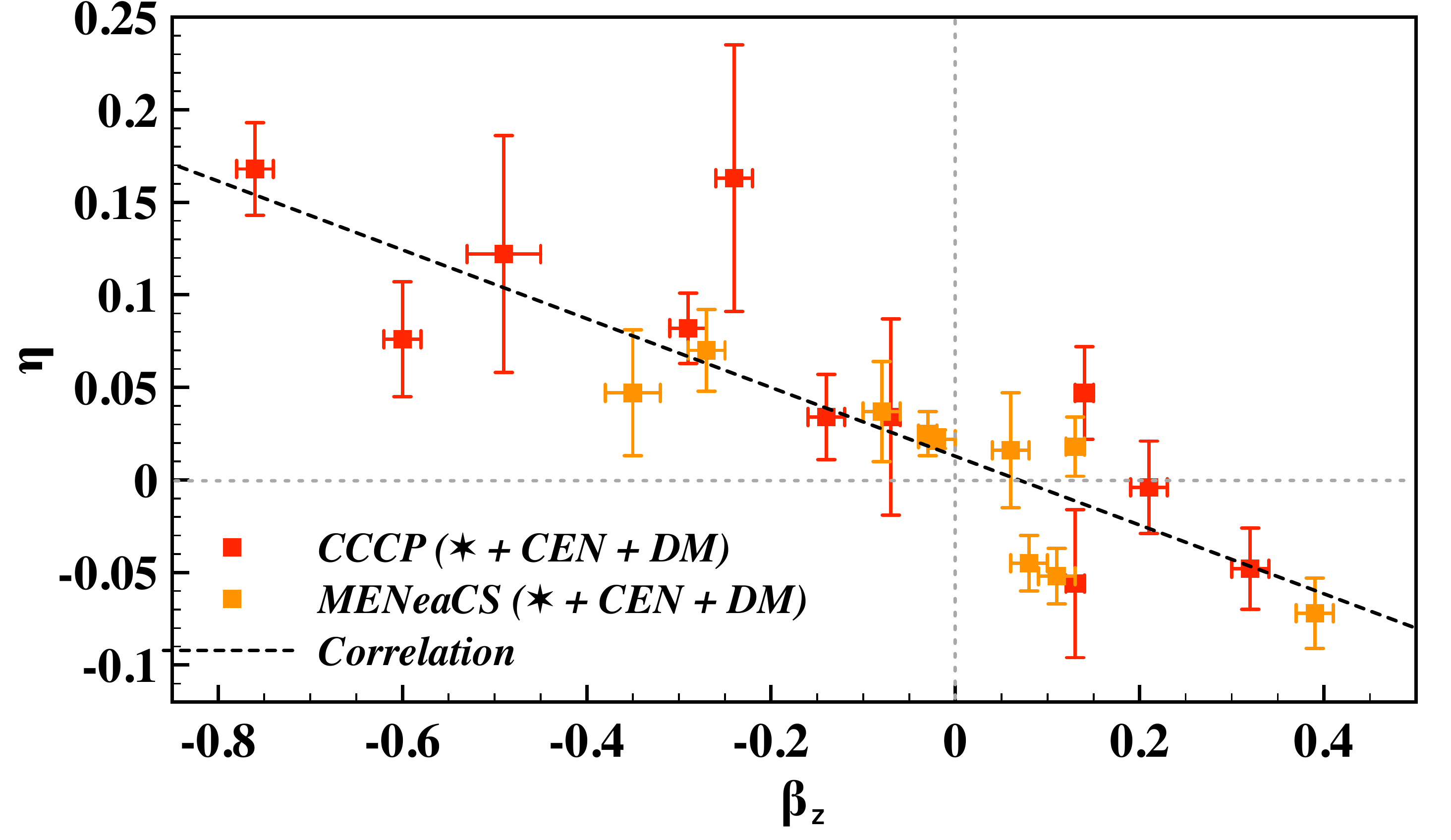}
   \caption{The correlation between best-fitting velocity anisotropy $\beta_{z}$ and velocity dispersion slope ($\eta$, as measured in \citet{Loubser2018}). We find a strong correlation between the two parameters, given by $\eta = (-0.185 \pm 0.028)\beta_{z} + (0.013 \pm 0.008)$, with an intrinsic scatter of 0.026, and correlation coefficient 0.929.}
\label{beta_sig}
\end{figure}

We plot our best-fitting $\beta_{z}$ parameters against the slope of the velocity dispersion profiles, $\eta$ from \citet{Loubser2018}, in Figure \ref{beta_sig}. The BCGs in Abell 2055 and Abell 963 are excluded as discussed in Section \ref{param_distrib}. Figure \ref{beta_sig} shows a strong correlation between best-fitting $\beta_{z}$ parameters (from the fit with all three mass components $\star$ + CEN + DM) against the slope of the velocity dispersion profiles. The large range of velocity anisotropy (see Figure \ref{Histograms}) that we derive corresponds to the diversity in the velocity dispersion profiles for our BCGs found in \citet{Loubser2018}. We again use the mixture model routine $\mathtt{linmix\_err}$ by \citet{Kelly2007} to fit the correlation taking errors on $\beta_{z}$ and $\eta$ into account. We find a slope = --0.186 $\pm$ 0.028, with an intrinsic scatter of 0.026, and correlation coefficient 0.929 (with a zero point = 0.013 $\pm$ 0.008).

Since $\beta_{z} = 1 - (\sigma^{2}_{z} / \sigma^{2}_{R})$, $0 < \beta_{z} < 1$ corresponds to radial anisotropy and $\beta_{z} < 0$ corresponds to tangential anisotropy. The trend in Figure \ref{beta_sig} is expected: For isothermal galaxies, the isotropic case is known to correspond to flat velocity dispersion profiles, whereas the radial anisotropy case corresponds to decreasing velocity dispersion gradients, and the tangential anisotropy case corresponds to rising velocity dispersion with radius \citep{Gerhard1993, Vandermarel1993, Rix1997, Gerhard1998, Thomas2007b}. The isotropic case is generally associated with $h_{4} = 0$, the radial anisotropic case with a positive $h_{4}$, and the tangential anisotropic case with a negative $h_{4}$, but all of our central measurements for $h_{4}$ are positive (see Figure \ref{h4_luminosity}). Positive $h_{4}$ values can also be expected if there are steep gradients in the circular velocity, regardless of isotropy/anisotropy \citep{Gerhard1993}, or stem from the superposition of two $\mathcal L_{\rm LOS}$, a narrower one of the stars feeling the potential of the galaxy, and a broader one, probing the potential of the cluster (as discussed in Section \ref{ICL}).

\subsubsection{Cylindrically- or Spherically-aligned Jeans Axisymmetric Models?}
\label{spherical}

As mentioned in Section \ref{sec:dynmasses}, there are some BCGs that can be classified as oblate, some as triaxial, and some as prolate \citep{Krajnovic2018}. We therefore also use the axisymmetric Jeans equations of stellar hydrodynamics under the assumption of an anisotropic (three-integral) velocity ellipsoid aligned with the spherical polar coordinate system \citep{Cappellari2020}. Comparisons between these two solutions (JAM with spherical polar coordinates, and JAM with cylindrical polar coordinates) allow for a robust assessment of the modelling results and dynamical parameters \citep{Cappellari2020}. Similar to the cylindrically-aligned JAM models (abbreviated throughout the paper as JAM), we adapt the spherically-aligned JAM models (abbreviated as JAM$_{\rm sph}$) for our purpose by modifying the models to fit our long-slit data, and to include a dark matter mass component derived from weak lensing results. 

\begin{figure*}
\centering
   \subfloat[Abell 646]{\includegraphics[scale=0.35]{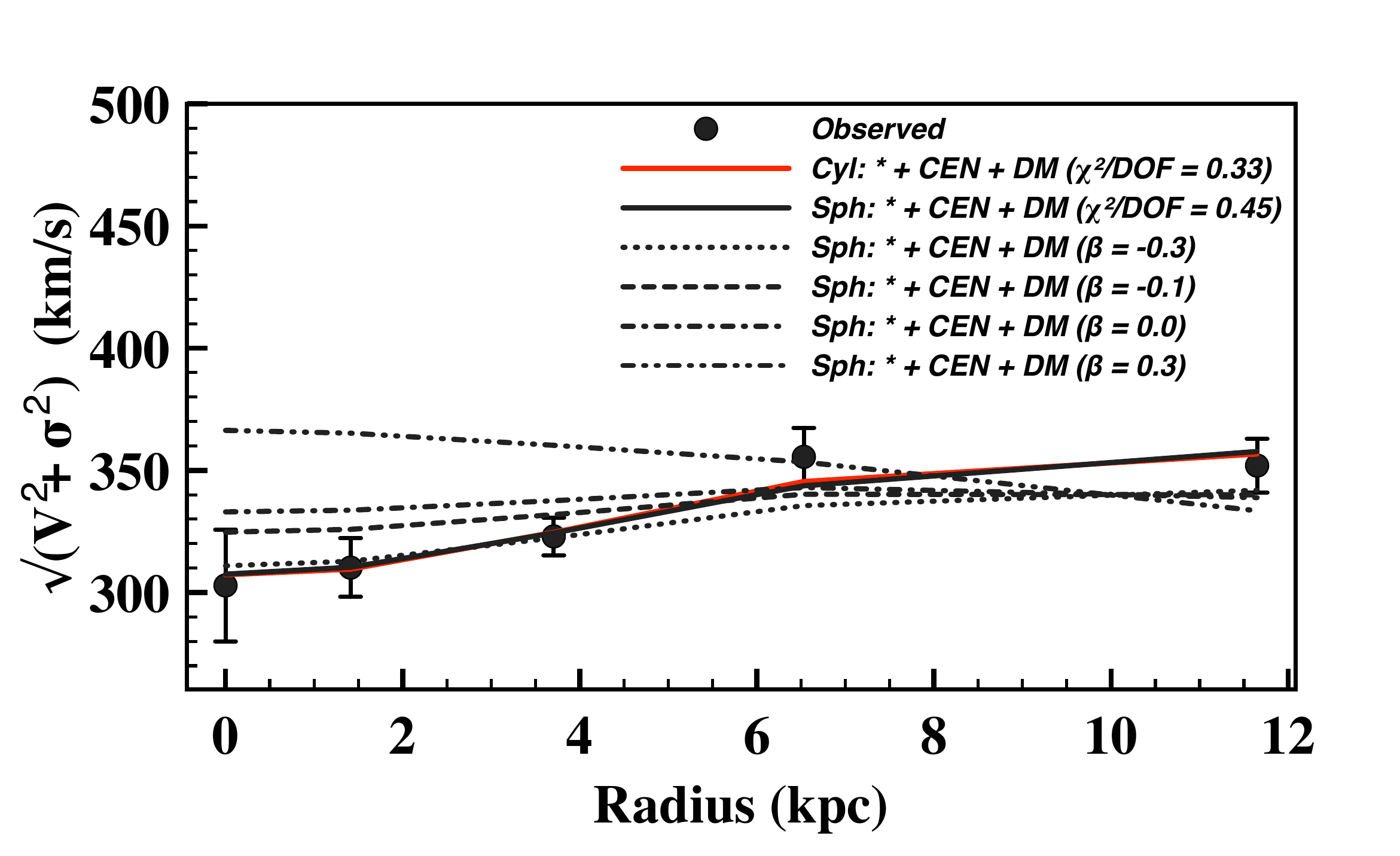}}
   \subfloat[Abell 2261]{\includegraphics[scale=0.35]{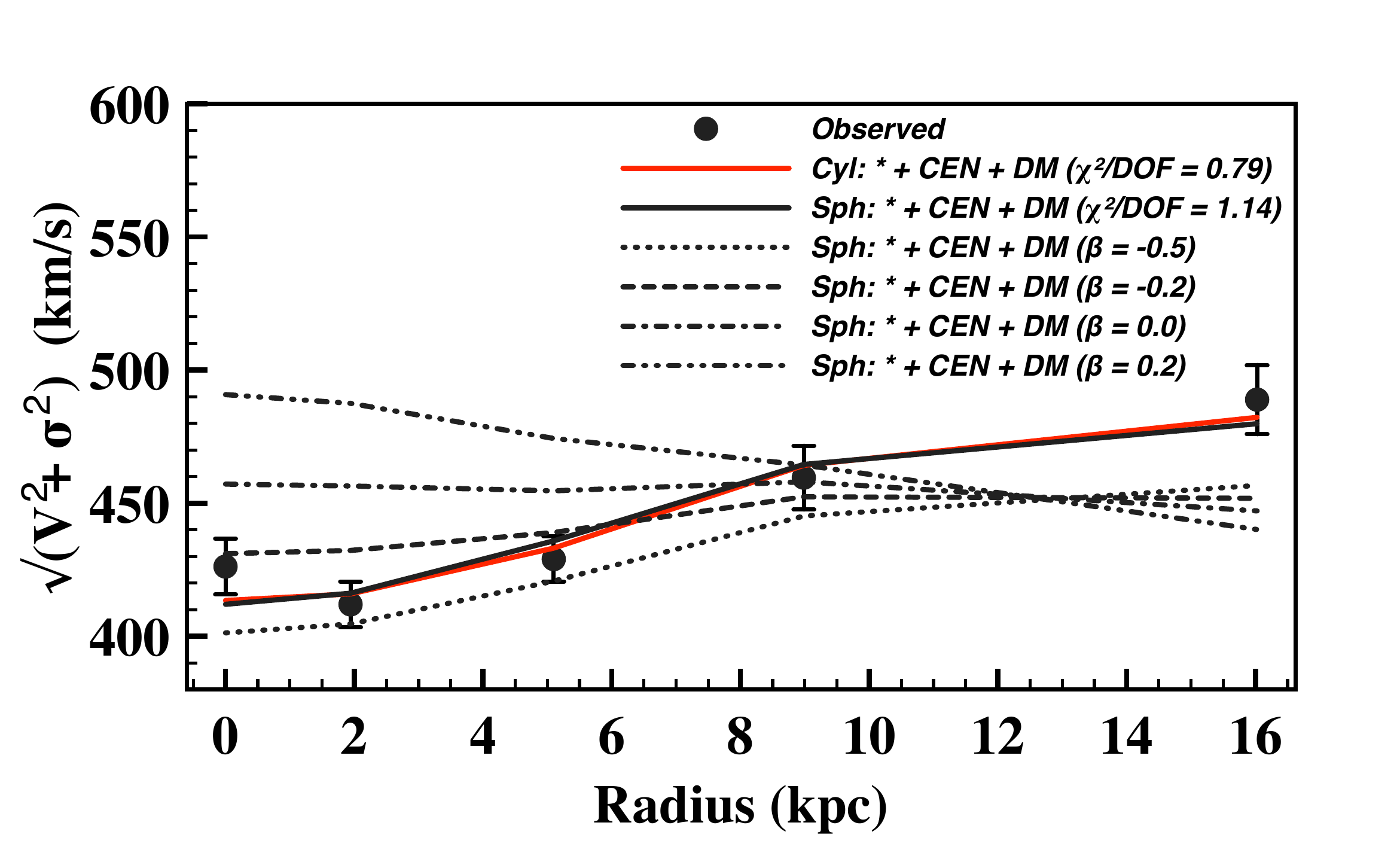}}\\
   \subfloat[Abell 68]{\includegraphics[scale=0.35]{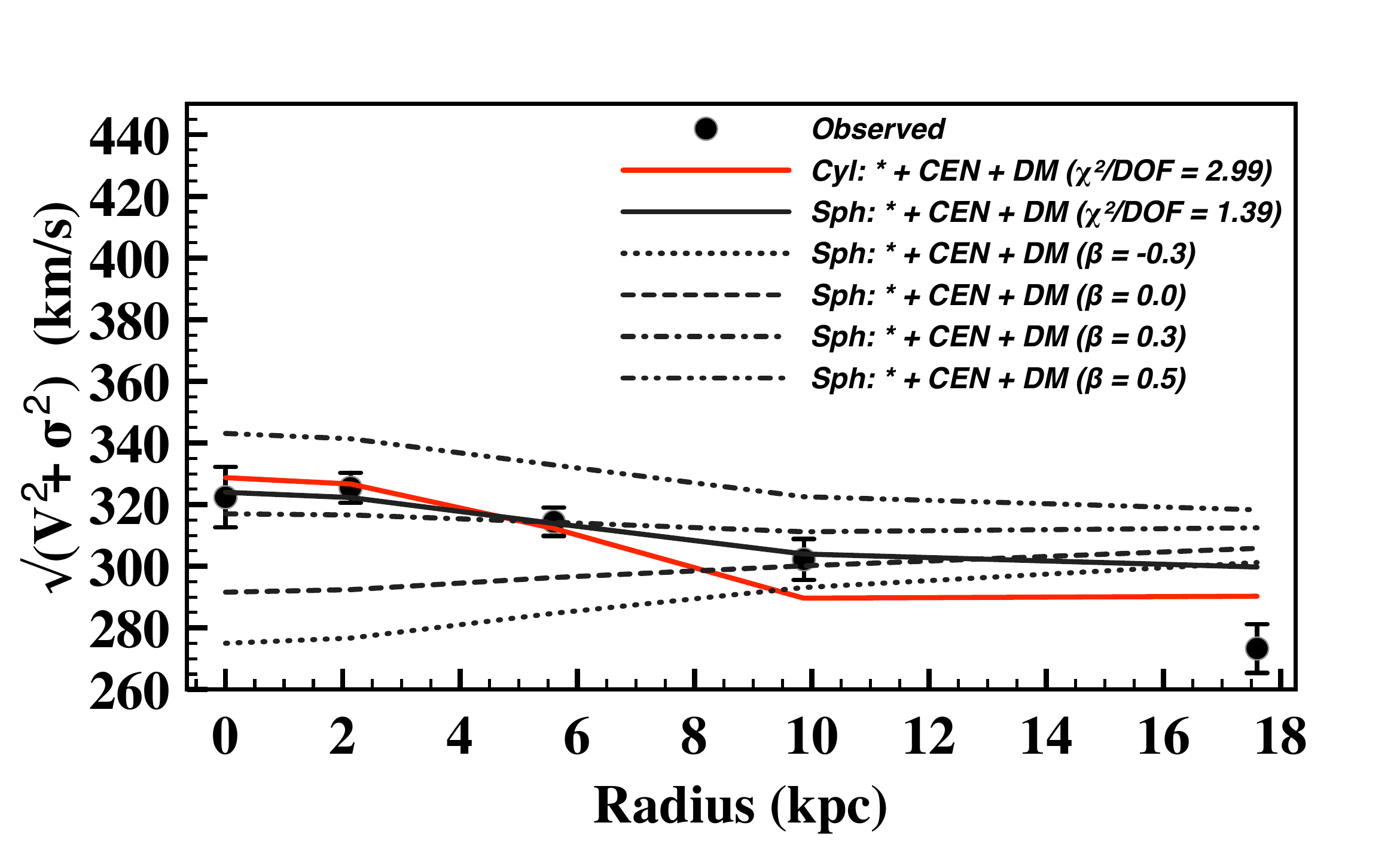}}
   \subfloat[MS1455+22]{\includegraphics[scale=0.35]{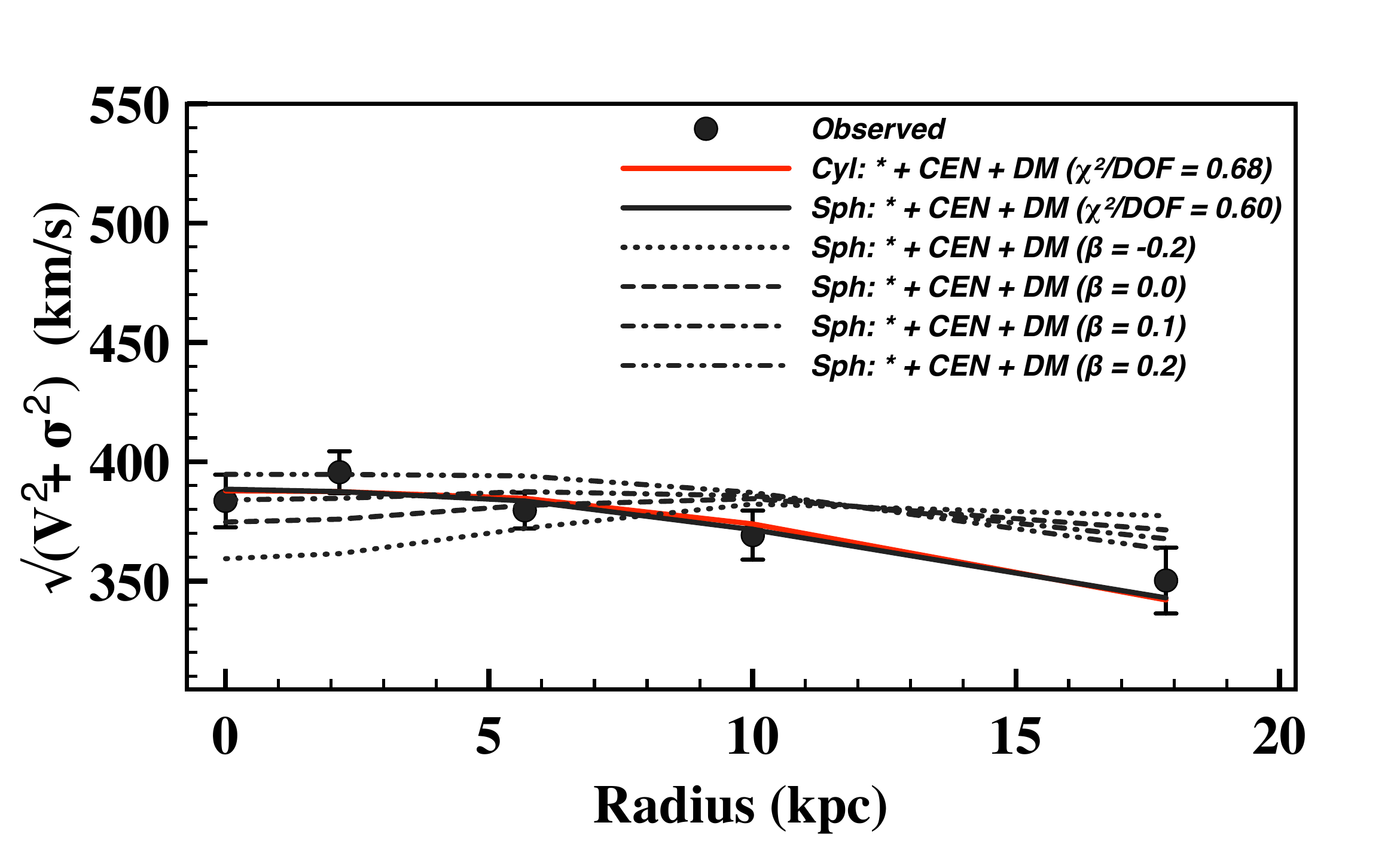}}\\
   \caption{In each case, the solid red line shows the best-fitting cylindrically-aligned JAM model, and the black solid line shows the best-fitting spherically-aligned JAM$_{\rm sph}$ model. In each case we also show JAM$_{\rm sph}$ models progressively changing $\beta$ from tangential anisotropy to radial anisotropy (where $\Upsilon_{\star \rm DYN}$ was kept constant at the best-fitting $\Upsilon_{\star \rm DYN}$ derived from the JAM cylindrically-aligned models) to show the change in the predicted $\nu_{\rm rms}$ slope as a function of $\beta$.}
\label{JAMspherical}
\end{figure*}

We show (Figure \ref{JAMspherical}) a direct comparison between the best-fitting JAM and JAM$_{\rm sph}$ models by using four BCGs (two with increasing velocity dispersion profiles and two with decreasing velocity dispersion profiles), whose characteristics are representative of our sample. We use mass models which include all three mass components (stellar, central and dark matter). For the JAM$_{\rm sph}$ models, the stellar anisotropy is defined as $\beta = 1 - (\sigma^{2}_{\theta} / \sigma^{2}_{R}$). We test whether the two solutions coincide in the isotropic limit ($\beta = \beta_{z} = 0$), and find negligible differences. We use four BCGs: Abell 646 and 2261 (rising velocity dispersion slopes) and Abell 68 and MS1455+22  (decreasing velocity dispersion slopes), where we found tangential and radial anisotropy, respectively, using cylindrically-aligned JAM models. In each case in Figure \ref{JAMspherical}, the solid red line shows the best-fitting cylindrically-aligned JAM model, and the black solid line shows the best-fitting spherically-aligned JAM$_{\rm sph}$ model. In each case we also show JAM$_{\rm sph}$ models progressively changing $\beta$ from tangential anisotropy to radial anisotropy (where $\Upsilon_{\star \rm DYN}$ was kept constant at the best-fitting $\Upsilon_{\star \rm DYN}$ derived from the JAM cylindrically-aligned models) to show the change in the predicted $\nu_{\rm rms}$ slope as a function of $\beta$. Since $\Upsilon_{\star \rm DYN}$ is a constant used to scale the stellar component contribution to the $\nu_{\rm rms}$ profile, modifying it moves the $\nu_{\rm rms}$ profile as a whole up or down. For these four examples, we also show the best-fitting parameters: $\beta_{z}$ and $\Upsilon_{\star \rm DYN}$ (JAM) vs $\beta$ and $\Upsilon_{\star \rm DYN}$ (JAM$_{\rm sph}$) in Table \ref{Jamtable}.

\begin{table*}
\caption{The best-fitting parameters: $\beta_{z}$ and $\Upsilon_{\star \rm DYN}$ (JAM) vs. $\beta$ and $\Upsilon_{\star \rm DYN}$ (JAM$_{\rm sph}$) for Abell 646, 2261, 68 and MS1455+22.}
\label{Jamtable}
\centering
\begin{tabular}{l|cc|cc}
\toprule
 & $\beta_{z}$ (JAM) &  $\Upsilon_{\star \rm DYN}$ (JAM) & $\beta$ (JAM$_{\rm sph}$) & $\Upsilon_{\star \rm DYN}$ (JAM$_{\rm sph}$) \\
$\star$ + CEN + DM  &  &  &  & \\
\midrule
Abell 646  & --0.27 $\pm$ 0.02 & 5.02 $\pm$ 0.10 & --0.60 $\pm$ 0.02 & 5.47 $\pm$ 0.14 \\
Abell 2261  & --0.29 $\pm$ 0.02 & 3.44 $\pm$ 0.06 & --0.61 $\pm$ 0.03 & 3.82 $\pm$ 0.10 \\
\midrule
Abell 68  & 0.32 $\pm$ 0.02 & 4.19 $\pm$ 0.19 & 0.52 $\pm$ 0.05 & 3.60 $\pm$ 0.10 \\
MS1455+22  & 0.13 $\pm$ 0.01 & 1.78 $\pm$ 0.05 & 0.30  $\pm$ 0.02  & 1.60 $\pm$ 0.03  \\
\bottomrule
\end{tabular}
\end{table*}

\bigskip

We find:

\begin{enumerate}
\item There are small differences in the $\chi^{2}$ (shown in Figure \ref{JAMspherical}) of the best-fitting JAM and JAM$_{\rm sph}$ models with neither model being significantly, or consistently, better or worse than the other.
\item For the JAM$_{\rm sph}$ models, we still find tangential anisotropy ($\beta < 0$) for BCGs with rising velocity dispersion profiles and radial anisotropy ($\beta > 0$) for decreasing velocity dispersion profiles. The (non-solid) black lines in Figure \ref{JAMspherical} illustrate how the $\nu_{\rm rms}$ slope change from increasing to decreasing corresponding to $\beta$ changing from tangential to radial anisotropy. 
\item As emphasised by \citet{Cappellari2020}, JAM$_{\rm sph}$ is characterized by a relative insensitivity of the model predictions to anisotropy. Our findings are in agreement, with the $\beta$ values being more extreme than $\beta_{z}$ (i.e. a bigger change from $\beta = 0$ is required to change the velocity dispersion slope to best fit the observed kinematics). The correlation of velocity dispersion slope with velocity anisotropy is therefore shallower for $\beta$ (JAM$_{\rm sph}$) than $\beta_{z}$ (JAM).
\item Corresponding to this systematic change in velocity anisotropy in JAM$_{\rm sph}$, there is a systematic change in best-fitting $\Upsilon_{\star \rm DYN}$, with $\Upsilon_{\star \rm DYN}$ being lower for decreasing $\nu_{\rm rms}$ profile BCGs (i.e. radial anisotropy, where $\beta$ is positive), and $\Upsilon_{\star \rm DYN}$ being higher for increasing $\nu_{\rm rms}$ profile BCGs (i.e. tangential anisotropy, where $\beta$ is negative). These changes are larger than the statistical error on the parameters. For example, the value of best-fitting velocity anisotropy can typically double (from JAM), corresponding to a change of ~10 to 15 per cent in best-fitting $\Upsilon_{\star \rm DYN}$.
\end{enumerate}

As discussed in Section 8.5 of \citet{Cappellari2020}, for external galaxies there is no straightforward answer regarding which model is preferable to use. The recommendation is not to favour one over the other one, but instead to use the two different assumptions on the alignment of the velocity ellipsoid to assess the sensitivity of the model results to the model assumptions.

\subsubsection{Contribution from Intracluster Light (ICL)}
\label{ICL}

\citet{Newman2013a}, who found rising velocity dispersion profiles for all seven of their BCGs, argue that the rising dispersions are not an artefact of the orbital distribution of the stars but reflect the genuine dynamical influence of the cluster potential. \citet{Bender2015} also argue that the increasing velocity profile and positive $h_{4}$ of the BCG NGC6166 are the result of the superposition of a galaxy and a cluster component in projection. As larger distances from the centre are probed, the cluster component becomes more important (in projection) and therefore the measured velocity dispersion increases. Therefore, the correlation between $\beta_{z}$ and $\eta$ shown in Figure \ref{beta_sig} could be driven by a sequence of decreasing importance of intracluster contamination: the observed steeply increasing velocity dispersions probe more the kinematics of the intracluster light that feels the cluster potential, as in the case of NGC6166.

To illustrate this using NGC6166, \citet{Bender2015} assume that the galaxy has a \citet{Sersic1968} brightness profile and a constant velocity dispersion of 300 km s$^{-1}$ at all radii, and that the halo also has a Sersic brightness profile and a constant velocity dispersion of 865 km s$^{-1}$ at all radii. This is an oversimplification, but should approximately fit the rising velocity dispersion profile. \citet{Bender2015} demonstrate that it fails, as the dispersion profile does not increase quickly enough outward. Modifying the assumed inner and outer dispersions, or using a Sersic-exponential decomposition for the surface brightness profiles did not help. They find that only using smaller Sersic indices for both components can account for the full rising velocity dispersion profile but is then inconsistent with the photometric profile. As a result, \citet{Bender2015} conclude that to explain the increasing velocity dispersion profile of their NGC6166, this contribution from the cluster component is not enough and tangential anisotropy is also needed.
  
We therefore test whether a two component stellar model (one component for the central galaxy with the galaxy velocity dispersion, and one component for the cluster halo/ICL with the cluster velocity dispersion) can account for the rising velocity dispersion profiles and whether tangential anisotropy is still needed. We assume that the components have independent Gaussian line-of-sight velocity distributions ($\mathcal L_{\rm LOS}$), and that the ICL is a dynamically hot component that is kinematically controlled by the gravitational potential of the cluster, i.e., unbound from the BCG. For this simplified scenario, 10 per cent of the light at 15 kpc needs to come from a 900 km s$^{-1}$ dispersion cluster component, and 90 per cent from a 300 km s$^{-1}$ dispersion galaxy component to cause the velocity dispersion to increase to 400 km s$^{-1}$ at 15 kpc (from 300 km s$^{-1}$). 

In photometric decompositions e.g. in \citet{Bender2015}, the outer component is sometimes interpreted as the ICL, assumed to be photometrically distinct. We note that in our photometric analysis, a single $R^{1/n}$ function is, in general, a good fit for our BCGs and it is not possible to physically distinguish an outer component. Many BCGs can not be decomposed into two photometric components. \citet{Kluge2020} present observations of 170 ($z<0.08$) BCGs. They found that 71 per cent of the BCG+ICL systems have surface brightness profiles that are well described by a single Sersic function whereas only 29 per cent require a double Sersic function to obtain a good fit. This is not uncommon, as the transition between inner and outer Sersic component is smooth so that any photometric decomposition is strongly degenerate \citep{Bender2015}.

Furthermore, for the BCGs that can be decomposed, there is currently no consensus of the contribution of the ICL to the integrated light. For NGC6166 \citep{Bender2015}, the ICL which they find to be more connected to the cluster than to the central galaxy, starts to dominate at $\sim$ 68 kpc. \citet{Gonzalez2005} present a detailed analysis of the surface brightness distribution of the BCG in each of 24 galaxy clusters at $0.03 < z < 0.13$. They use two-component profile fitting to model the surface brightness out to 300 kpc for each BCG, comparing $R^{1/4}$ \citep{deVaucouleurs1948}, $R^{1/n}$ and double $R^{1/4}$ models. They find that their envelope-to-total flux ratios (within 300 kpc) is around 0.9 but can be as low as 0.4 (their figure 7). \citet{Zibetti2005} find lower values for the flux contribution of the envelope component. They analyse the spatial distribution and colour of the ICL in 683 clusters of galaxies between $z = 0.2$ and 0.3, selected from the Sloan Digital Sky Survey (SDSS-DR1). They find that the ICL contributes 10.9 per cent to the total cluster light and the central galaxy contributes 21.9 per cent. This is equivalent to an envelope-to-total flux ratio of $\sim$ 33$\pm$6 per cent (in 500 kpc, using an averaged surface brightness profile). However, as pointed out in \citet{Kluge2020}, \citet{Zibetti2005} fit only one analytic function to the inner light profile and calculate the ICL as the excess light above it, thus excluding a contribution from an outer profile to the inner regions, resulting in a lower ICL fraction. An overview of the derived ICL fractions and the limiting depths of various BCG photometric surveys can be found in Table 3 of \citet{Kluge2020}, and it illustrates the large intrinsic scatter of photometrically determined ICL fractions.

However, some studies suggest that the outer photometric component is not the ICL component. \citet{Kluge2020} suggest that the outer Sersic component is unlikely to trace the dynamically hot ICL since BCG+ICL systems grow at present epoch predominantly in their outskirts. This is supported by results from numerical simulations. In simulations the BCG+ICL system is decomposed by fitting a double Maxwell distribution to the particle velocities. They find that the component with the higher characteristic velocity does not correlate with the ``photometrically" determined ICL \citep{Puchwein2010, Rudick2011, Cui2014, Remus2017}. \citet{Kluge2020} speculate that the two-component structure of the light profiles might be nothing more than a result of the recent accretion events and a photometric decomposition into two Sersic functions is likely to be unphysical.

In summary, for an ICL stellar envelope component more connected to the cluster than to the galaxy, the velocity dispersion gradients (Figure \ref{beta_sig}) can be a sequence of decreasing ICL contribution at 15 kpc. It is unlikely that it contributes approximately 10 per cent to the integrated light at 15 kpc, and that is solely responsible for the increasing velocity dispersion profile. However, due to the fact that our BCGs can be fit with a single Sersic function, and that a photometrically decomposed outer component might not represent the ICL velocity component, the possibility that 10 per cent of the light at 15 kpc can come from a cluster component can not be excluded from our data. 

\subsubsection{Comparison to other elliptical galaxies}
\label{OtherEs}

Figure \ref{beta_sig} shows the range of velocity anisotropy that we find for our BCGs, including $\beta_{z} < 0$ for the BCGs with rising velocity dispersion profiles. Using spherical models\footnote{\citet{Kronawitter2000} and \citet{Gerhard2001} use spherical models, where velocity anisotropy is defined as $\beta = 1 - (\sigma^{2}_{\phi} / \sigma^{2}_{R})$. The relation between $\beta$, and $\beta_{z}$ as used for our axisymmetric (cylindrically-aligned) models, is given in \citet{Cappellari2007}. For the axisymmetric (spherically-aligned) models in Section \ref{spherical}, $\beta = 1 - (\sigma^{2}_{\theta} / \sigma^{2}_{R}$).}, \citet{Kronawitter2000} and \citet{Gerhard2001} studied the dynamics of 21 luminous, slowly rotating, mostly round elliptical galaxies and found, on average, mild radial anisotropy with $\beta = 1 - (\sigma^{2}_{\phi} / \sigma^{2}_{R}) \sim 0$ to 0.35, but only one of their ellipticals show an increasing velocity dispersion profile over the radial range that we investigate here. They found two exceptions, NGC4486B and NGC4636, where $\beta$ is between --0.4 and --0.6. The first, NGC4486B, is a close companion of M87 and tidally disturbed, although it also suffers from poorer quality data than the other galaxies. The second, NGC4636, shows a rising velocity dispersion gradient. \citet{Cappellari2007}, in their SAURON data, also found two exceptions ($\beta_{z} < 0$) in the elliptical galaxies NGC4473 and NGC4550, both with increasing velocity dispersion gradients along the major axis. 

In summary, the norm for other elliptical galaxies is to have radial anisotropy, but there are known exceptions with tangential anisotropy, and those often show increasing velocity dispersion gradients. The 12 BCGs in our sample with rising velocity dispersion profiles all have $\beta_{z} <0$ (Figure \ref{beta_sig}).

For completeness, we also investigate the effect of a non-constant $\Upsilon_{\star \rm DYN}$ or $\beta_{z}$, and whether the rising velocity dispersion can be due to variable parameters instead of a tangential stellar velocity anisotropy or a significant contribution from the ICL, in Section \ref{ML_Dyn} and \ref{additionh4}.

\subsection{A variable stellar mass-to-light ratio ($\Upsilon_{\star \rm DYN}$) or anisotropy ($\beta_{z}$)?}
\label{ML_Dyn}

Several BCGs have velocity dispersion profiles that first decrease with radius before rising again (e.g. Abell 267, 383, 611, 644, 754, 1991, 2029 and 2420 as shown in Appendix \ref{DynMods}). The parameter $\Upsilon_{\star \rm DYN}$ is the constant stellar mass-to-light ratio that best fits the observed kinematics over our (short) radial range. We now explore whether a variable $\Upsilon_{\star \rm DYN}(r)$ or $\beta_{z}(r)$ (variable with radius along the major axis) can explain the kinematic profiles of these BCGs. To illustrate this, we use Abell 2029 as an example.  

\begin{figure}
\centering
     \subfloat{\includegraphics[scale=0.32]{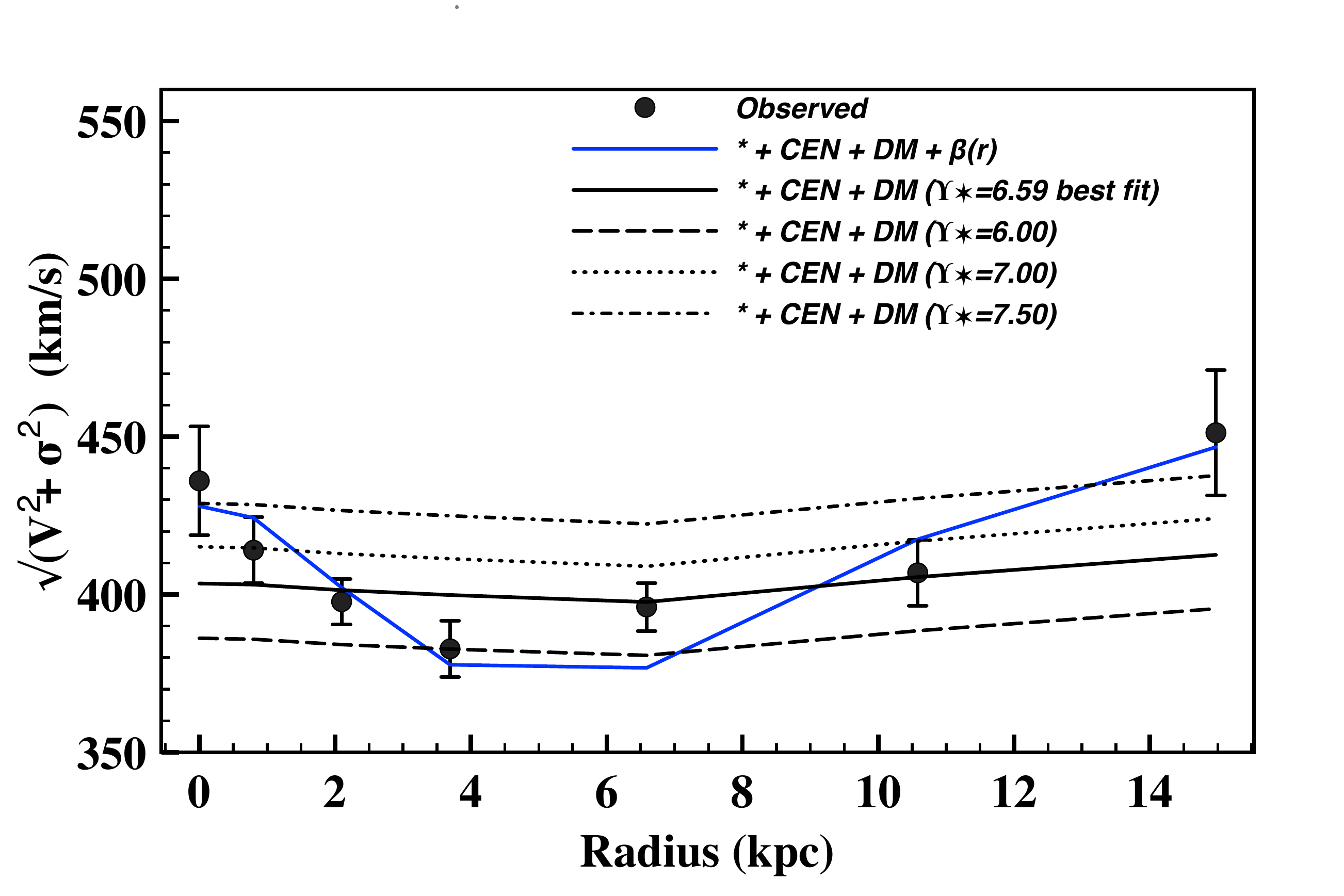}}
   \caption{We show the model fit for all three mass components ($\star$ + CEN + DM) for Abell 2029, where the best-fitting value for a constant $\Upsilon_{\star \rm DYN}$ (solid line), and lines for other constant values for $\Upsilon_{\star \rm DYN}$, are shown in black. We also show a fit for a variable $\beta_{z}(r)$ from radial to tangential anisotropy (in blue). A similar fit could be achieved by first decreasing and then increasing $\Upsilon_{\star \rm DYN}(r)$.}
\label{A267_varML}
\end{figure}

For this BCG, Figure \ref{A267_varML} shows the model fit when all three mass components ($\star$ + CEN + DM) are used, and the resulting best-fitting value for a constant $\Upsilon_{\star \rm DYN}=$ 6.59 (solid black line, $\chi^{2}/DOF=2.47$). We also show a fit for a variable $\beta_{z}(r)$ ($\beta_{z}$=[0.6, 0.3, --0.05, --0.15, --0.30], with $\Upsilon_{\star \rm DYN}=7.0$, where the linear radial increments are for the different Gaussians used in the stellar mass description, shown with a blue line, $\chi^{2}/DOF=1.88$). We also show lines for constant $\Upsilon_{\star \rm DYN}$ at 6.0, 7.0 and 7.5, and we see that a similar fit could be achieved by varying $\Upsilon_{\star \rm DYN}(r)$ from e.g. 7.5 down to 6.0 and again up to 7.5. 

A variable $\Upsilon_{\star \rm DYN}$ is neither unrealistic nor unreasonable. Out of the 32 BCGs, we find from our stellar population analysis a variable $\Upsilon_{\star \rm POP}(r)$ in 13 cases (Loubser et al.,\ in prep). However, from the analysis we find that the stellar mass-to-light ratio derived for Abell 2029 from stellar populations $\Upsilon_{\star \rm POP}$ is constant between our central aperture (0 -- 5 kpc) and the outer aperture (5 -- 15 kpc) within the errors, with an average value of 4.08 $\pm$ 1.04 (for a Salpeter IMF). It is also unlikely that the mass-to-light ratio will decrease and then again increase sharply over this radial range. 

It is therefore possible that a non-constant $\Upsilon_{\star \rm DYN}$ or $\beta_{z}$ can improve the model fit to the observed kinematics for a subset of BCGs, but we find it more likely that, for these BCGs, $\beta_{z}$ is radially variable (changing from radial to tangential) than a non-constant $\Upsilon_{\star}$ that is decreasing then increasing. 

\subsection{Interpreting the central $h_{4}$ measurements}
\label{additionh4} 

\begin{figure}
\centering
   \subfloat{\includegraphics[scale=0.37]{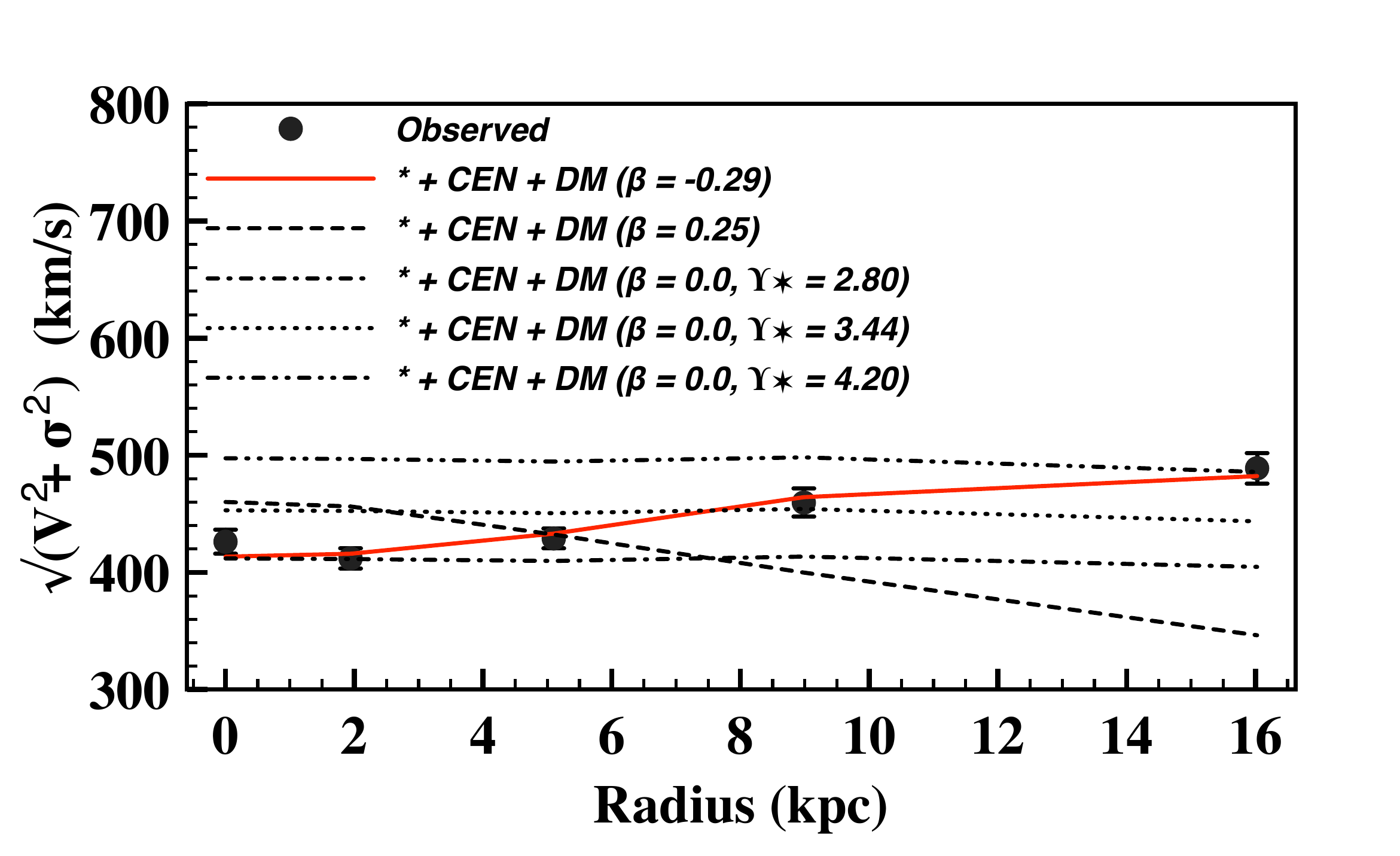}}
   \caption{Fits to the observed kinematics, illustrated for the BCG in Abell 2261 for $\beta_{z}$ negative, positive or zero.}
\label{A2261_varML}
\end{figure}

Generally, positive $\beta_{z}$ (radial anisotropy) is associated with negative velocity dispersion gradients (i.e. decreasing with increasing radius) and negative $\beta_{z}$ (tangential anisotropy) is associated with increasing velocity dispersion, similar to what we find in Figure \ref{beta_sig}. However, radial anisotropy has also been associated with positive $h_{4}$ measurements. From the result that all our central $h_{4}$ measurements are positive, one might have naively expected that all the BCGs have radial anisotropy. However, a tangential $\beta_{z}$ ($<$0) describes the observed kinematics better (see e.g. the solid red line in Figure \ref{A2261_varML}) for the BCGs with negative velocity dispersion gradients. This prompted us to investigate whether, if $\beta_{z}$ is forced to be $\geq 0$, a (monotonically) variable $\Upsilon_{\star \rm DYN}$ can describe the observed kinematics. 

Here, we show the best-fitting model for the BCG in Abell 2261, as an example. If we consider the $\star$ + CEN + DM fit for the BCG in Abell 2261, the best-fitting $\beta_{z}$ is negative ($\beta_{z}$ = --0.29), with $\Upsilon_{\star \rm DYN}$ = 3.44 ($\chi^{2}/DOF$ = 0.79, solid red line in Figure \ref{A2261_varML}). If we, for illustration, force $\beta_{z}$ = 0.25, then the best-fitting $\Upsilon_{\star \rm DYN}$ = 2.80, yields a worse fit at $\chi^{2}/DOF$ = 61.01 (black dashed line). To fit the kinematics profile with $\beta_{z} = 0$, $\Upsilon_{\star \rm DYN}$ would have to increase from 2.80 in the centre to 4.20 for the outer two data points (a combination of the dotted and dot-dashed lines in Figure \ref{A2261_varML}). 

For Abell 2261 we find constant $\Upsilon_{\star \rm POP}$ in our stellar population analysis. So, it is more likely that the rising velocity dispersion profile is a result of a constant $\beta_{z}$ that is tangential, than a variable $\Upsilon_{\star}(r)$. For the 12 BCGs with rising velocity dispersion profiles (and a best-fitting tangential $\beta_{z}$) in Figure \ref{beta_sig}, eight have a constant $\Upsilon_{\star}$ from the stellar population analysis, and tangential anisotropy, or a significant contribution from the ICL, is necessary to describe the kinematic profiles.

The positive measurements of $h_{4}$ must result from gradients in the circular velocity curves or stem from the superposition of two $\mathcal L_{\rm LOS}$, a narrower one of the stars feeling the potential of the galaxy, and a broader one probing the potential of the cluster (as discussed in Section \ref{ICL}).

\section{Conclusions}
\label{conclusions}

We use spatially-resolved long-slit spectroscopy for 32 BCGs, and present the first comprehensive measurements of the Gauss-Hermite higher order velocity moments $h_{3}$ and $h_{4}$ for a large sample of BCGs. We then model the stellar and dynamical mass of BCGs located in 25 massive clusters, using the Multi-Gaussian Expansion (MGE) and an adapted Jeans Anisotropic Method (JAM) for an axisymmetric (cylindrically-aligned) case, deriving the stellar mass-to-light ratio $\Upsilon_{\star \rm DYN}$, and anisotropy $\beta_{z}$. We also add the dark matter mass derived from weak lensing results, to investigate the remarkable diversity in their velocity dispersion slopes found in \citet{Loubser2018}. Our fits to the observed kinematics are restricted to within $<$20 kpc from the galaxy centre, and we perform various tests to illustrate the robustness of the data measurements and modelling results in Section \ref{data} and \ref{higherorder}, and in the Appendices. Our main results may be summarised as follows:

\medskip
\begin{enumerate}
\item We find positive $h_{4}$ values for all of our BCGs and BGGs. If we exclude the six BCGs with young stellar components, we find $\langle h_{4} \rangle = 0.049 \pm 0.004$ and $\langle h_{3} \rangle = 0.011 \pm 0.004$ for the BCG sample. We do not find a significant correlation between $h_{3}$ or $h_{4}$ vs $M_{K}$ over four magnitudes in luminosity (Figure \ref{h4_luminosity}), or with central velocity dispersion ($\sigma_{0}$) or velocity dispersion slope ($\eta$) of the BCGs, or $M_{500}$ of the host clusters. 

\item We find that adding a central mass component or a fixed dark matter halo mass component does not give a significantly better or worse fit to the observed kinematics than just the stellar mass component alone, at least over our limited radial range. Adding the halo mass component decreases $\Upsilon_{\star \rm DYN}$ on average by 8.3 $\pm$ 2.9 per cent over our kinematic range, and increases $\beta_{z}$ by on average 0.04. 

\item From our test of the effect of the uncertainties in the weak lensing masses and concentration parameter on the best-fitting parameters in Appendix \ref{Delta}, we find that our dynamical modelling is robust against the dark matter distribution or the value used for the concentration parameter. We include a dark matter halo mass component, but we cannot use the dynamical modelling to constrain the dark matter distribution by leaving it as a free parameter.

\item We plot our best-fitting $\beta_{z}$ parameters (for the $\star$ + CEN +DM mass component fits) against the slope of the velocity dispersion profiles \citep{Loubser2018} in Figure \ref{beta_sig}. This figure shows a strong, negative correlation between the best-fitting $\beta_{z}$ parameters and the slope of the velocity dispersion profiles (slope = --0.186 $\pm$ 0.028, with an intrinsic scatter of 0.026, and correlation coefficient 0.929). In other words, BCGs with rising velocity dispersion have tangential anisotropy, and with decreasing velocity dispersion have radial anisotropy. The norm is for elliptical galaxies to have radial anisotropy, but there are exceptions with tangential anisotropy, and those often show increasing velocity dispersion gradients. However, the correlation between $\beta_{z}$ and velocity dispersion gradient shown in Figure \ref{beta_sig} could be driven by a sequence of decreasing importance of intracluster contamination, as discussed in Section \ref{ICL}. 

\item We also compare our results to JAM models for an axisymmetric, spherically-aligned (JAM$_{\rm sph}$) case, and find that neither model is significantly, or consistently, better or worse than the other. For the JAM$_{\rm sph}$ models, we still find tangential anisotropy ($\beta < 0$) for BCGs with rising velocity dispersion profiles and radial anisotropy ($\beta > 0$) for decreasing velocity dispersion profiles. JAM$_{\rm sph}$ is characterized by a relative insensitivity of the model predictions to anisotropy, and the correlation of velocity dispersion slope with velocity anisotropy is therefore shallower for $\beta$ (JAM$_{\rm sph}$) than $\beta_{z}$ (JAM). There is however a systematic change (at the 10 -- 15 per cent level) in best-fitting $\Upsilon_{\star \rm DYN}$, with $\Upsilon_{\star \rm DYN}$ being lower for decreasing $\nu_{\rm rms}$ profile BCGs (i.e. radial anisotropy, where $\beta$ is positive), and $\Upsilon_{\star \rm DYN}$ being higher for increasing $\nu_{\rm rms}$ profile BCGs (i.e. tangential anisotropy, where $\beta$ is negative). Using both JAM and JAM$_{\rm sph}$ allows us to assess the sensitivity of the model results to the model assumptions.

\item The isotropic case has also been associated with $h_{4} = 0$, the radial anisotropic case with a positive $h_{4}$, and the tangential anisotropic case with a negative $h_{4}$, and all of our central measurements for $h_{4}$ are positive. Our BCG results encompass a range of velocity anisotropy (from radial to tangential), and in cases with tangential anisotropy (typically associated with negative $h_{4}$), the positive measurements of $h_{4}$ must primarily result from gradients in the circular velocity curves \citep{Gerhard1993}, or from the superposition of two $\mathcal L_{\rm LOS}$ due to the contribution of the ICL.

\item Several BCGs have non-monotonic velocity dispersion profiles that first decrease with radius before increasing again (e.g.\ Abell 267, 383, 611, 644, 754, 1991, 2029 and 2420 as shown in Appendix \ref{DynMods}). We therefore explore the extent to which a variable $\Upsilon_{\star \rm DYN}(r)$ or $\beta_{z}(r)$ could explain the kinematic profiles of these BCGs. We find that a radially-changing $\beta_{z}(r)$ (from radial to tangential) can fit those observed kinematic profiles, and that a similar fit could be achieved by varying $\Upsilon_{\star \rm DYN}(r)$, first decreasing then increasing from the centre outwards. A variable $\Upsilon_{\star \rm DYN}$ is neither unrealistic nor unreasonable. For the 32 BCGs, we find from our stellar population analysis a variable $\Upsilon_{\star \rm POP}(r)$ for 13 of these cases (Loubser et al.,\ in prep). It is therefore possible that a non-constant $\Upsilon_{\star \rm DYN}$ or $\beta_{z}$ can improve the model fit to the observed kinematics for a subset of BCGs, but it is more likely that, for these BCGs, $\beta_{z}$ is radially variable (changing from radial to tangential) than a non-constant $\Upsilon_{\star}$ that is decreasing then increasing over this radial range. 

\item Since all our central $h_{4}$ measurements are positive (typically associated with radial anisotropy), we investigate whether, if $\beta_{z}$ is forced to be $> 0$, a (monotonically) variable $\Upsilon_{\star \rm DYN}$ can possibly describe the rising velocity dispersion profiles, instead of tangential anisotropy or a significant contribution from the ICL. To fit the rising kinematics profile with a non-negative value for $\beta_{z}$ requires a radially-variable $\Upsilon_{\star \rm DYN}$ (see example in Figure \ref{A2261_varML}). However, eight of the 12 BCGs with rising velocity dispersion profiles (and a best-fitting tangential $\beta_{z}$) have a constant $\Upsilon_{\star}$ from the stellar population analysis (Loubser et al.,\ in prep). 

So it is likely that the rising velocity dispersion profiles in most of the BCGs are a result of a constant $\beta_{z}$ that is tangential, or a significant contribution from the ICL component in projection (or a combination of the two). For a small number of BCGs, a monotonically increasing $\Upsilon_{\star \rm DYN}$ can also contribute to the rising velocity dispersion profiles.
\end{enumerate}

Despite the fact that most BCGs are located in a similar, special environment in the centres of X-ray luminous clusters, they exhibit differences in their stellar populations with those BCGs residing at the centres of cool-core clusters forming stars, as opposed to the passive evolution of BCGs hosted by non-cool core clusters \citep{Sarazin1983, Bildfell2008, Pipino2009, Loubser2013, Donahue2015, Loubser2016}. However, even within the star-forming or non star-forming populations of BCGs, they exhibit remarkable diversity in their stellar kinematics, particularly their velocity dispersion profiles \citep{Loubser2018}. The velocity dispersion profile slopes correlate with $K$-band luminosity \citep{Loubser2018}, and from the detailed dynamical modelling presented in this paper, we find that the diversity also corresponds to a very large range of velocity anisotropy, and stellar mass-to-light ratios. These properties illustrate that BCGs are not the homogeneous class of objects they are often assumed to be.

In future, it would be interesting to see how these results extend and compare to the central group galaxies (see \citealt{Loubser2018}). To simultaneously fit the observed second moment of velocity $\nu_{\rm RMS}$, as well as $h_{4}$, IFU data and modelling the stellar orbits using Schwarzschild modelling for triaxial systems will be necessary. 

\section*{Acknowledgements}

We thank the the anonymous referee for thorough, constructive comments. This research was enabled, in part, by support provided by the bilateral funding agreement between the National Research Foundation (NRF) of South Africa, and the Netherlands Organisation for Scientific Research (NWO) to SIL and HH. SIL is aided by a Henri Chr\'etien International Research Grant administered by the American Astronomical Society. AB acknowledges support from NSERC (Canada) through the Discovery Grant program. AB would also like to thank the Centre for Space Research, North-West University, South Africa for hospitality during the summer of 2019. HH acknowledges support from the European Research Council FP7 grant number 279396. YMB acknowledges funding from the EU Horizon 2020 research and innovation programme under Marie Sk{\l}odowska-Curie grant agreement 747645 (ClusterGal) and the NWO through VENI grant 016.183.011. EOS acknowledges support from the National Aeronautics and Space Administration (NASA) through Chandra Awards GO6-17121X and GO6-17122X, issued by the Chandra X-ray Observatory Center, which is operated by the Smithsonian Astrophysical Observatory on behalf of NASA under contract NAS8-03060. 

Based, in part, on observations obtained at the Gemini Observatory, which is operated by the Association of Universities for Research in Astronomy, Inc., under a cooperative agreement with the NSF on behalf of the Gemini partnership: the National Science Foundation (United States), the National Research Council (Canada), CONICYT (Chile), Ministerio de Ciencia, Tecnolog\'{i}a e Innovaci\'{o}n Productiva (Argentina), and Minist\'{e}rio da Ci\^{e}ncia, Tecnologia e Inova\c{c}\~{a}o (Brazil). Based, in part, on observations obtained at the Canada-France-Hawaii Telescope (CFHT) which is operated by the National Research Council of Canada, the Institut National des Sciences de l'Univers of the Centre National de la Recherche Scientifique of France, and the University of Hawaii. This research used the facilities of the Canadian Astronomy Data Centre operated by the National Research Council of Canada with support from the Canadian Space Agency. 

Any opinion, finding and conclusion or recommendation expressed in this material is that of the author(s) and the NRF does not accept any liability in this regard.

\section*{Data availability}

The data underlying this article will be shared on reasonable request to the corresponding author.




\bibliographystyle{mnras}
\bibliography{References} 



\appendix

\section{Central \lowercase{$h_{4}$} measurements: comparison to the MASSIVE survey}
\label{h4}

\begin{table}
\centering
\caption{Comparison of the central $h_{4}$ measurements of the four BGGs in common with the MASSIVE study to their central measurements of $h_{4}$.}
\label{h4_h4}
\begin{tabular}{lcc}
\toprule
Galaxy  & Our measurement & \citet{Veale2017a} \\
\midrule
NGC0410 & 0.051$\pm$0.015 & 0.041 \\
NGC0777 & 0.056$\pm$0.033 & 0.051 \\
NGC1060 & 0.045$\pm$0.028 & 0.055 \\
NGC0315 & 0.057$\pm$0.037 & 0.052  \\
\bottomrule
\end{tabular}
\end{table}	

We compare our central $h_{4}$ measurements of the four BGGs in common with the MASSIVE study to their central measurements of $h_{4}$ in Table \ref{h4_h4}, and find good agreement. 

\section{MGE and masking the images}
\label{masking}

\begin{figure}
\centering
         \subfloat[Abell 611 before]{\includegraphics[height=2.4cm, width=2.8cm, trim = 0mm 82mm 0mm 72mm, clip]{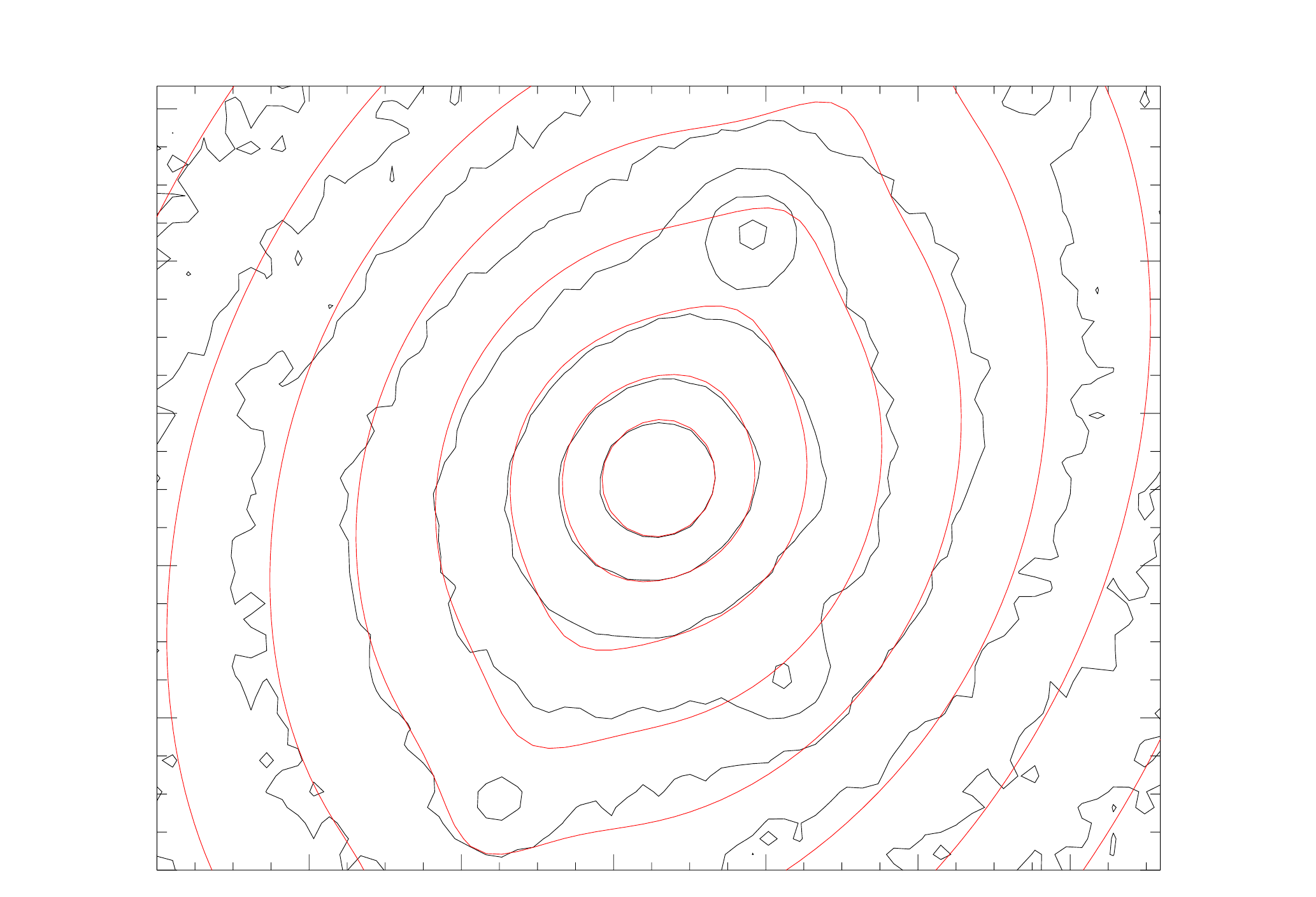}}
         \subfloat[Abell 611 after]{\includegraphics[height=2.2cm, width=2.6cm, trim = 0mm 67mm 0mm 67mm, clip]{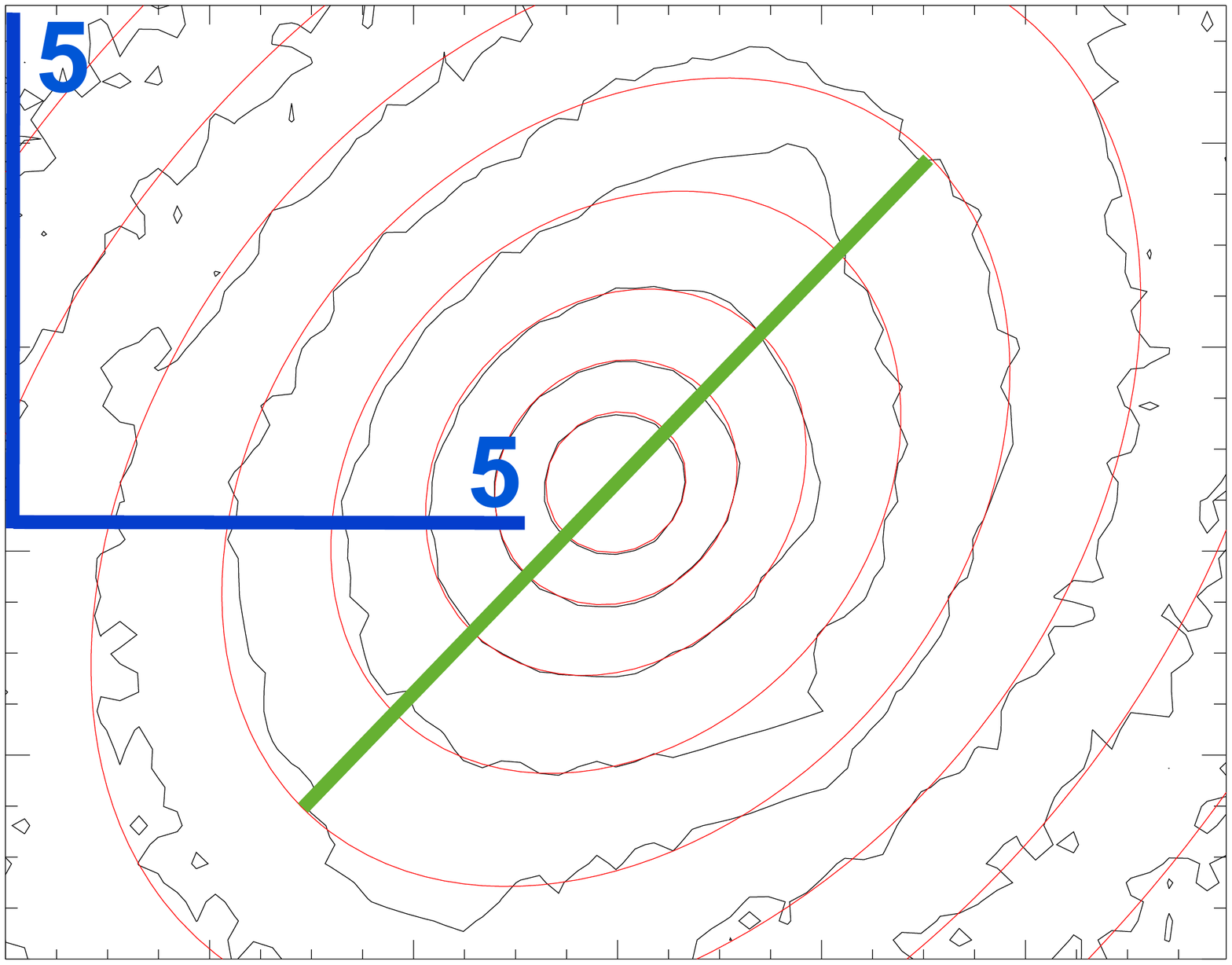}}\\
         \subfloat[Abell 1763 before]{\includegraphics[height=2.6cm, width=2.8cm, trim = 0mm 70mm 0mm 70mm, clip]{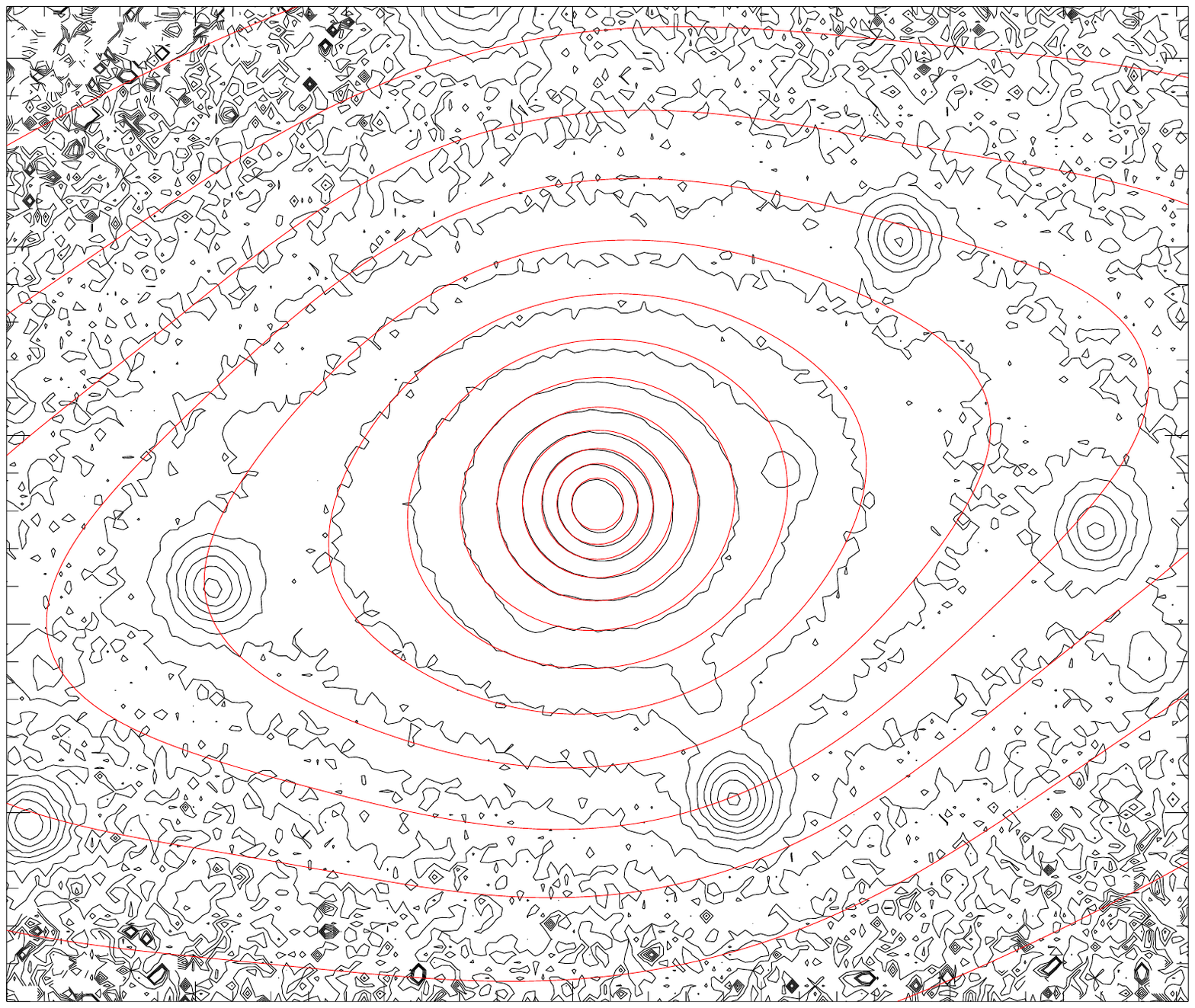}}
         \subfloat[Abell 1763 after]{\includegraphics[height=2.4cm, width=2.6cm, trim = 0mm 50mm 0mm 66mm, clip]{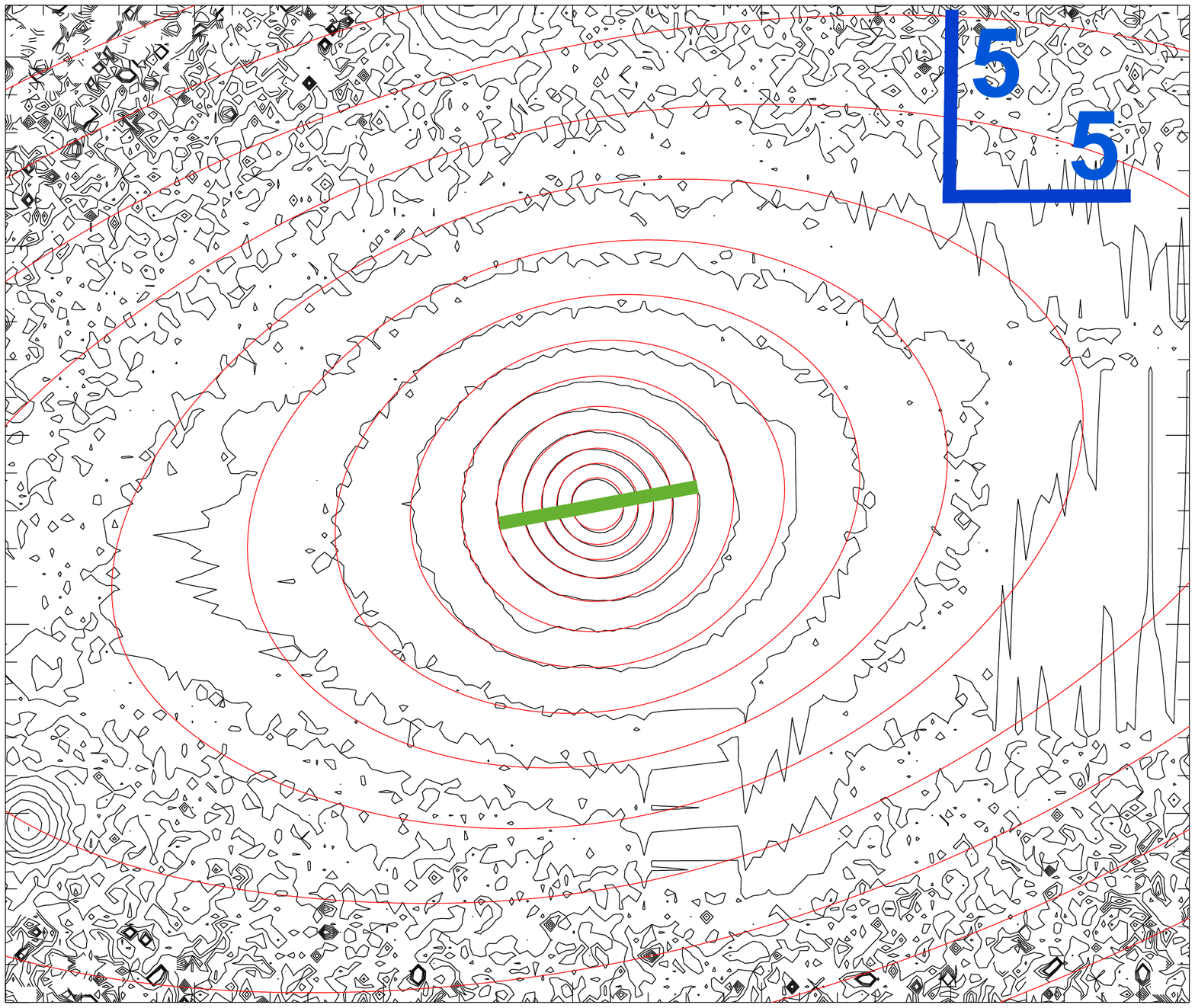}}\\
         \subfloat[MS14 before]{\includegraphics[height=2.6cm, width=2.8cm, trim = 0mm 70mm 0mm 70mm, clip]{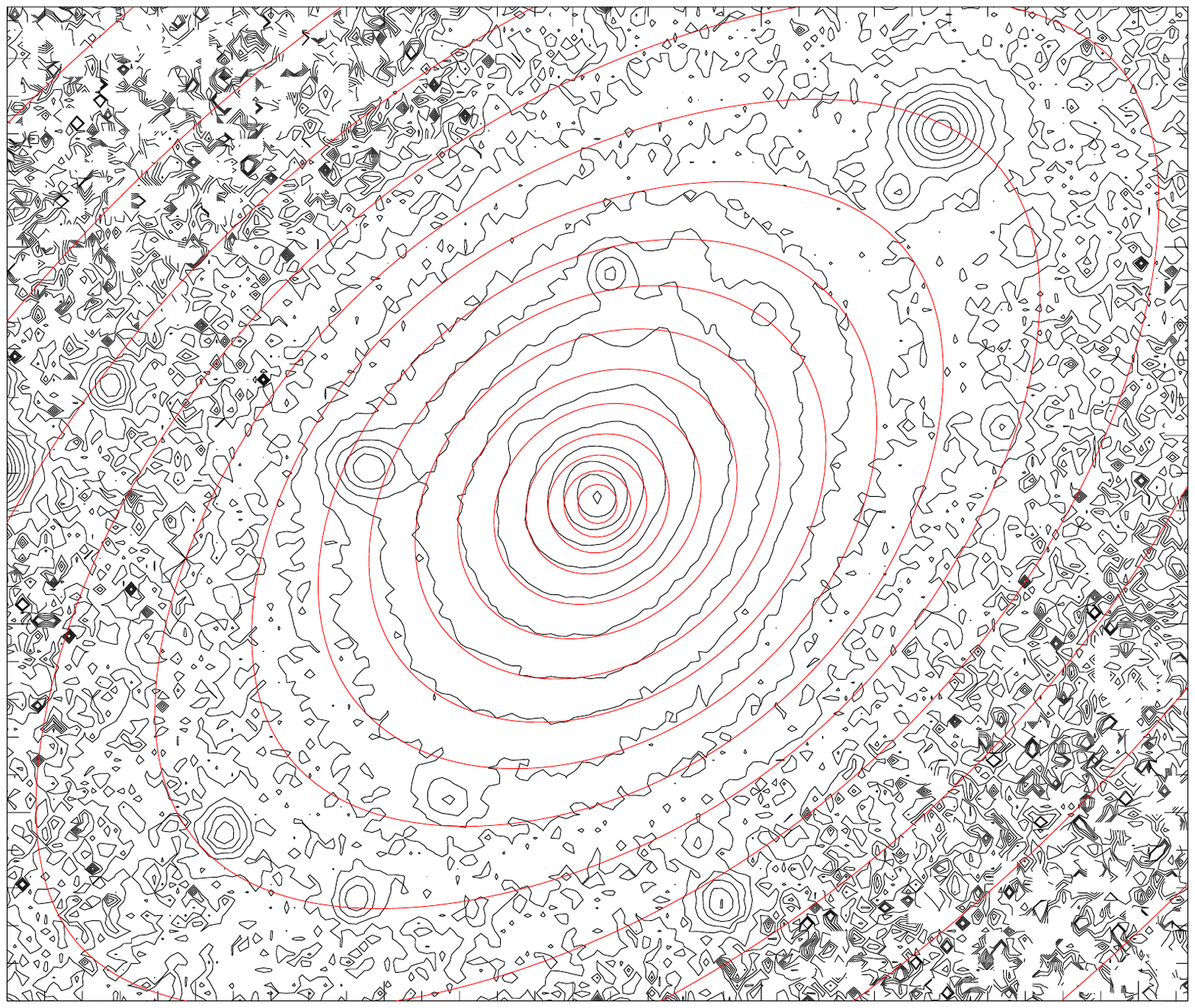}}
         \subfloat[MS14 after]{\includegraphics[height=2.4cm, width=2.6cm, trim = 0mm 50mm 0mm 62mm, clip]{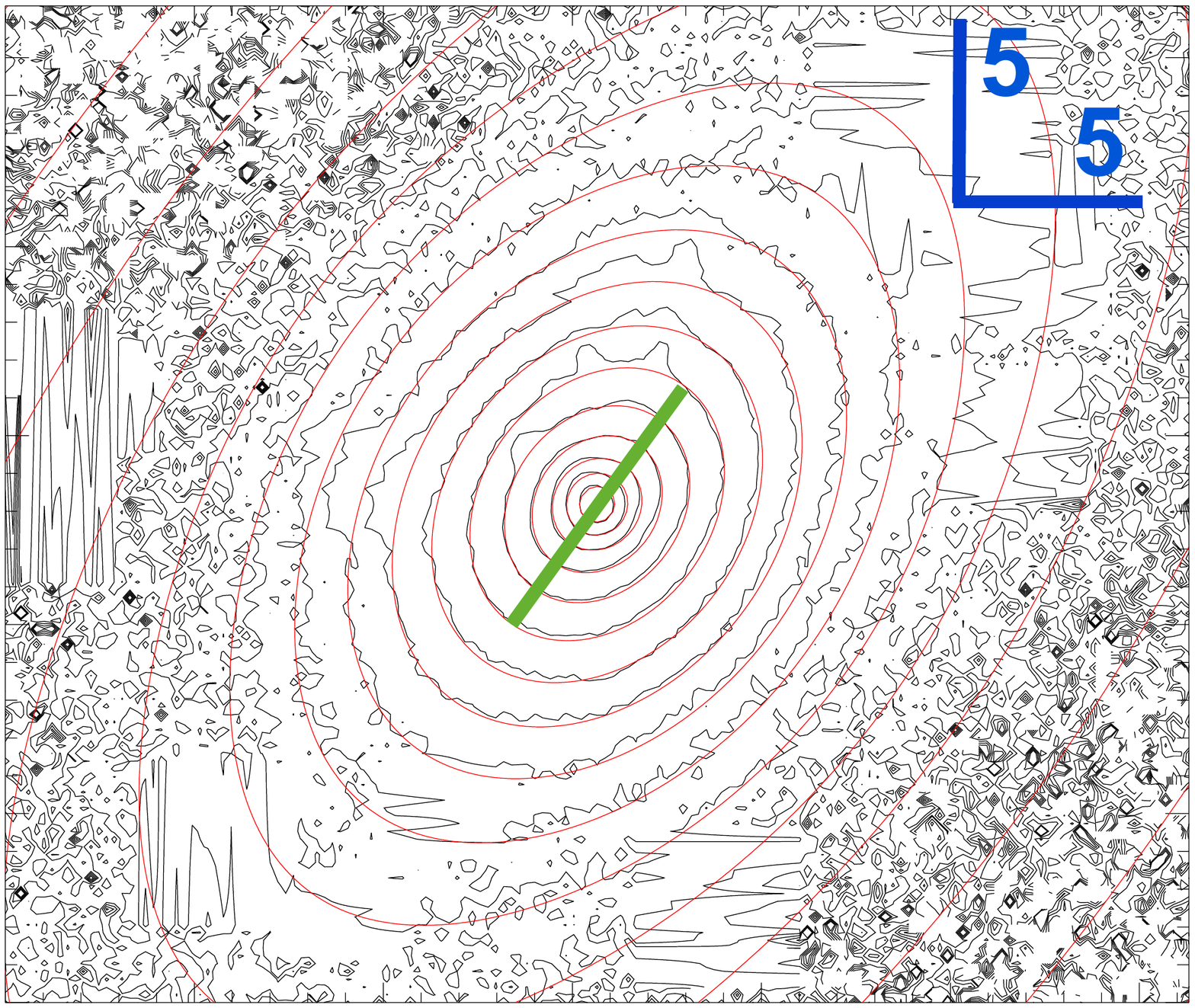}}\\
   \caption{MGE: before and after examples where the $r$-band images were masked to improve the MGE model fitting.}
\label{fig:MGE3}
\end{figure}

\begin{figure}
\centering
         \subfloat[Abell 990]{\includegraphics[height=2.4cm, width=2.4cm, trim = 22mm 88mm 10mm 100mm, clip]{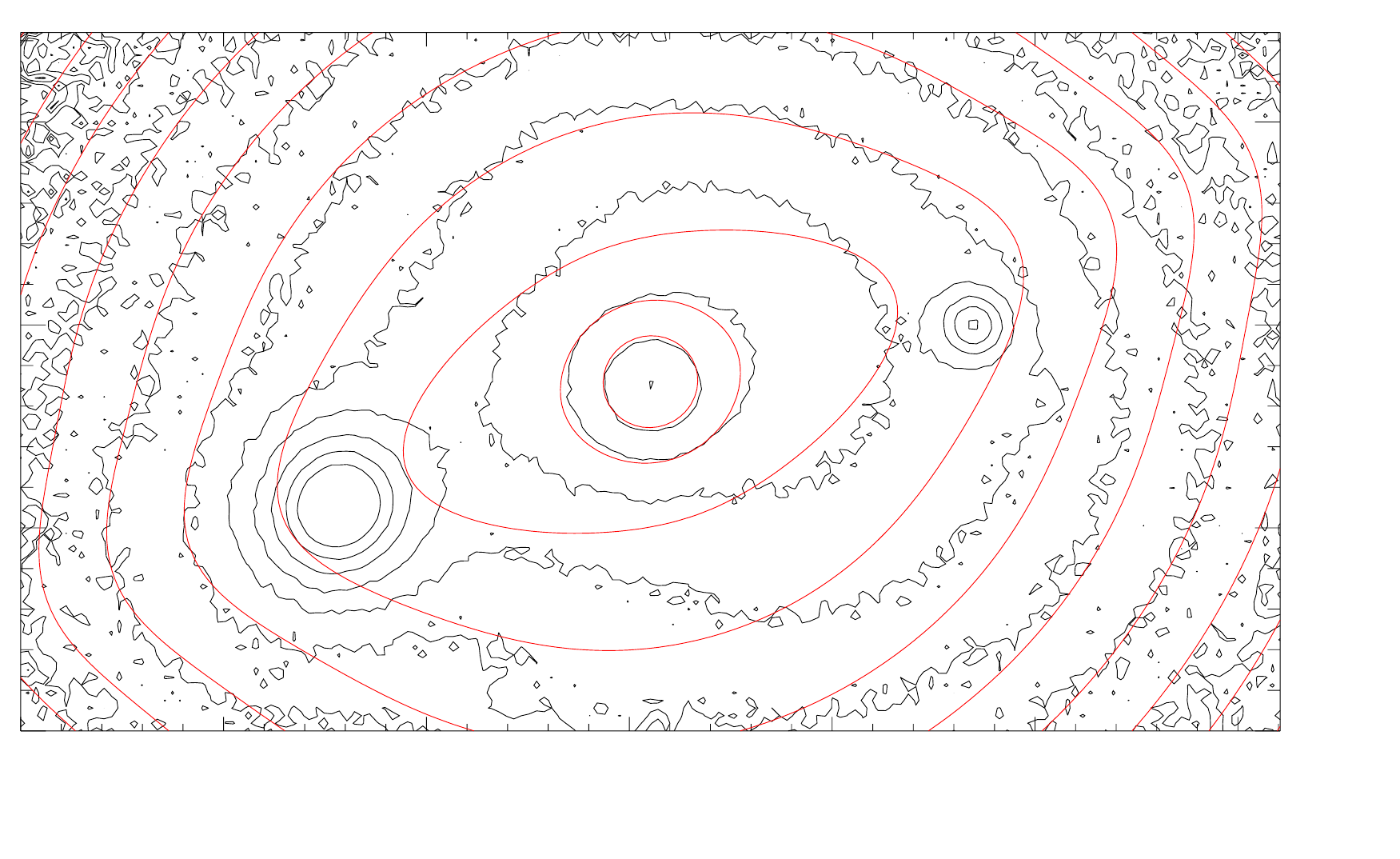}}
         \subfloat[Abell 1835]{\includegraphics[height=2.4cm, width=2.4cm, trim = 22mm 70mm 10mm 90mm, clip]{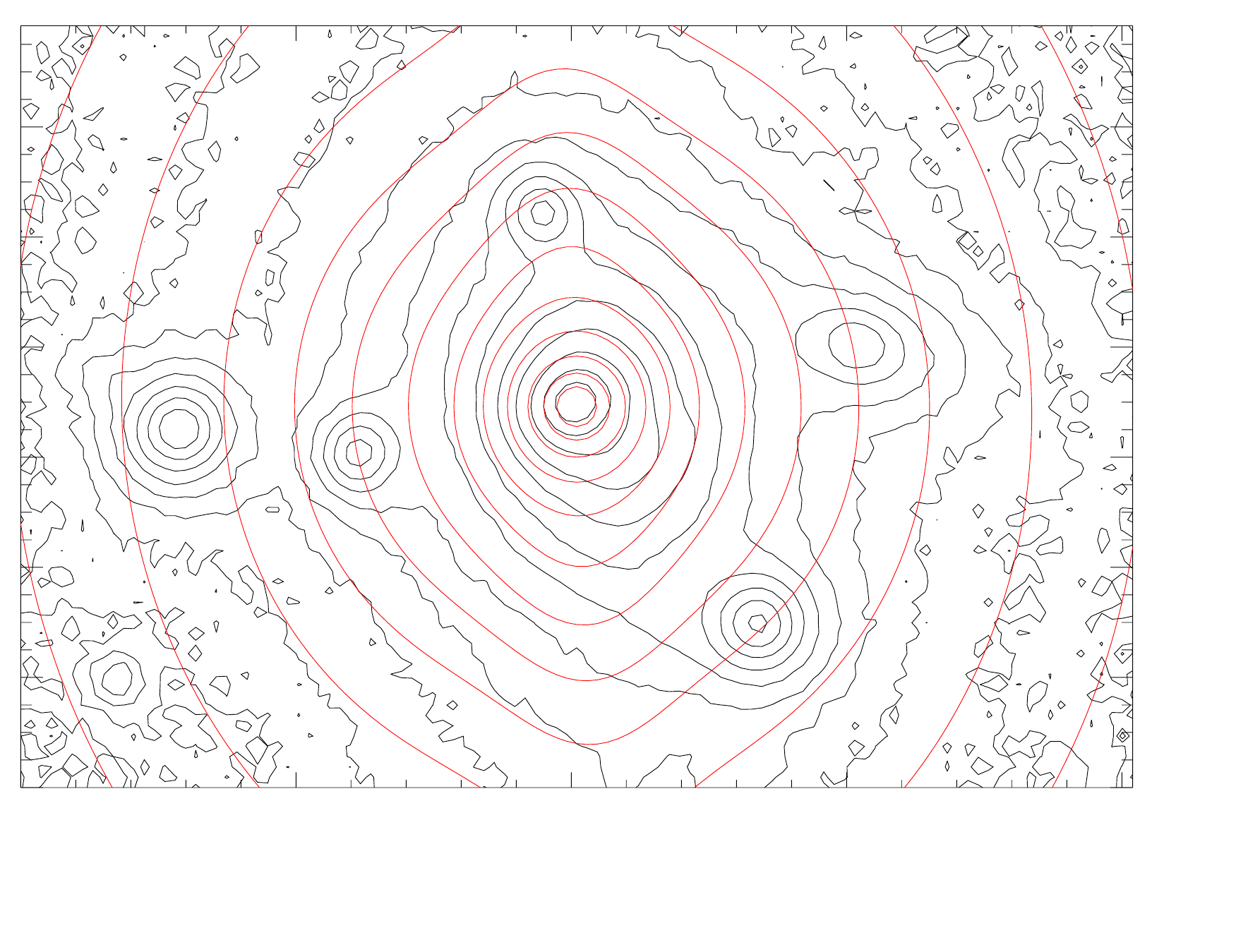}}\\
         \subfloat[Abell 2104]{\includegraphics[height=2.4cm, width=2.4cm, trim = 22mm 70mm 10mm 90mm, clip]{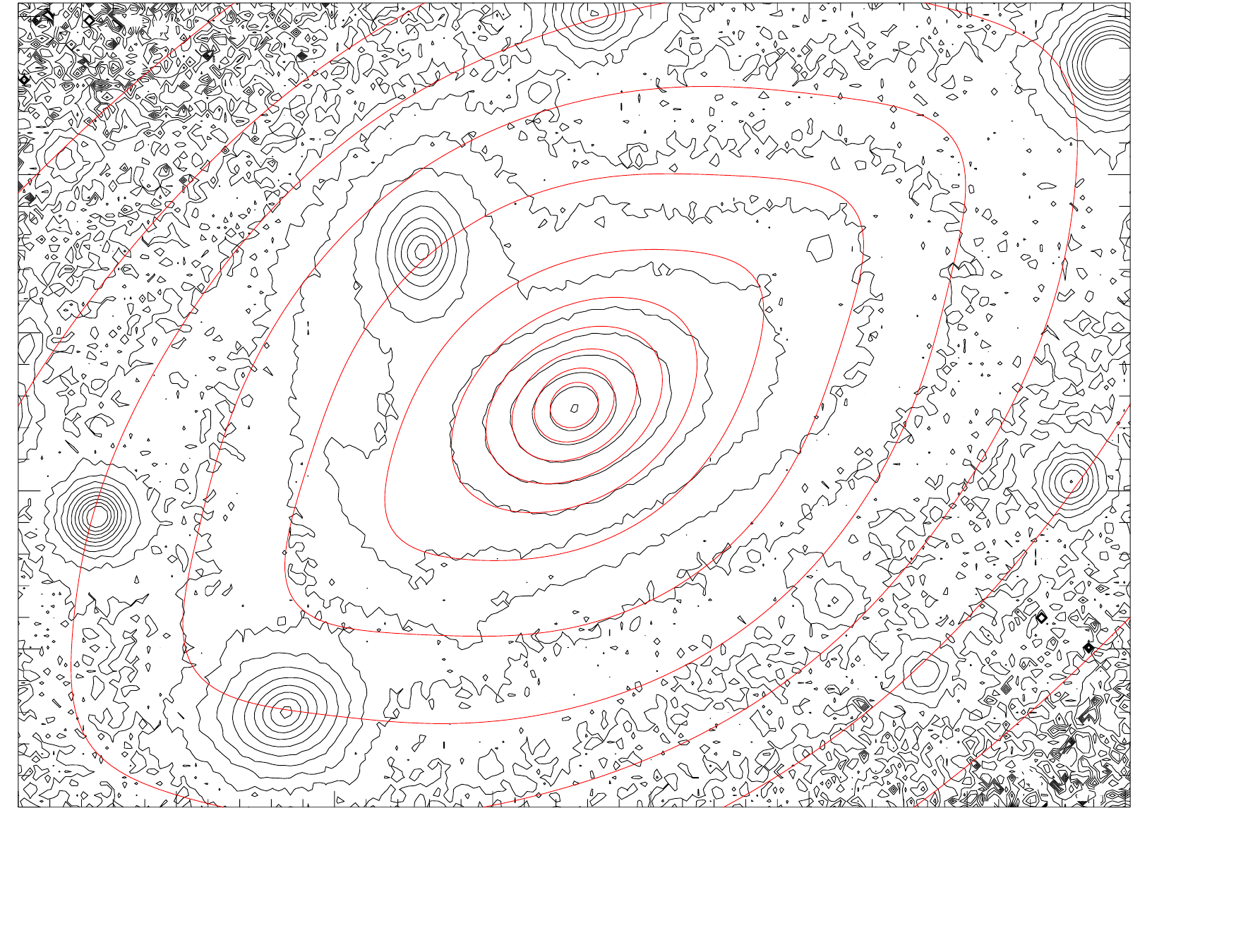}}
         \subfloat[Abell 2390]{\includegraphics[height=2.4cm, width=2.4cm, trim = 22mm 55mm 10mm 80mm, clip]{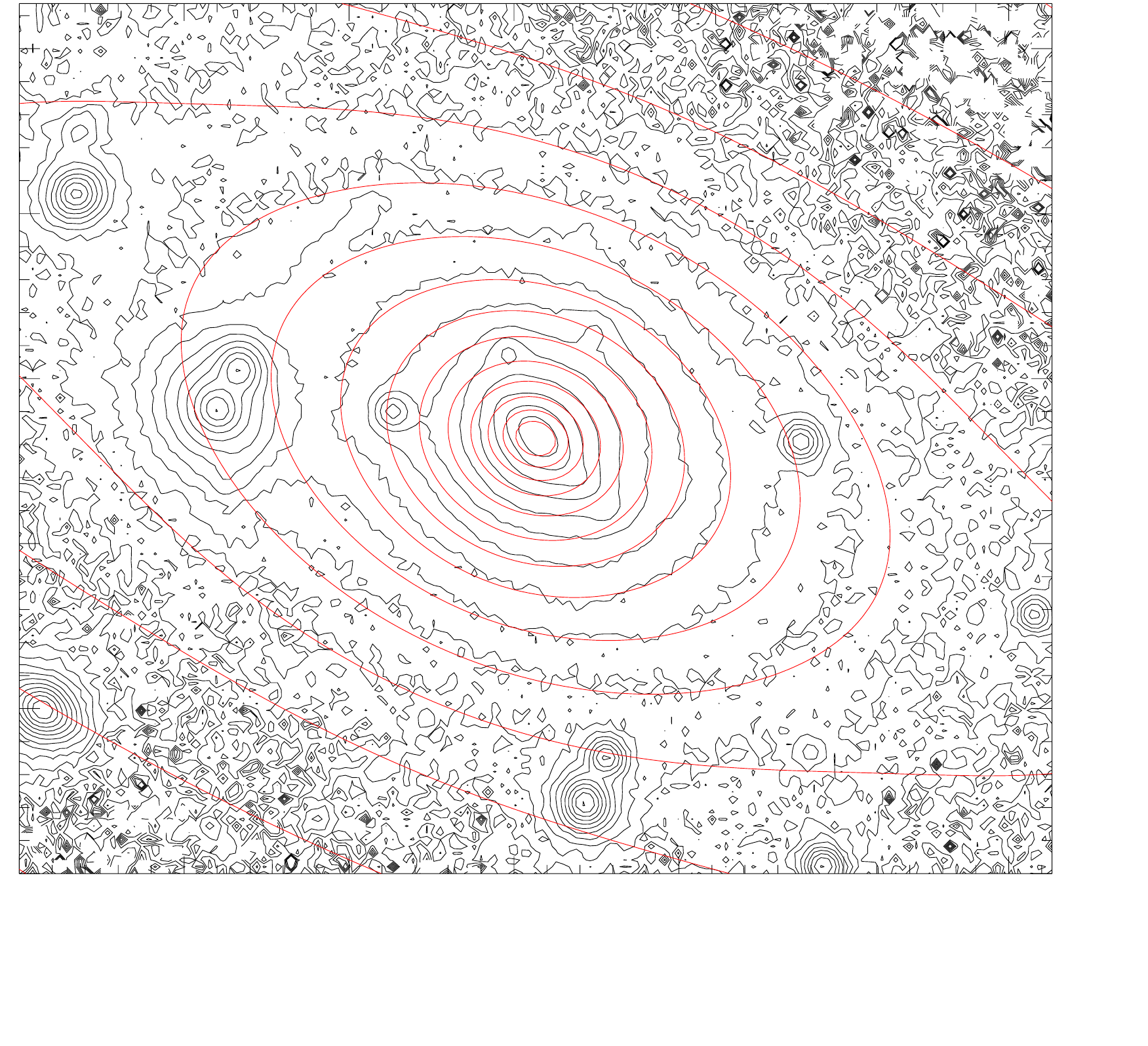}}\\
          \caption{MGE: Cases where the features in the $r$-band images were particularly problematic, and could not be completely removed. These are not used in the dynamical modelling.}
\label{fig:MGE4}
\end{figure}

\begin{figure}
\centering
         \subfloat[Abell 586]{\includegraphics[height=2.5cm, width=2.7cm, trim = 0mm 64mm 0mm 60mm, clip]{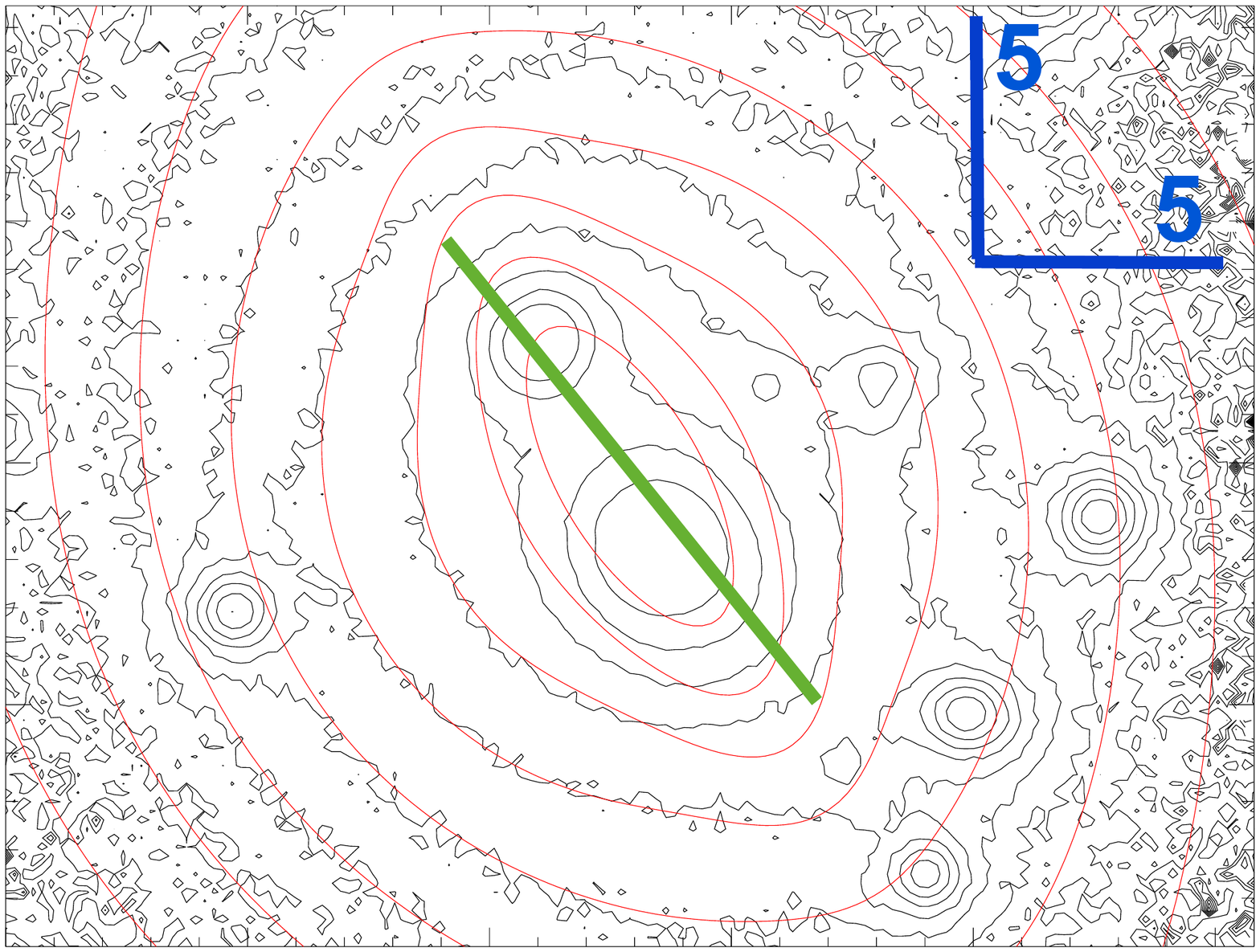}}
         \subfloat[MS0440+02]{\includegraphics[height=2.5cm, width=2.5cm, trim = 0mm 56mm 0mm 52mm, clip]{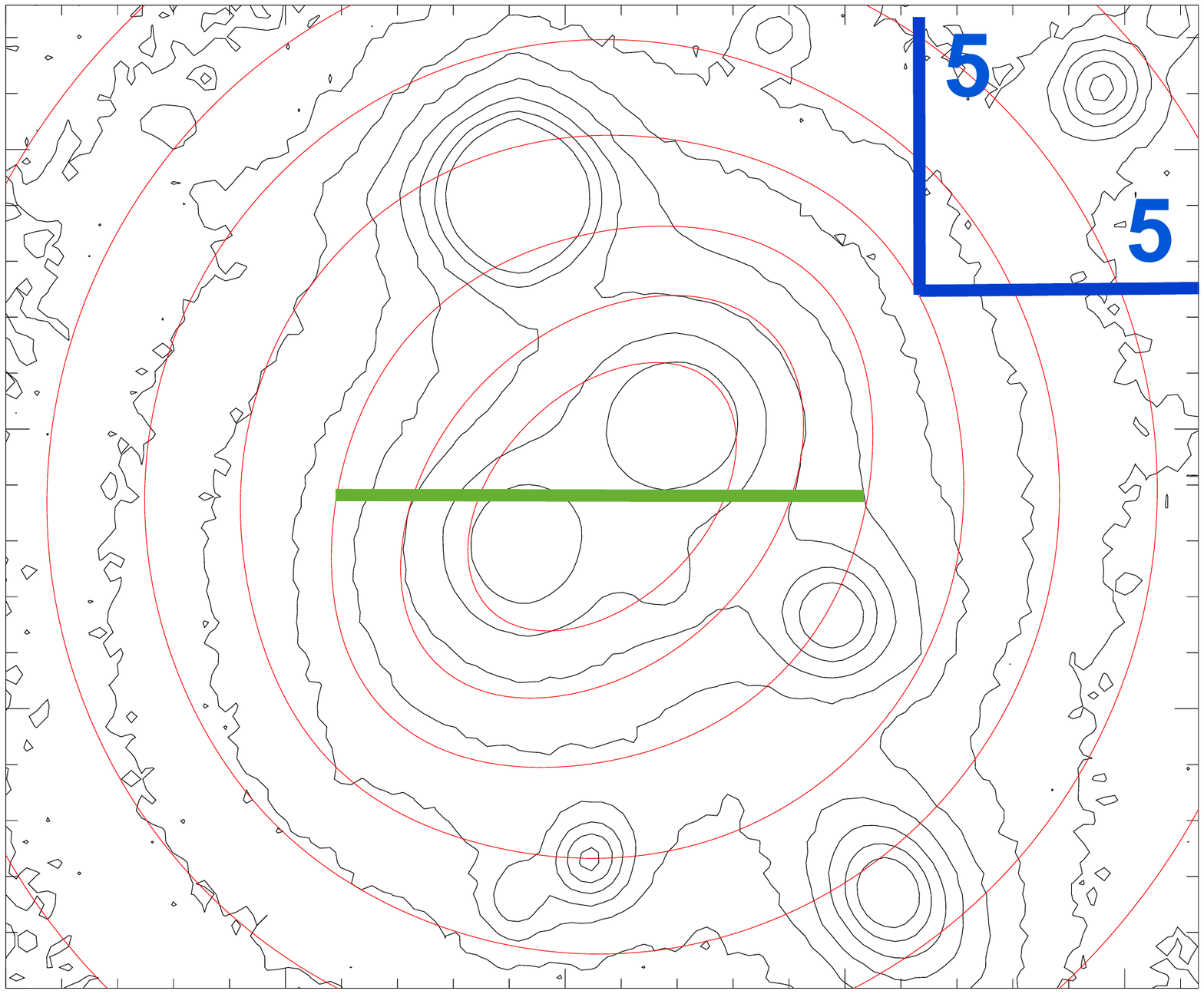}}
         \subfloat[MS0906+11]{\includegraphics[height=2.5cm, width=2.5cm, trim = 0mm 55mm 0mm 52mm, clip]{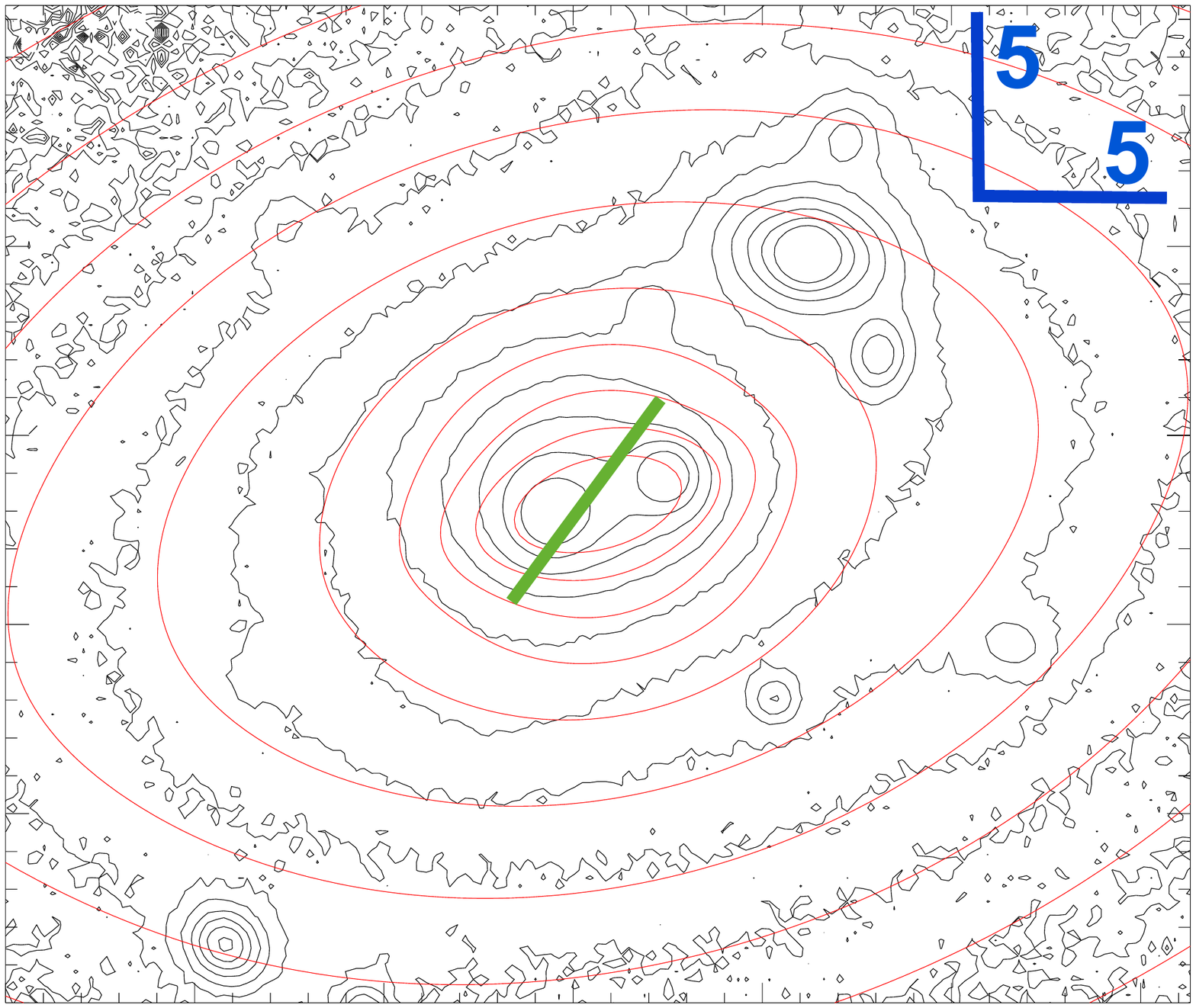}}\\
                  \caption{MGE: the BCGs with double nuclei. These are not used for the dynamical modelling.}
\label{fig:MGE5}
\end{figure}

We mask some of the images to eliminate the influence of sources in and around the central region of the BCG. We discuss and give examples of four types of cases below:

\begin{enumerate}

\item BCGs where no masking was necessary: Abell 68, 644, 646, 754, 1689, 2029, 2055, 2259, and 2420. All of these are used for dynamical modelling.

\item Cases where some masking were performed for line-of-sight features in the outer areas of the BCG, and therefore do not influence the results: Abell 267, 383, 611, 780, 963, 1650, 1763, 1795, 1942, 1991, 2050, 2142, 2261, 2319, 2537 and MS1455+22. We give examples of some of these objects, before and after masking in Figure \ref{fig:MGE3}. All of these are used for dynamical modelling.

\item Cases where masking was particularly problematic with foreground features in the BCG: Abell 990, 1835, 2104, and 2390. These are shown in Figure \ref{fig:MGE4}, and not used for the dynamical modelling.

\item Cases where the BCG has a double nucleus: Abell 586, MS0440+02, and MS0906+11. These are very hard to accurately fit using the MGE formalism and a single set of Gaussians, and they are shown in Figure \ref{fig:MGE5} and not used for the dynamical modelling.

\end{enumerate}

\begin{figure*}
\centering
         \subfloat[Abell 68]{\includegraphics[height=2.9cm, width=2.9cm,  trim = 0mm 40mm 0mm 40mm, clip]{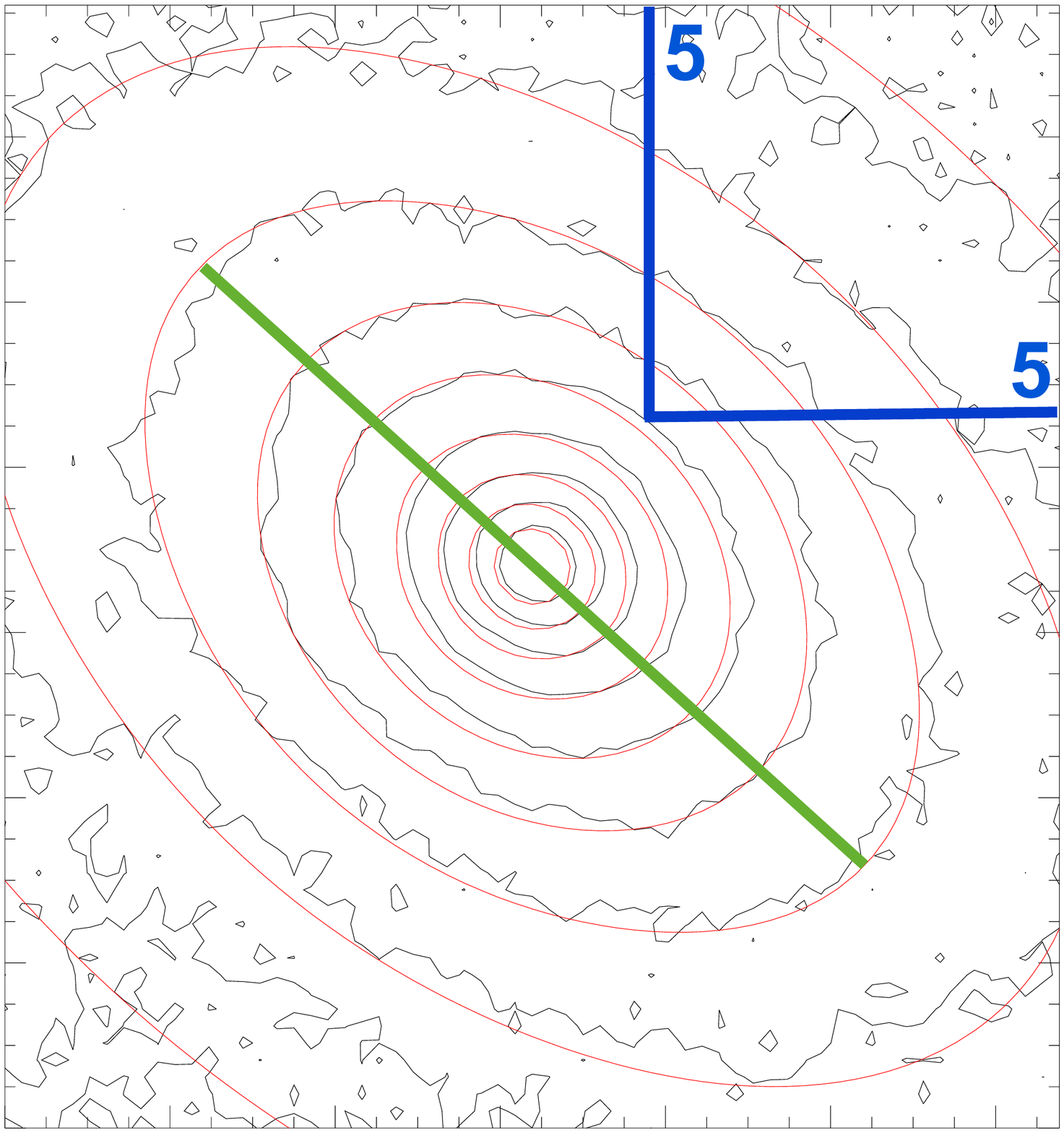}}
         \subfloat[Abell 267]{\includegraphics[height=2.9cm, width=3.2cm,  trim = 0mm 62mm 0mm 62mm, clip]{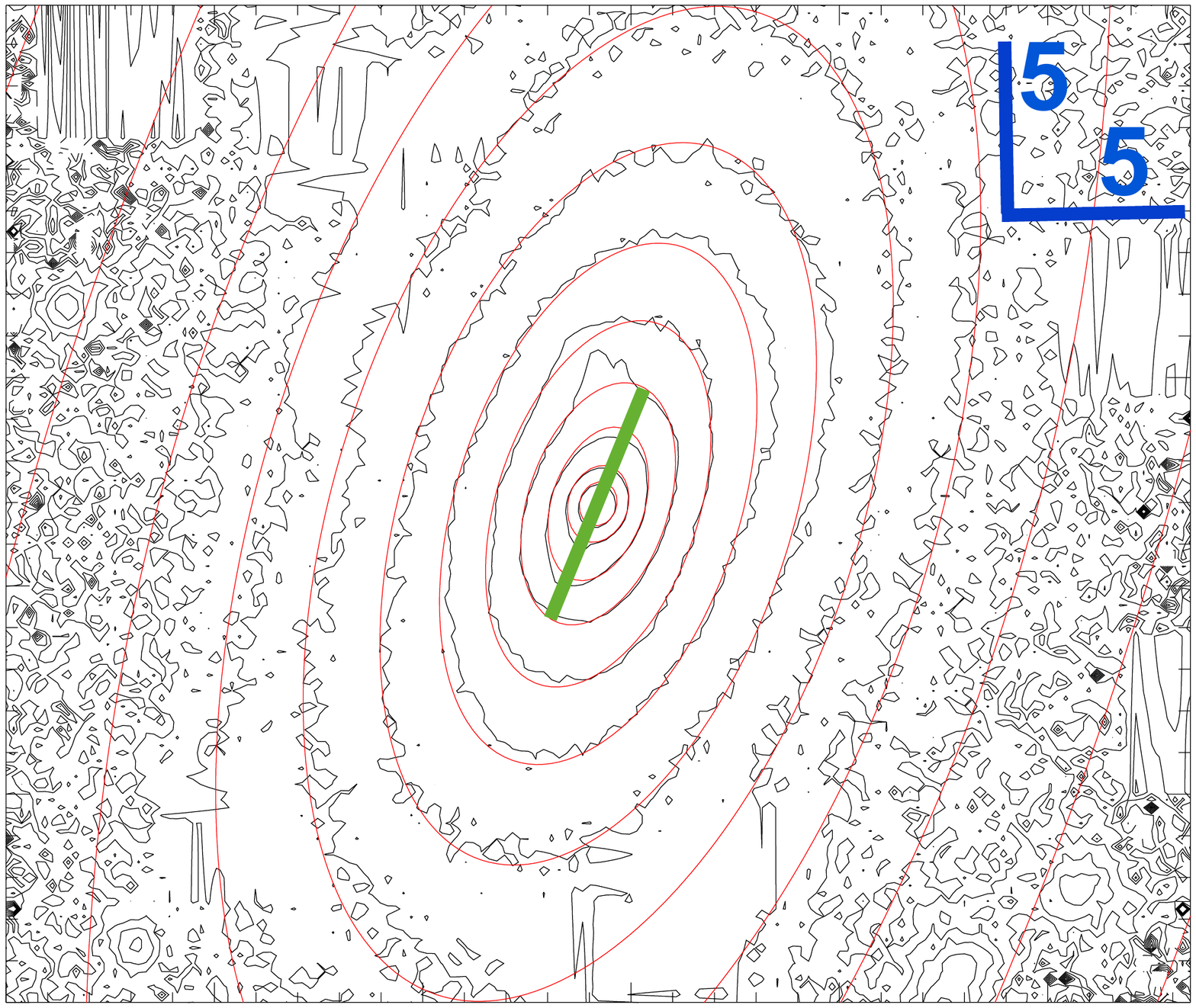}}
         \subfloat[Abell 383]{\includegraphics[height=2.9cm, width=3.0cm,  trim = 0mm 62mm 0mm 62mm, clip]{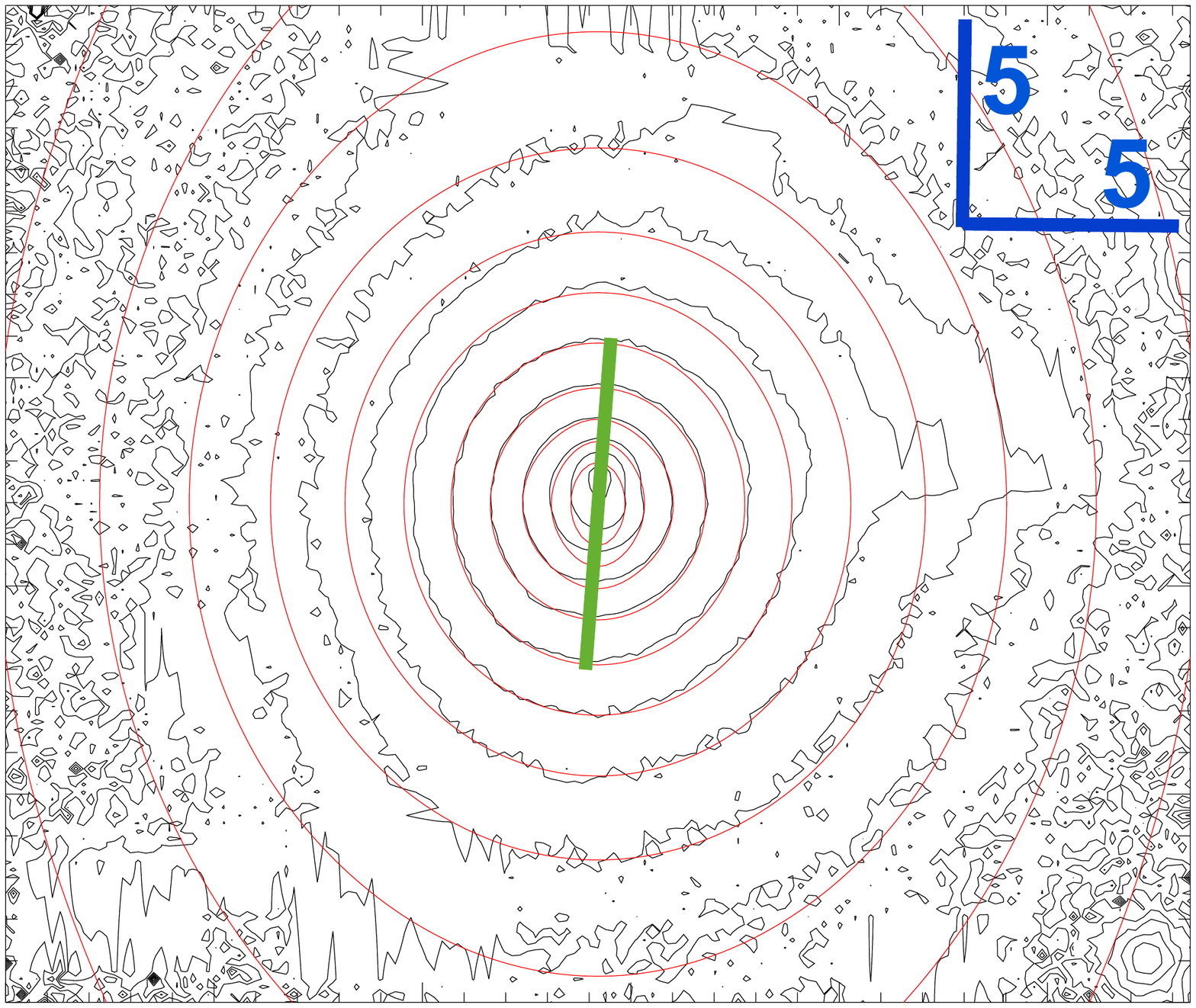}}
         \subfloat[Abell 611]{\includegraphics[height=2.9cm, width=3.5cm,  trim = 0mm 67mm 0mm 67mm, clip]{A611_nuclear_slit3.pdf}}
          \subfloat[Abell 644]{\includegraphics[height=2.9cm, width=4.8cm,  trim = 0mm 90mm 0mm 90mm, clip]{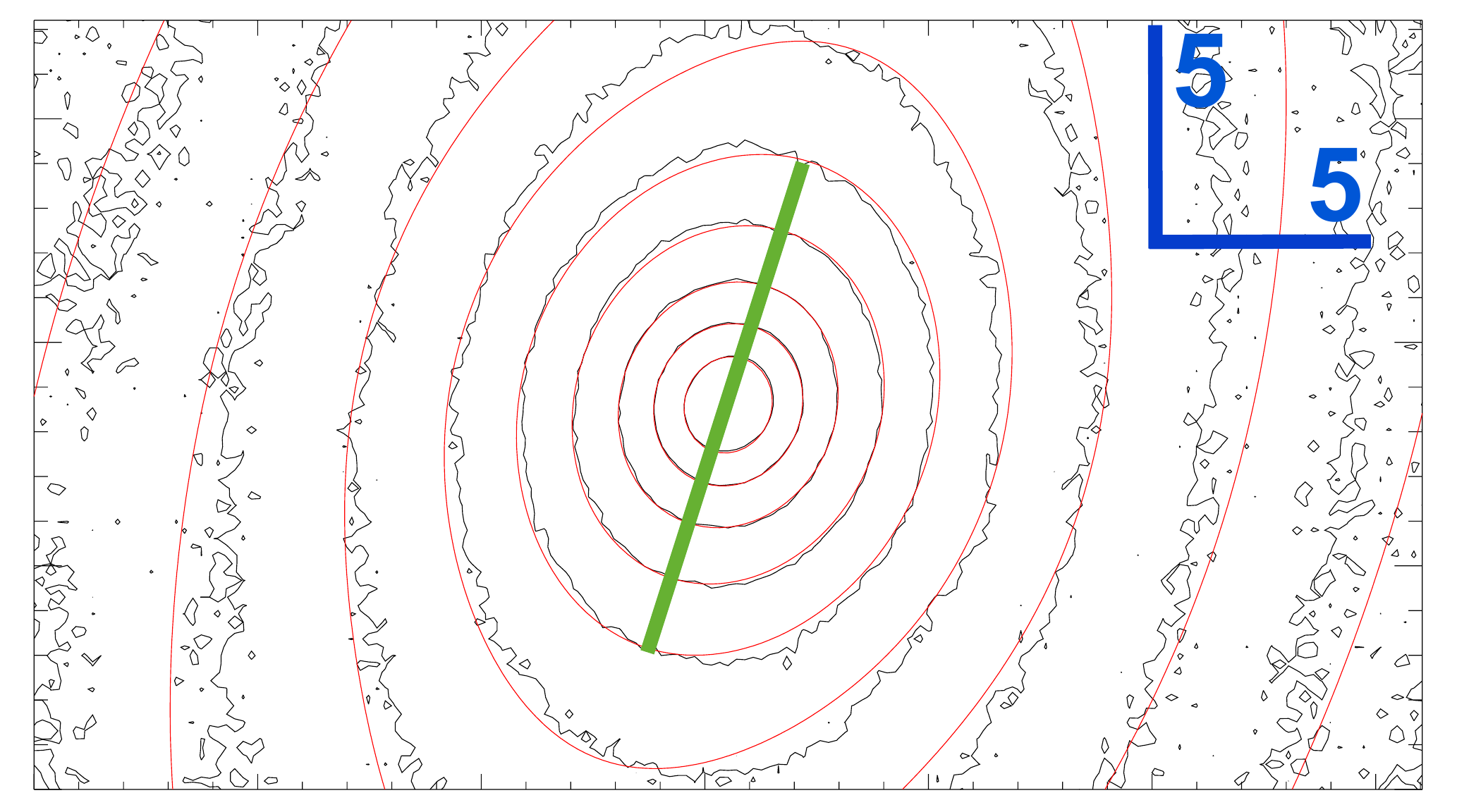}}\\
           \subfloat[Abell 646]{\includegraphics[height=2.9cm, width=4.6cm,  trim = 0mm 90mm 0mm 90mm, clip]{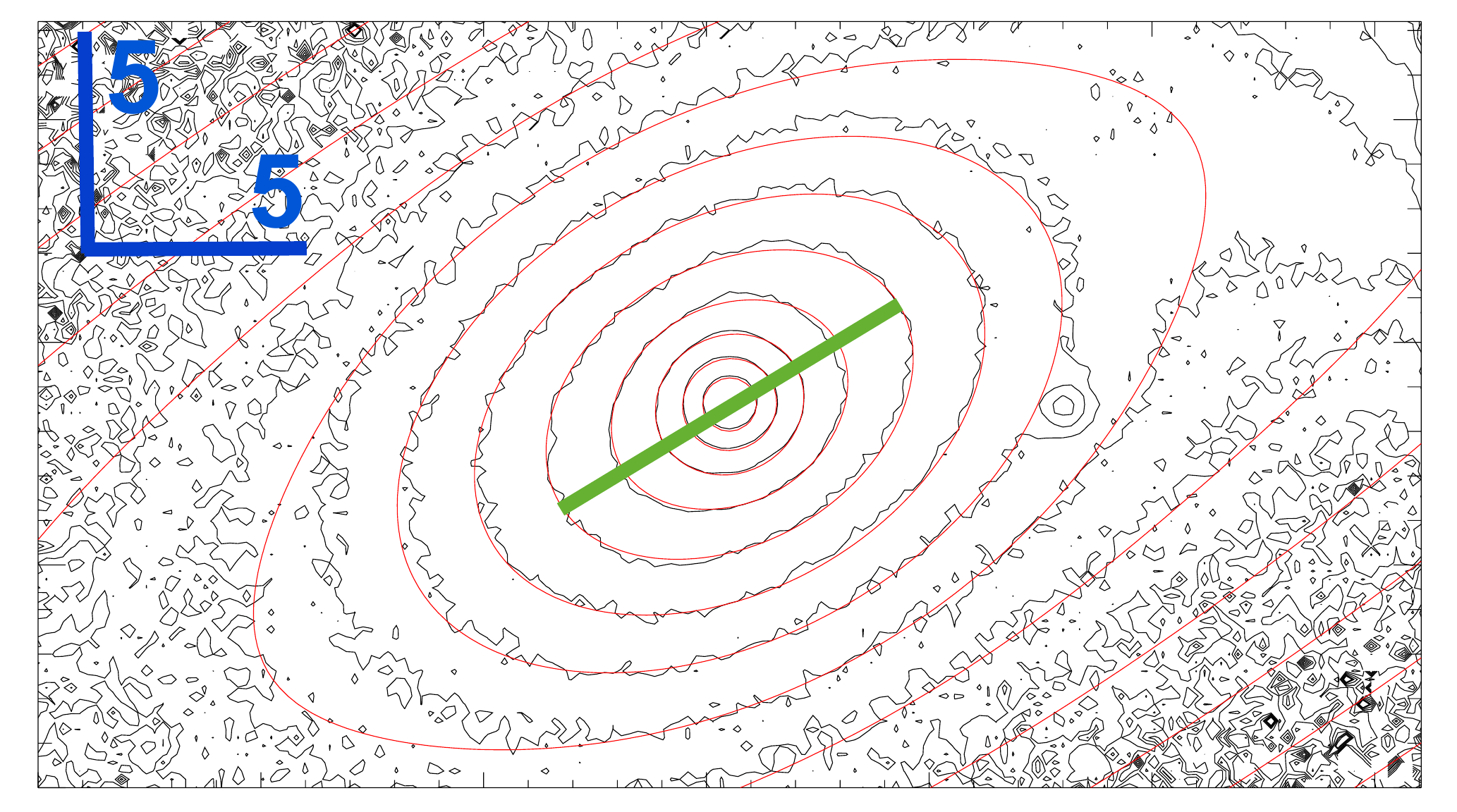}}
            \subfloat[Abell 754]{\includegraphics[height=2.9cm, width=4.9cm,  trim = 0mm 90mm 0mm 90mm, clip]{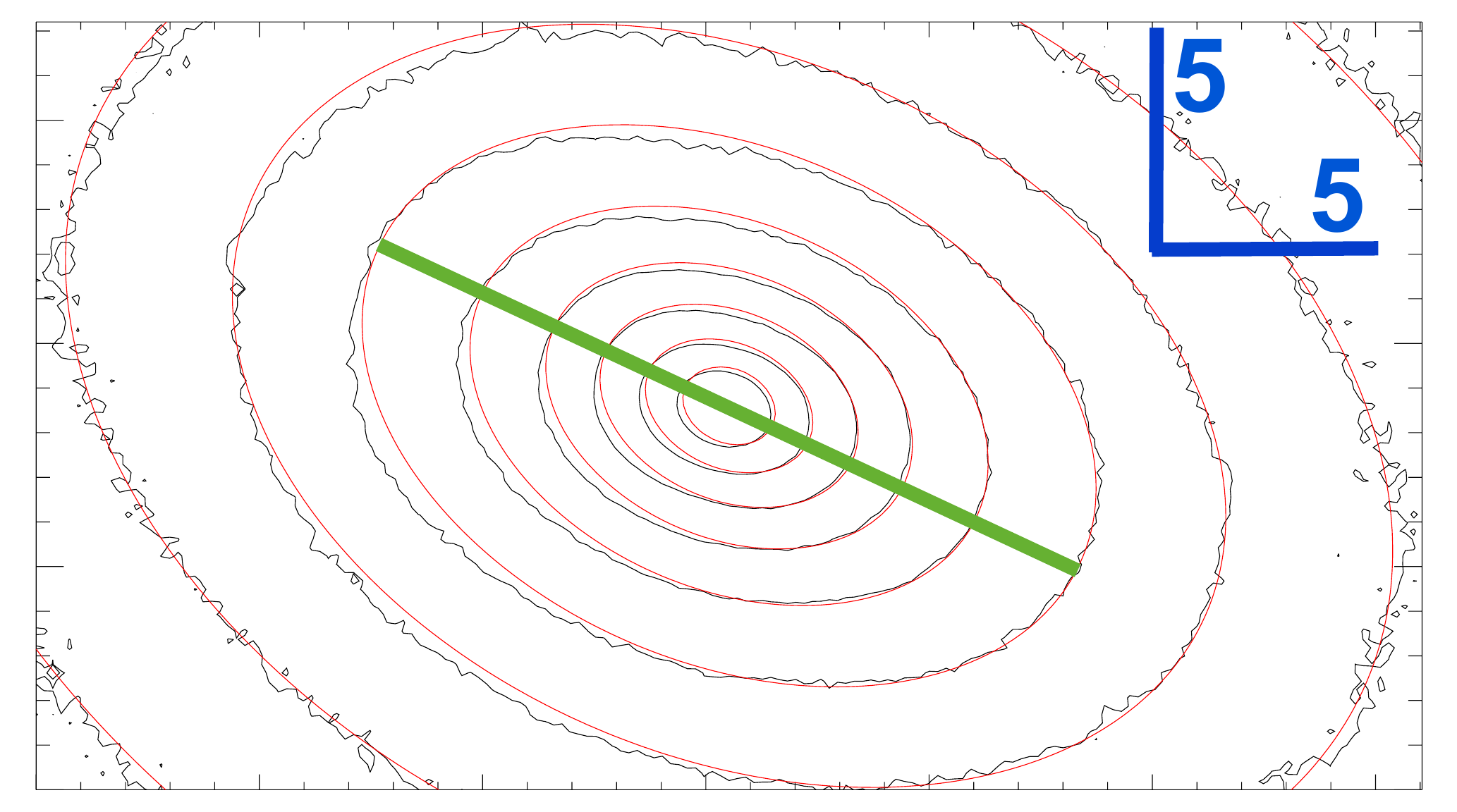}}
             \subfloat[Abell 780]{\includegraphics[height=2.9cm, width=4.9cm,  trim = 0mm 90mm 0mm 90mm, clip]{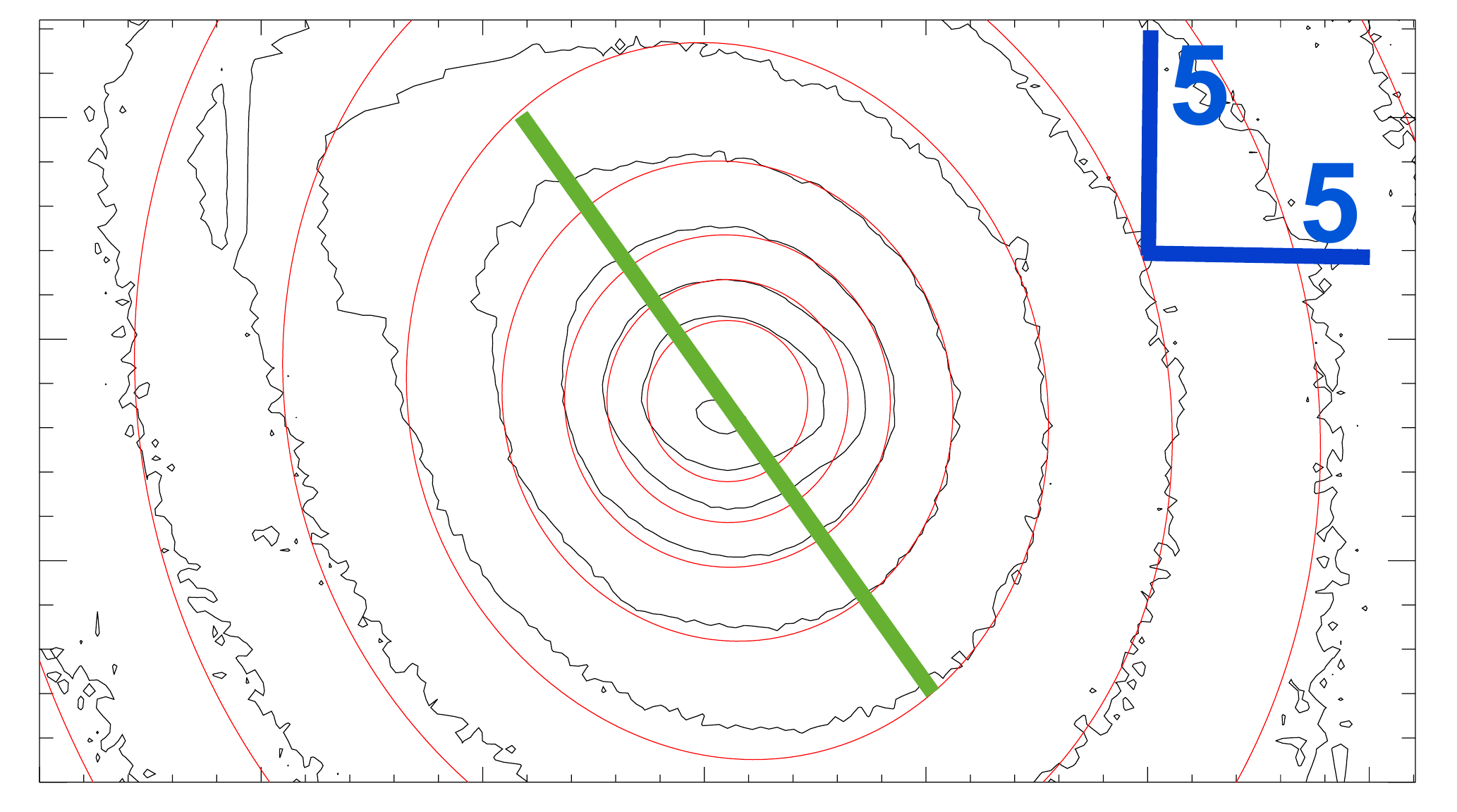}}
              \subfloat[Abell 963]{\includegraphics[height=2.9cm, width=3.0cm,  trim = 0mm 58mm 0mm 60mm, clip]{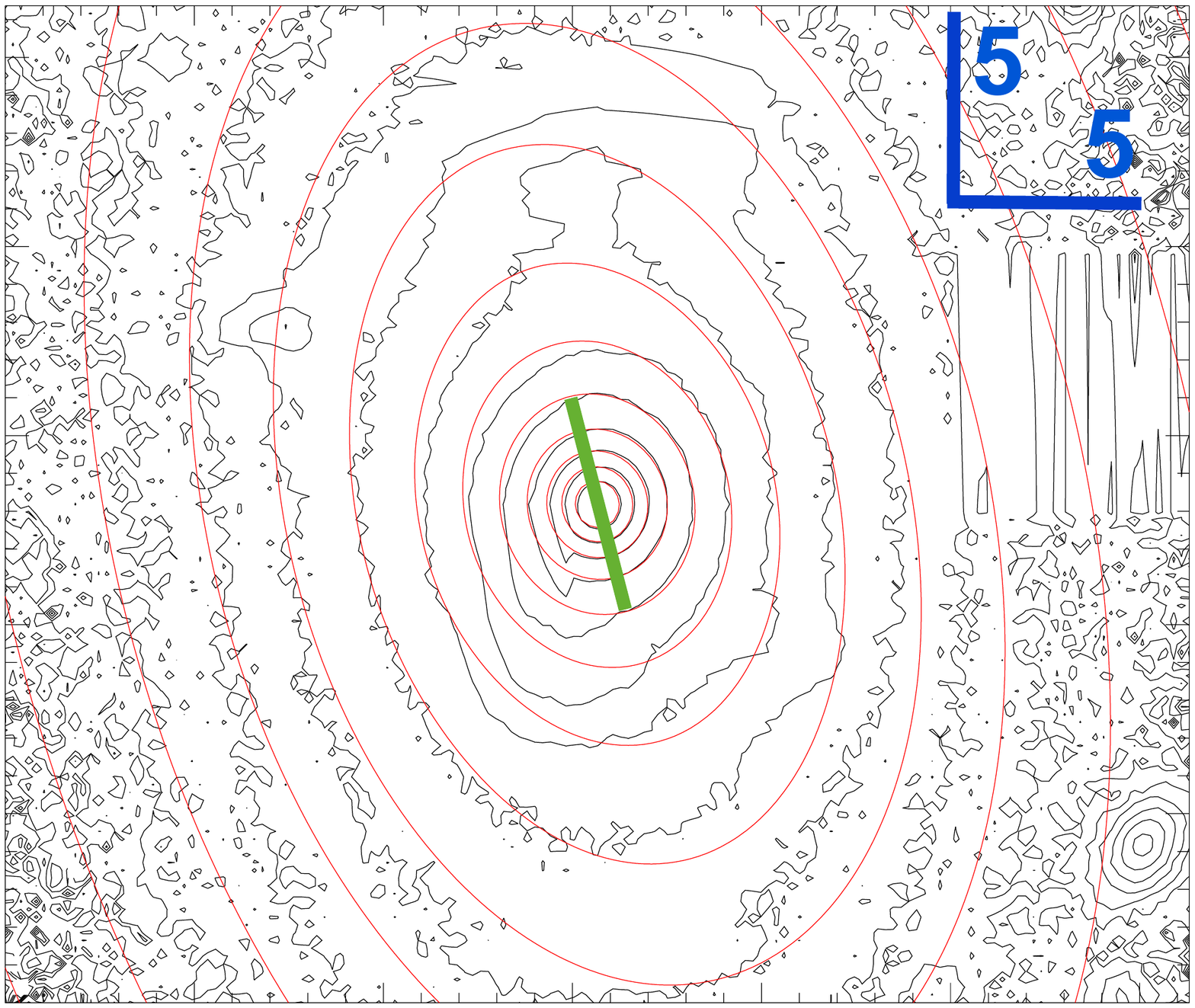}}\\
               \subfloat[Abell 1650]{\includegraphics[height=2.9cm, width=4.9cm,  trim = 0mm 90mm 0mm 90mm, clip]{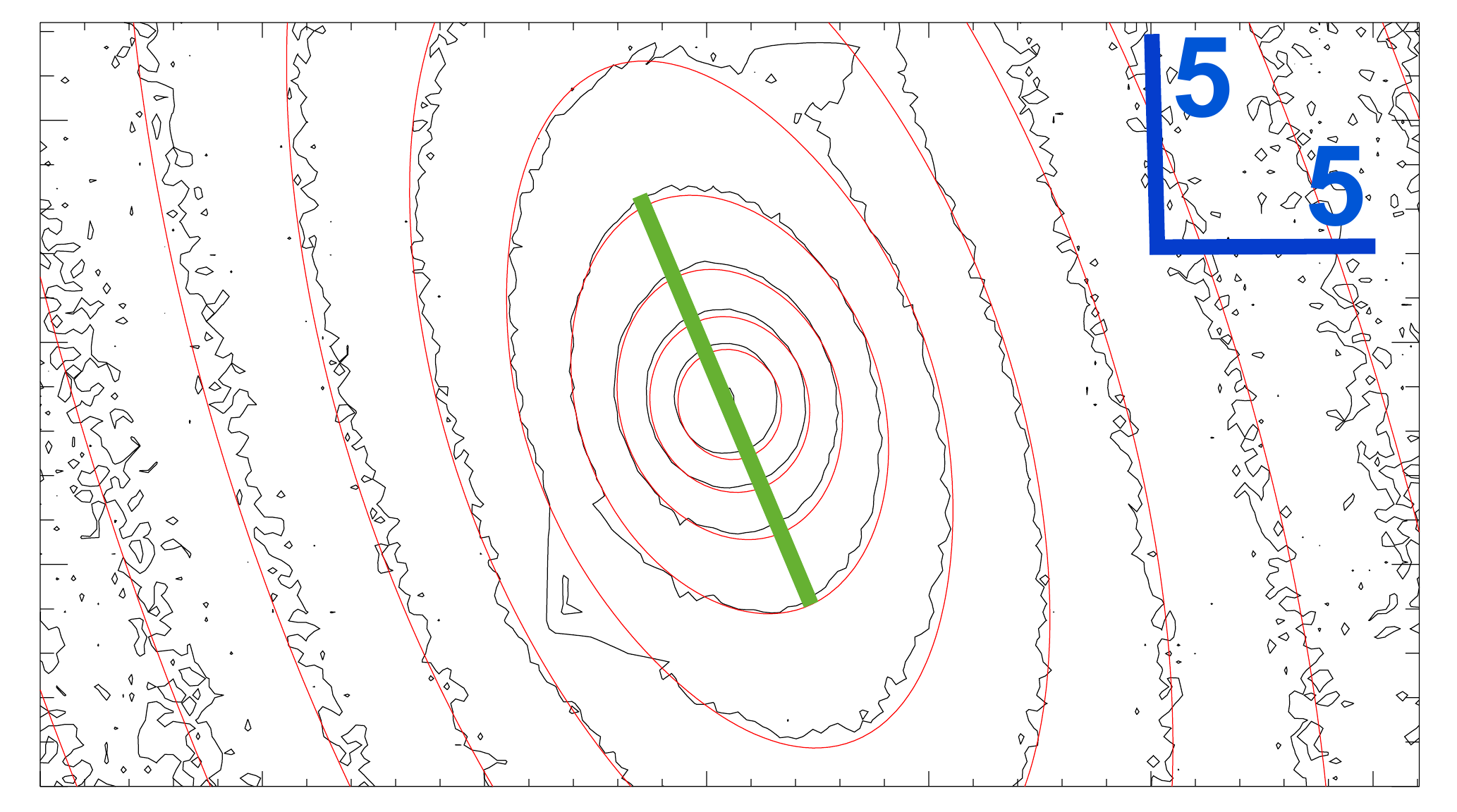}}
                \subfloat[Abell 1689]{\includegraphics[height=2.9cm, width=3.3cm,  trim = 0mm 65mm 0mm 65mm, clip]{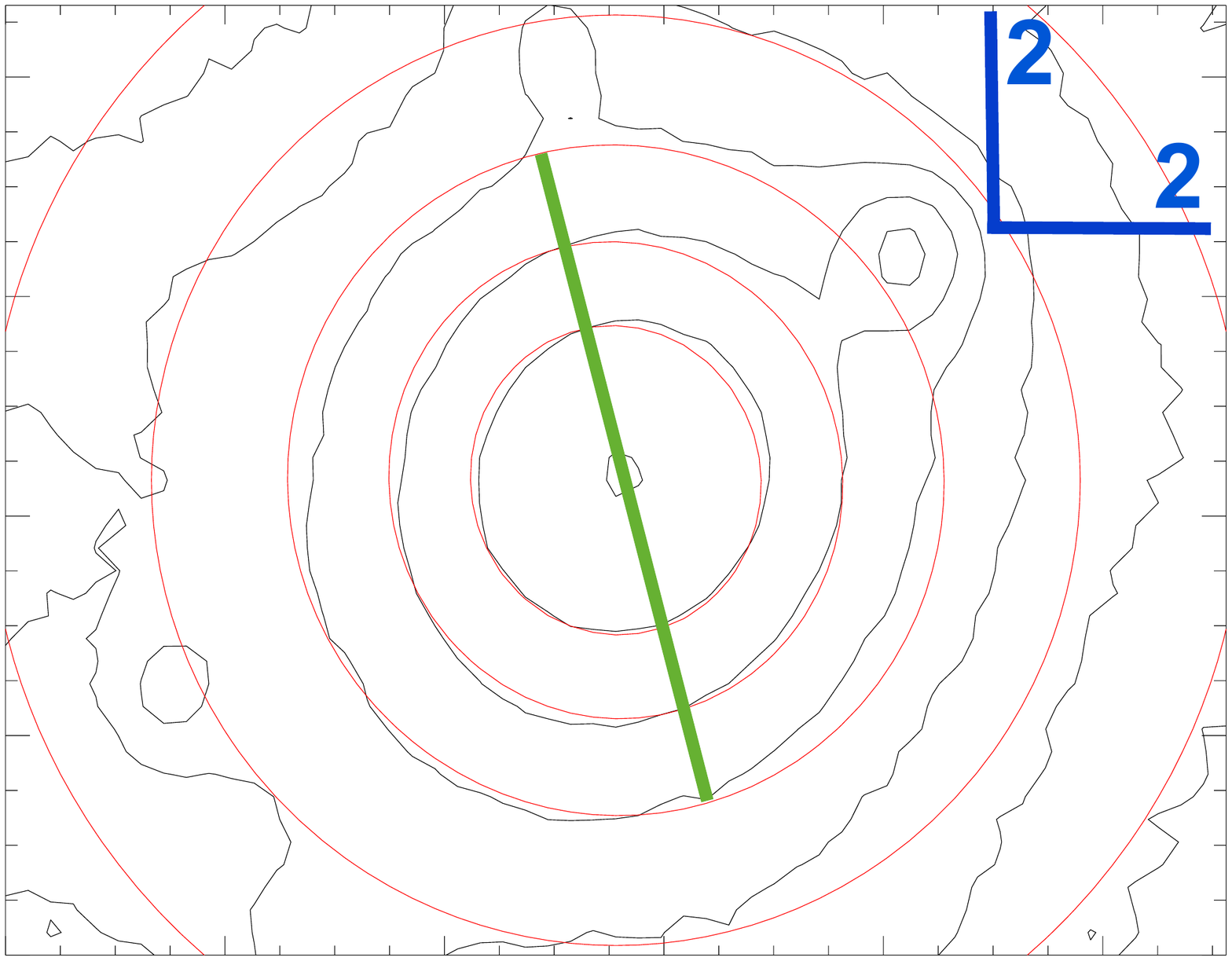}}
                 \subfloat[Abell 1763]{\includegraphics[height=2.9cm, width=3.3cm,  trim = 0mm 60mm 0mm 60mm, clip]{A1763_nuclear_slit3.pdf}}
                  \subfloat[Abell 1795]{\includegraphics[height=2.9cm, width=4.9cm,  trim = 0mm 90mm 0mm 90mm, clip]{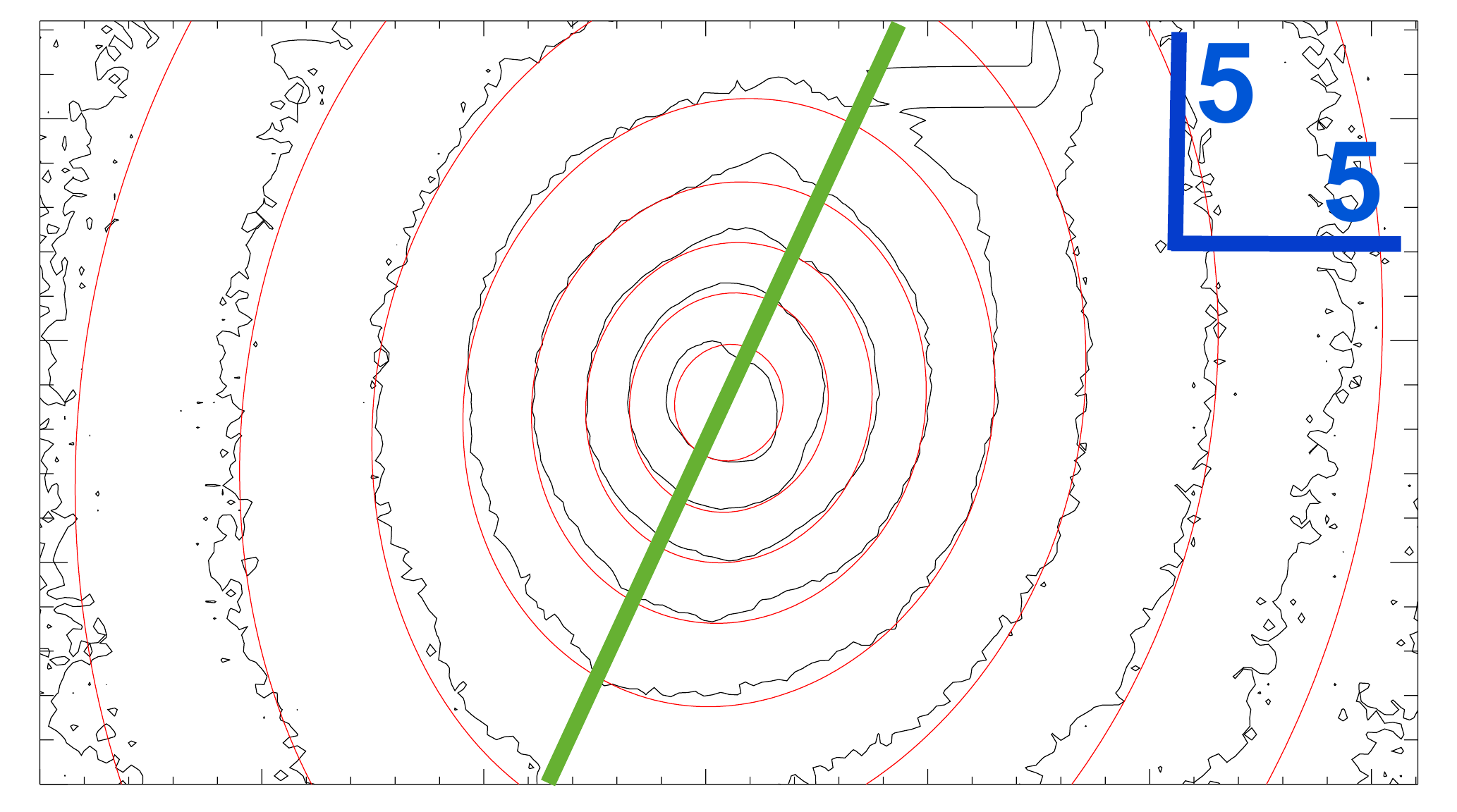}}\\
                   \subfloat[Abell 1942]{\includegraphics[height=2.9cm, width=4.5cm,  trim = 0mm 82mm 0mm 82mm, clip]{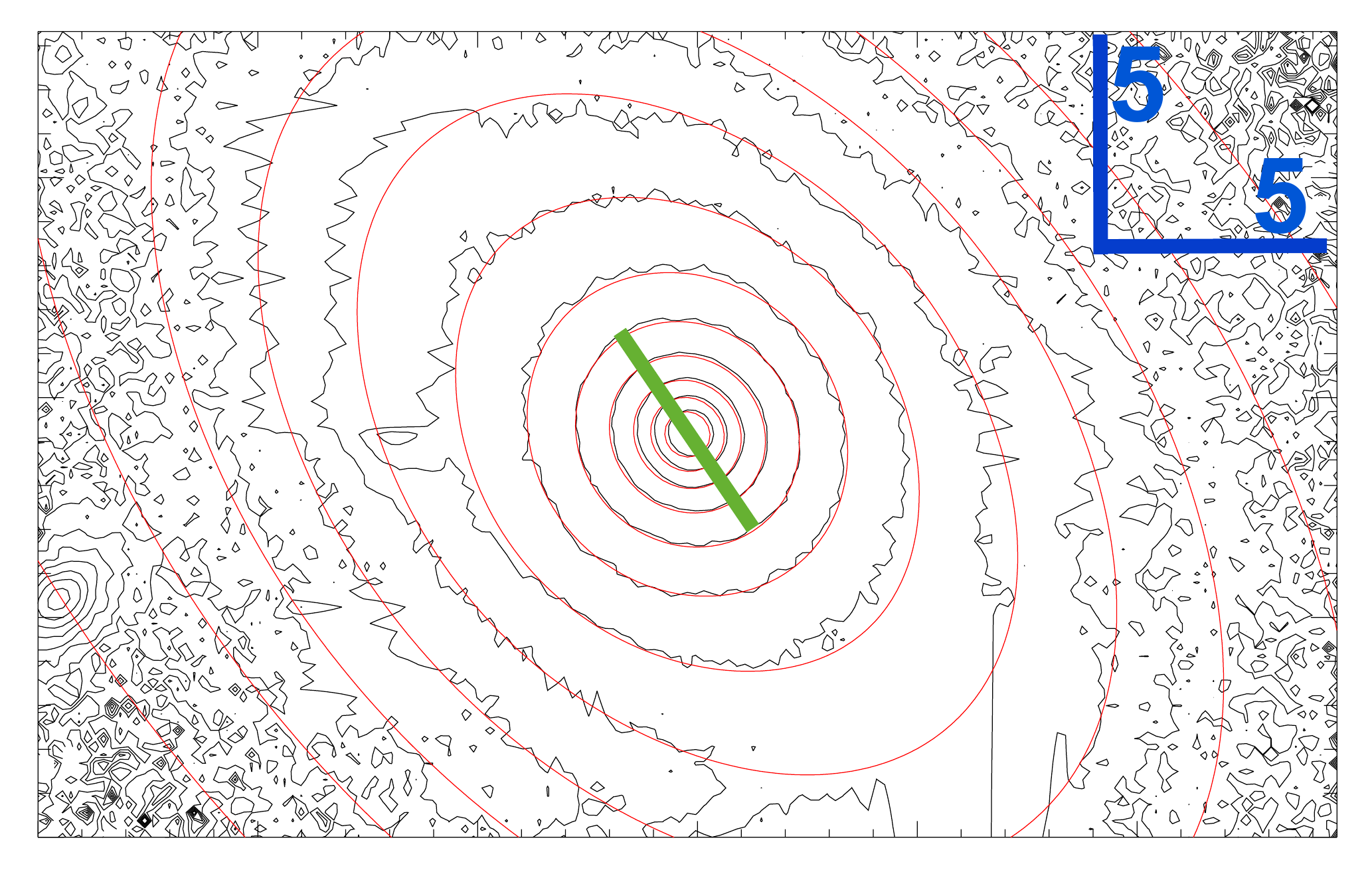}}
                    \subfloat[Abell 1991]{\includegraphics[height=2.9cm, width=4.5cm,  trim = 0mm 90mm 0mm 90mm, clip]{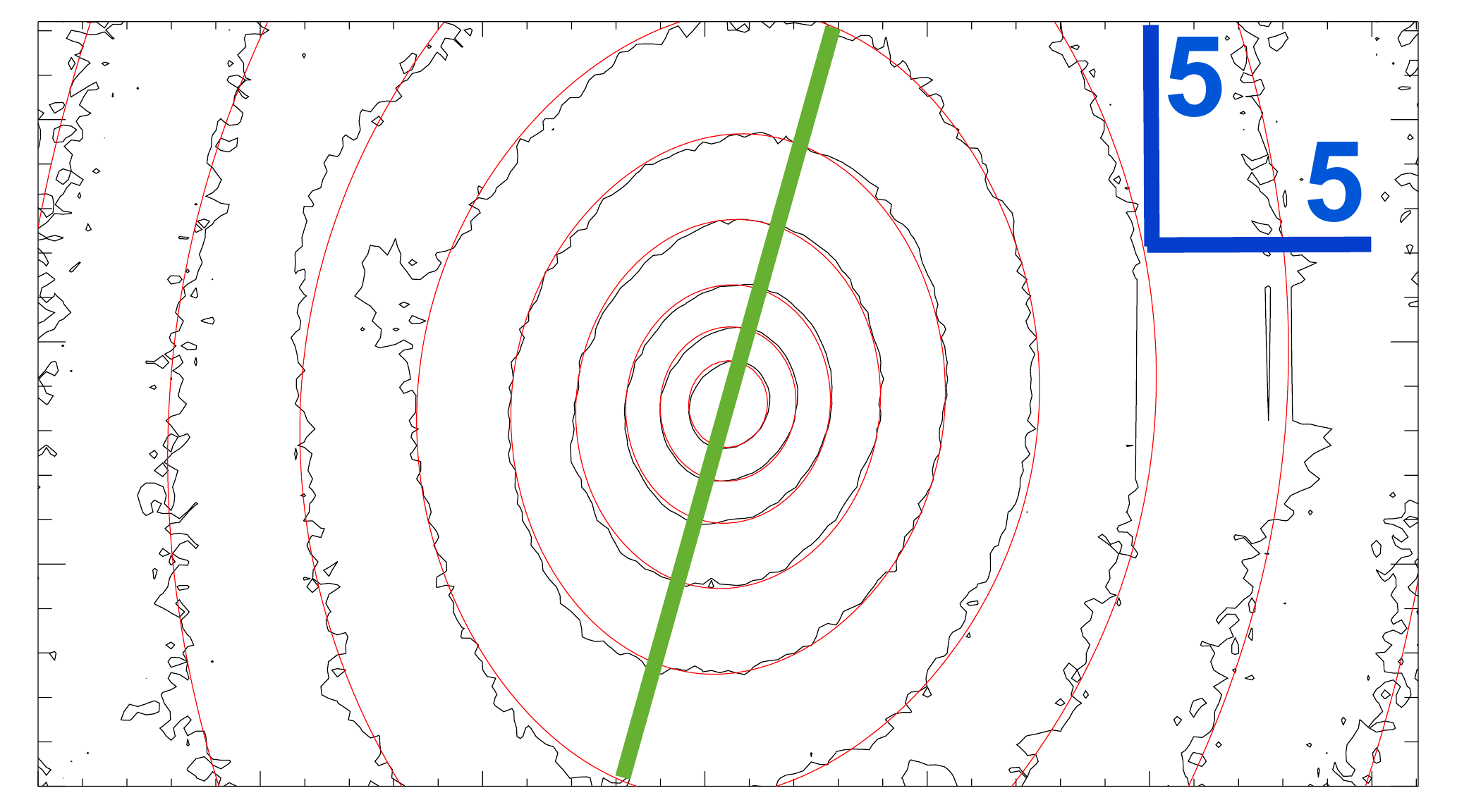}}
                     \subfloat[Abell 2029]{\includegraphics[height=2.9cm, width=4.5cm, trim = 0mm 90mm 0mm 90mm, clip]{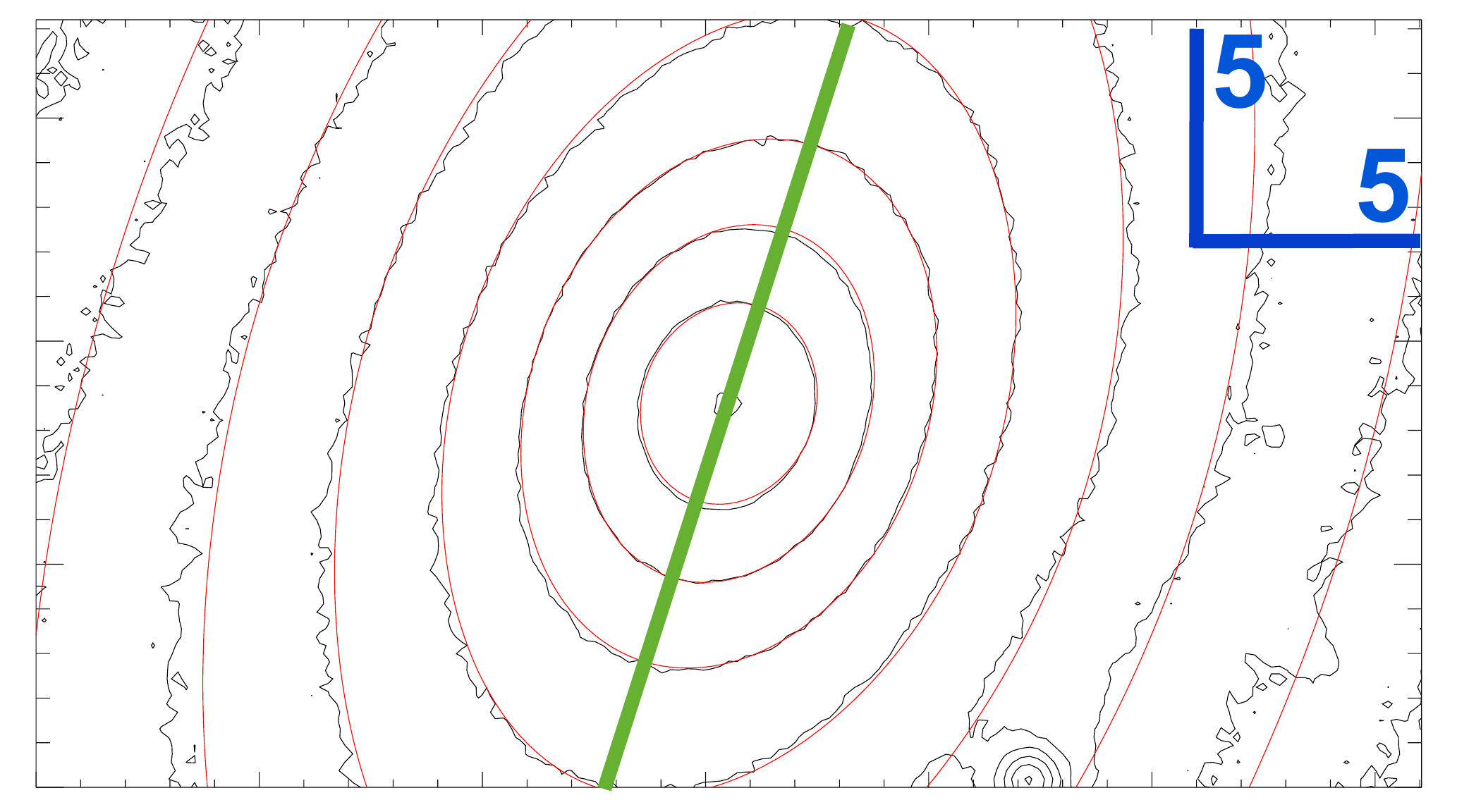}}
                      \subfloat[Abell 2050]{\includegraphics[height=2.9cm, width=4.5cm,  trim = 0mm 90mm 0mm 90mm, clip]{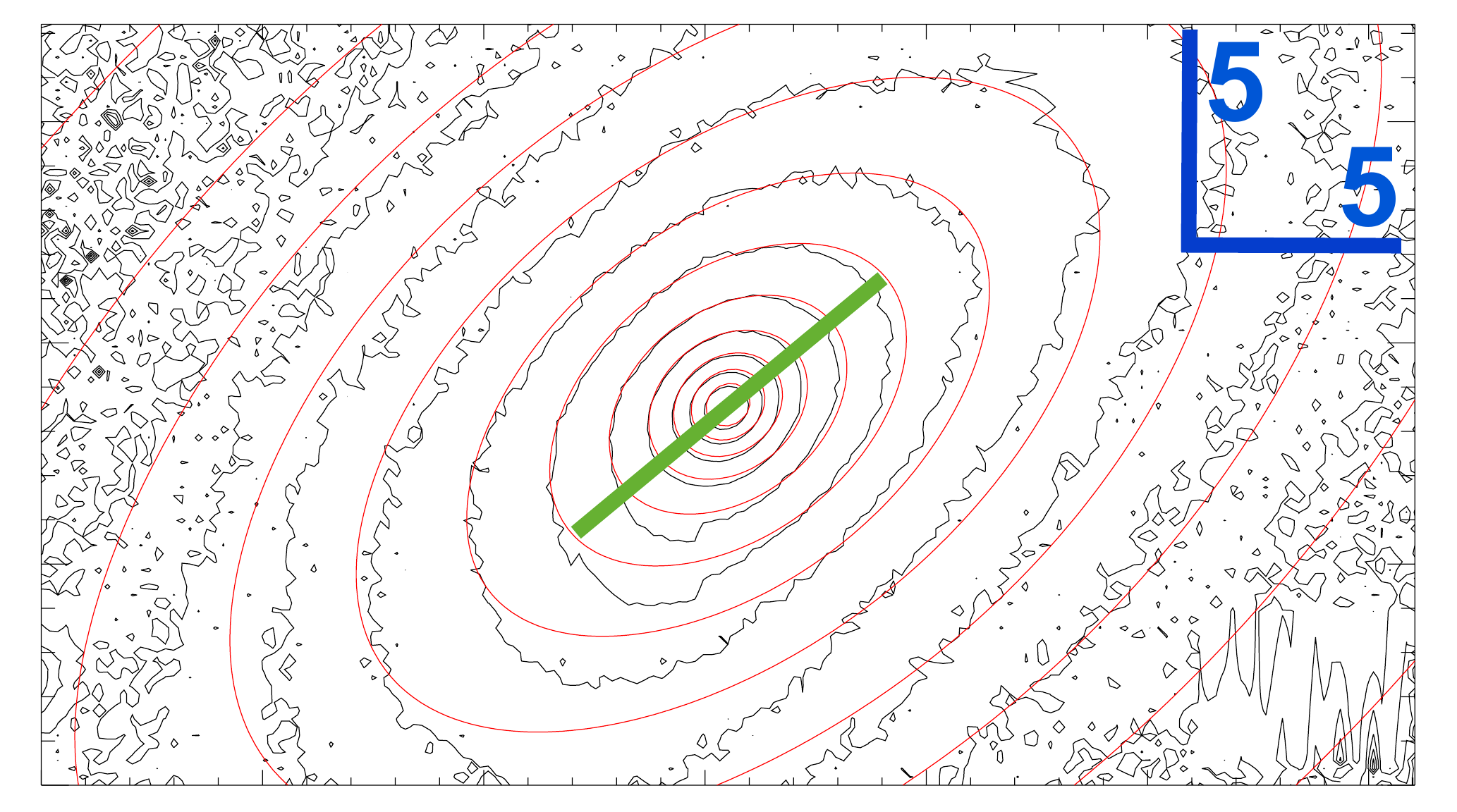}}\\
                       \subfloat[Abell 2055]{\includegraphics[height=2.9cm, width=4.9cm,  trim = 0mm 90mm 0mm 90mm, clip]{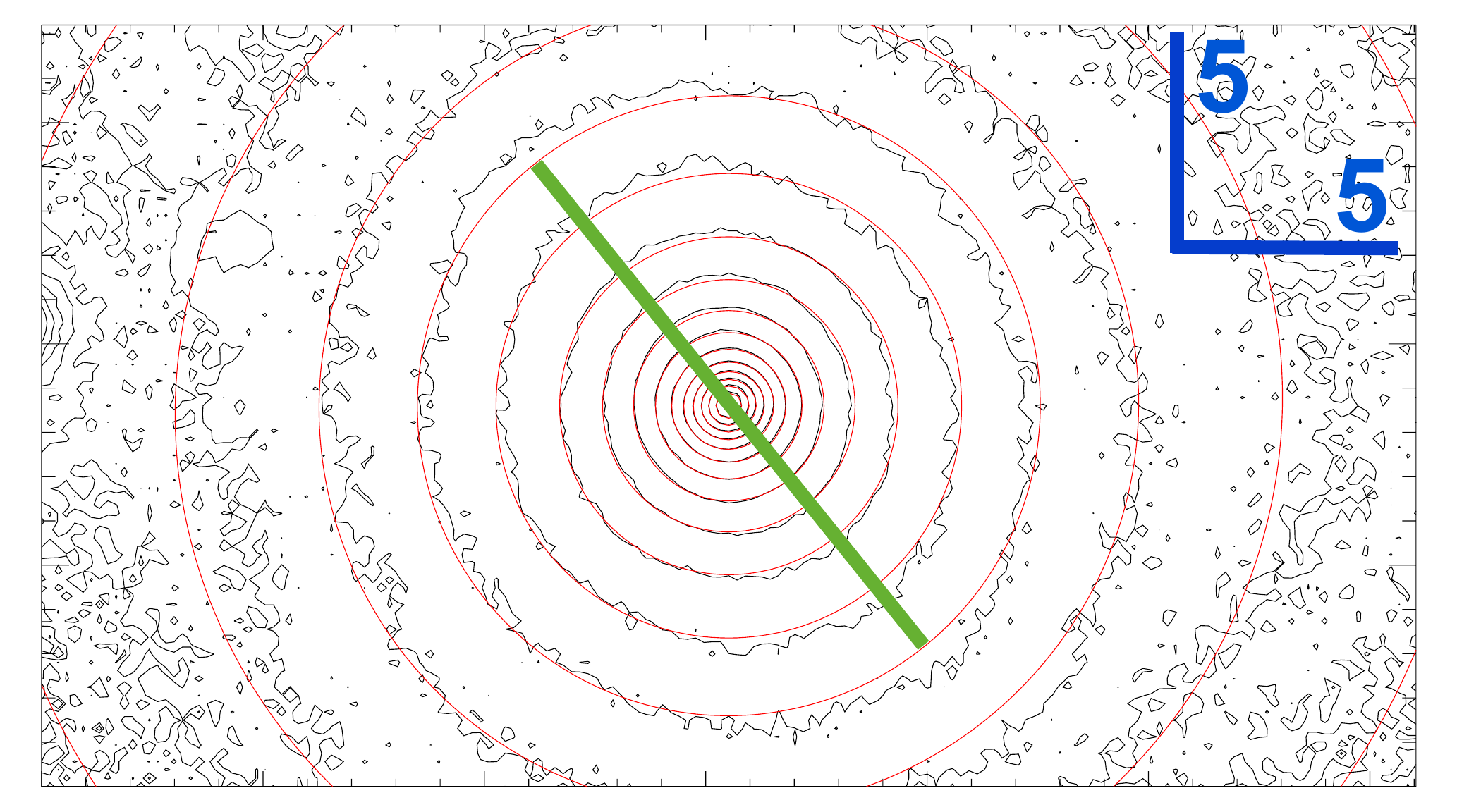}}
                        \subfloat[Abell 2142]{\includegraphics[height=2.9cm, width=4.9cm,  trim = 0mm 90mm 0mm 90mm, clip]{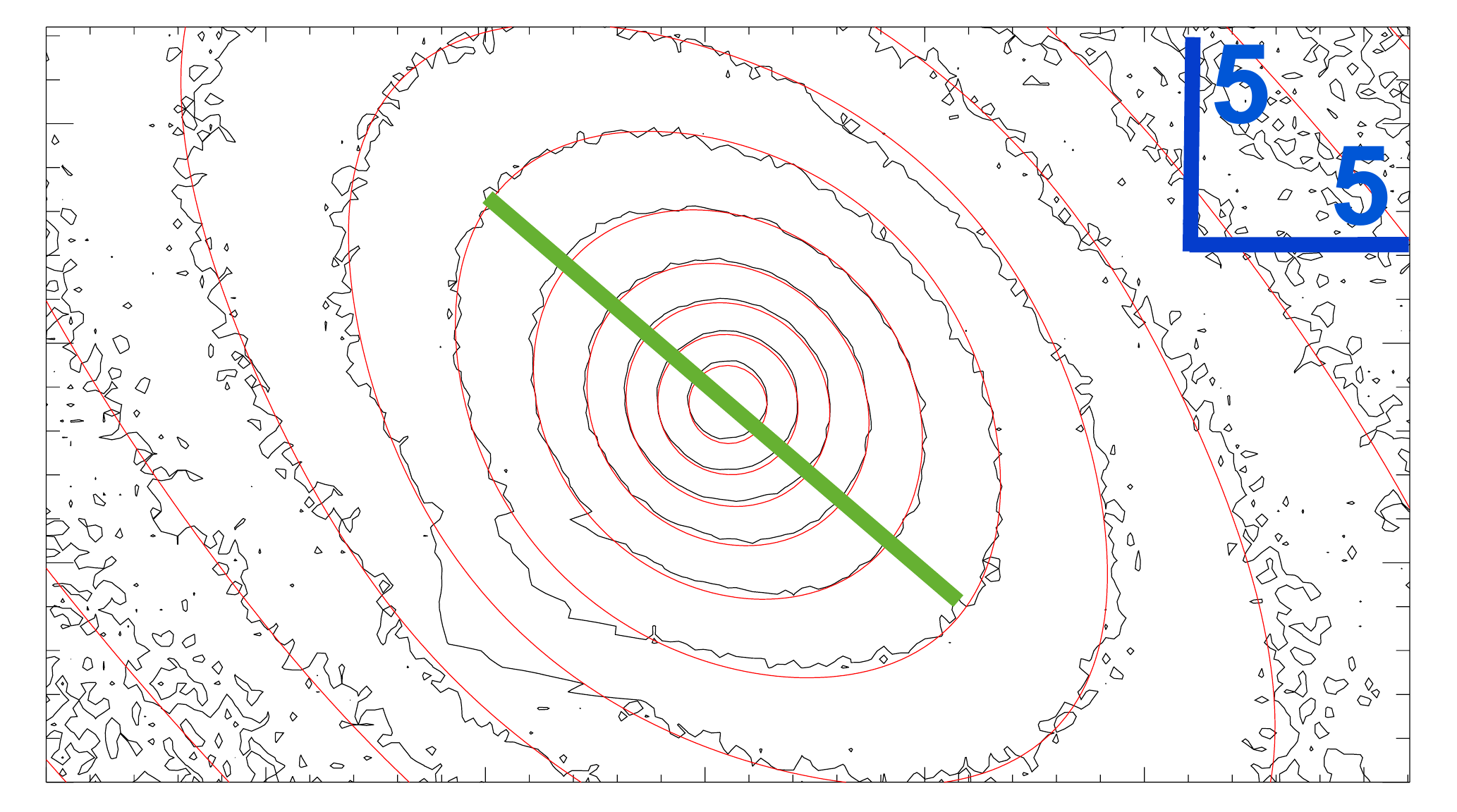}}
                         \subfloat[Abell 2259]{\includegraphics[height=2.9cm, width=3.9cm,  trim = 0mm 88mm 0mm 88mm, clip]{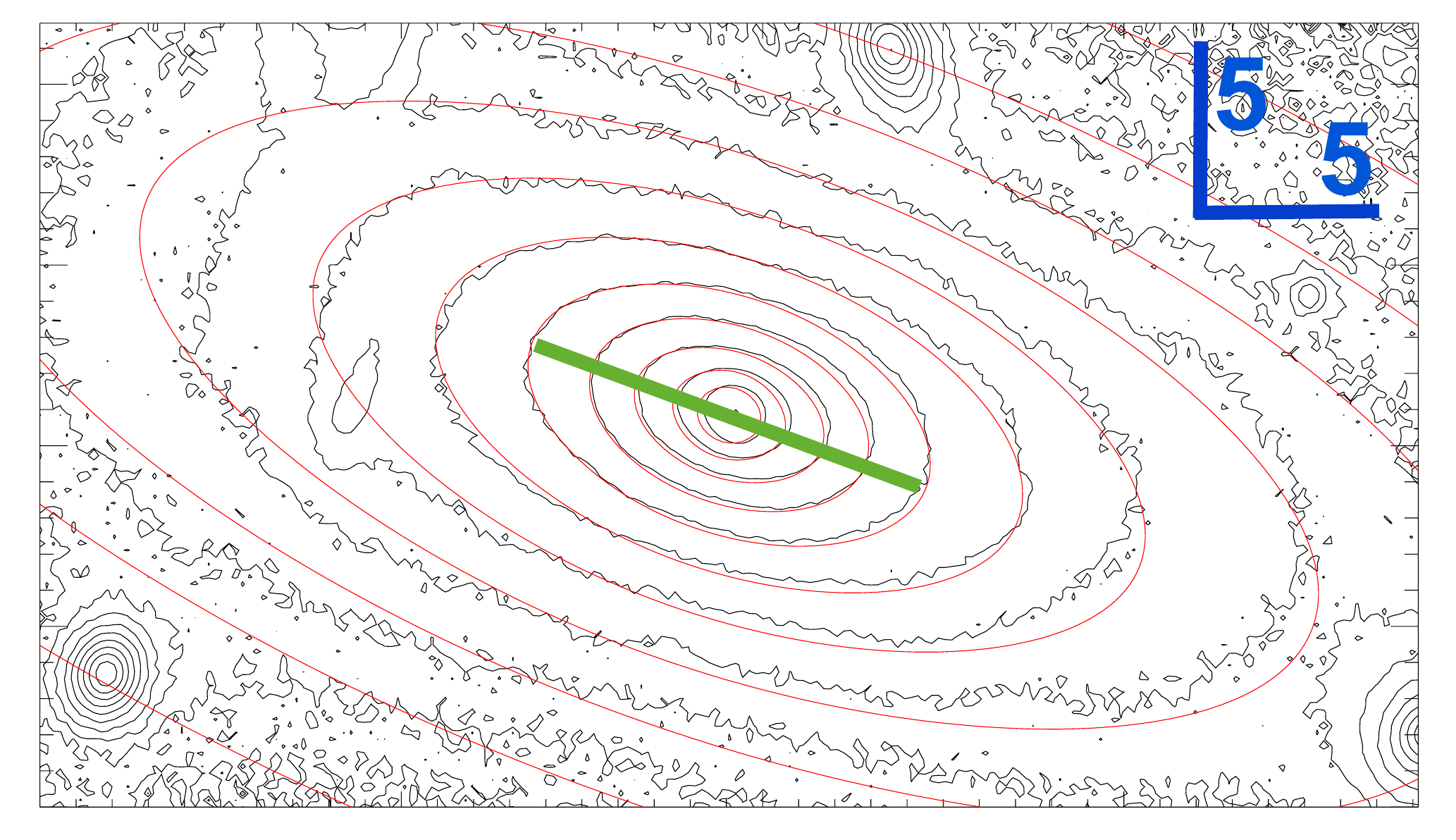}}
                          \subfloat[Abell 2261]{\includegraphics[height=2.9cm, width=3.2cm,  trim = 0mm 65mm 0mm 65mm, clip]{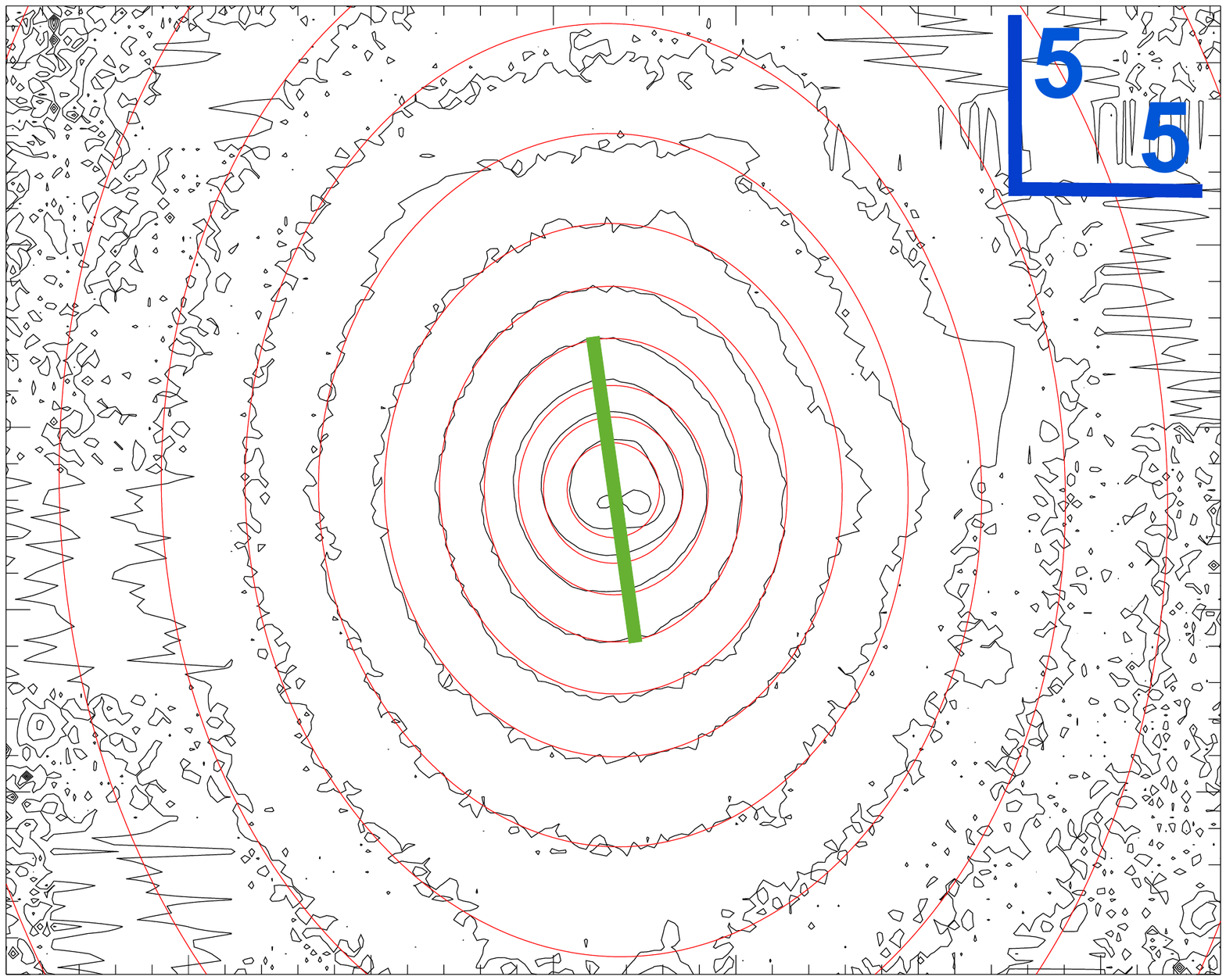}}\\
                           \subfloat[Abell 2319]{\includegraphics[height=2.9cm, width=4.4cm,  trim = 0mm 90mm 0mm 90mm, clip]{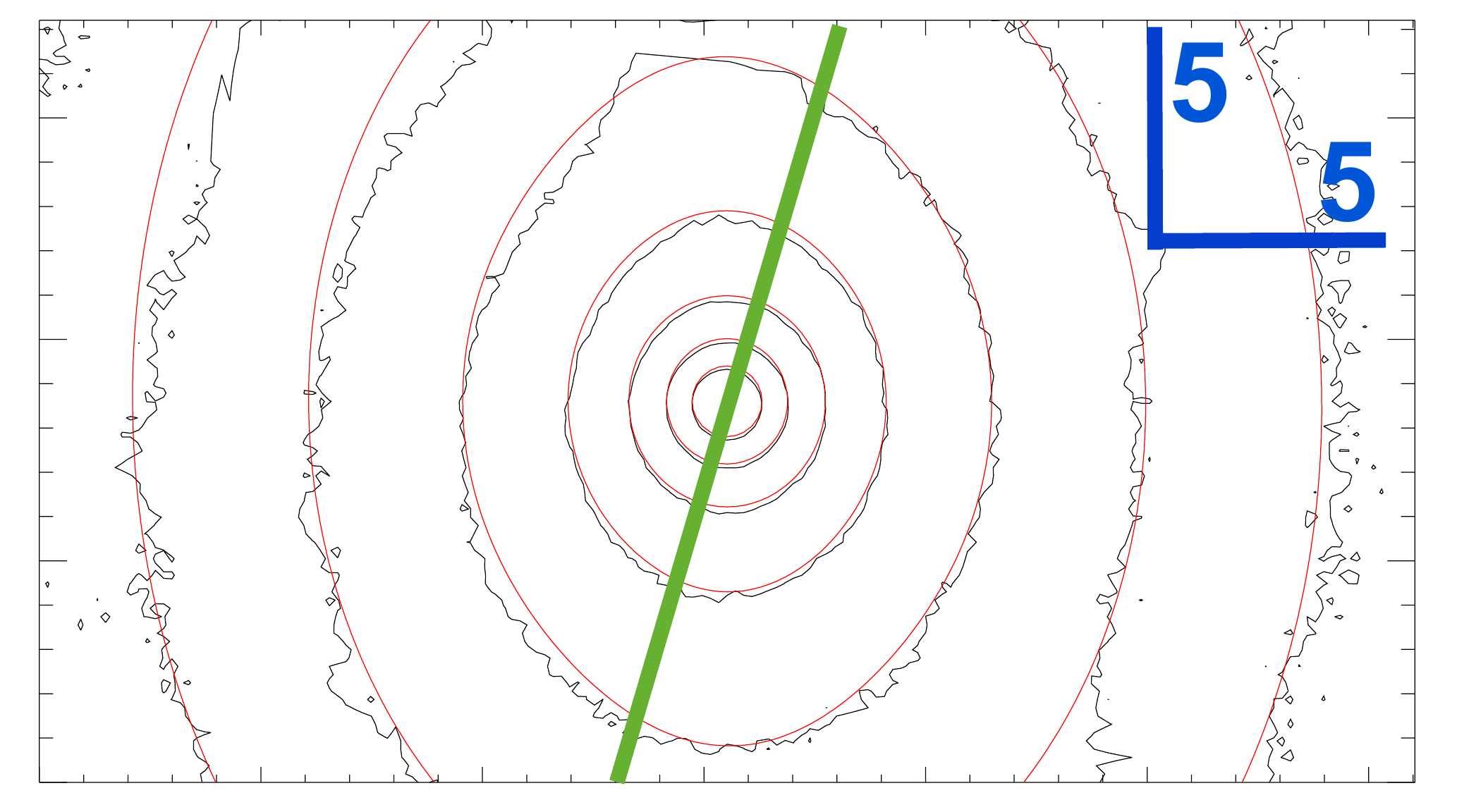}}
                            \subfloat[Abell 2420]{\includegraphics[height=2.9cm, width=4.6cm,  trim = 0mm 90mm 0mm 90mm, clip]{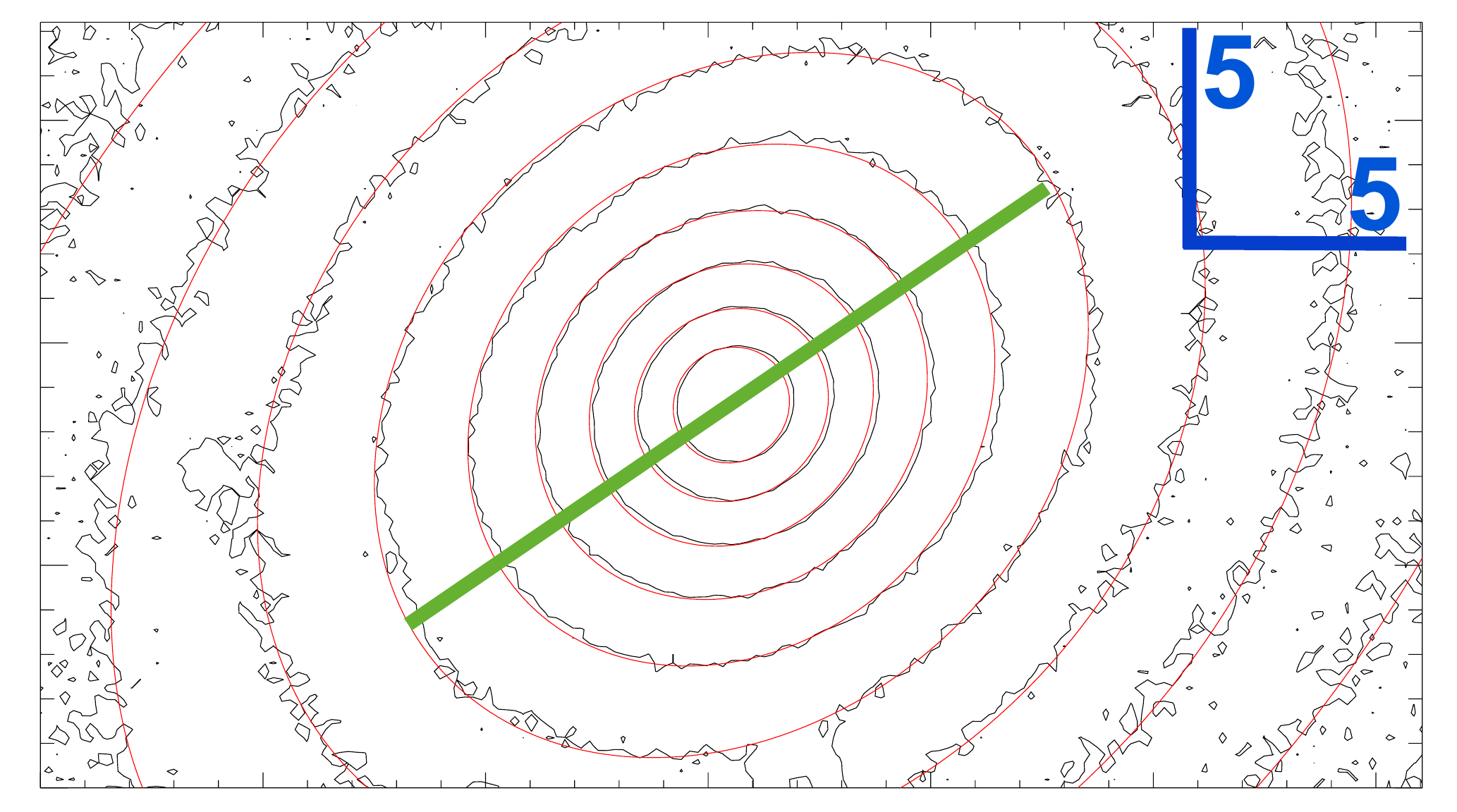}}
                             \subfloat[Abell 2537]{\includegraphics[height=2.9cm, width=4.4cm,  trim = 0mm 80mm 0mm 80mm, clip]{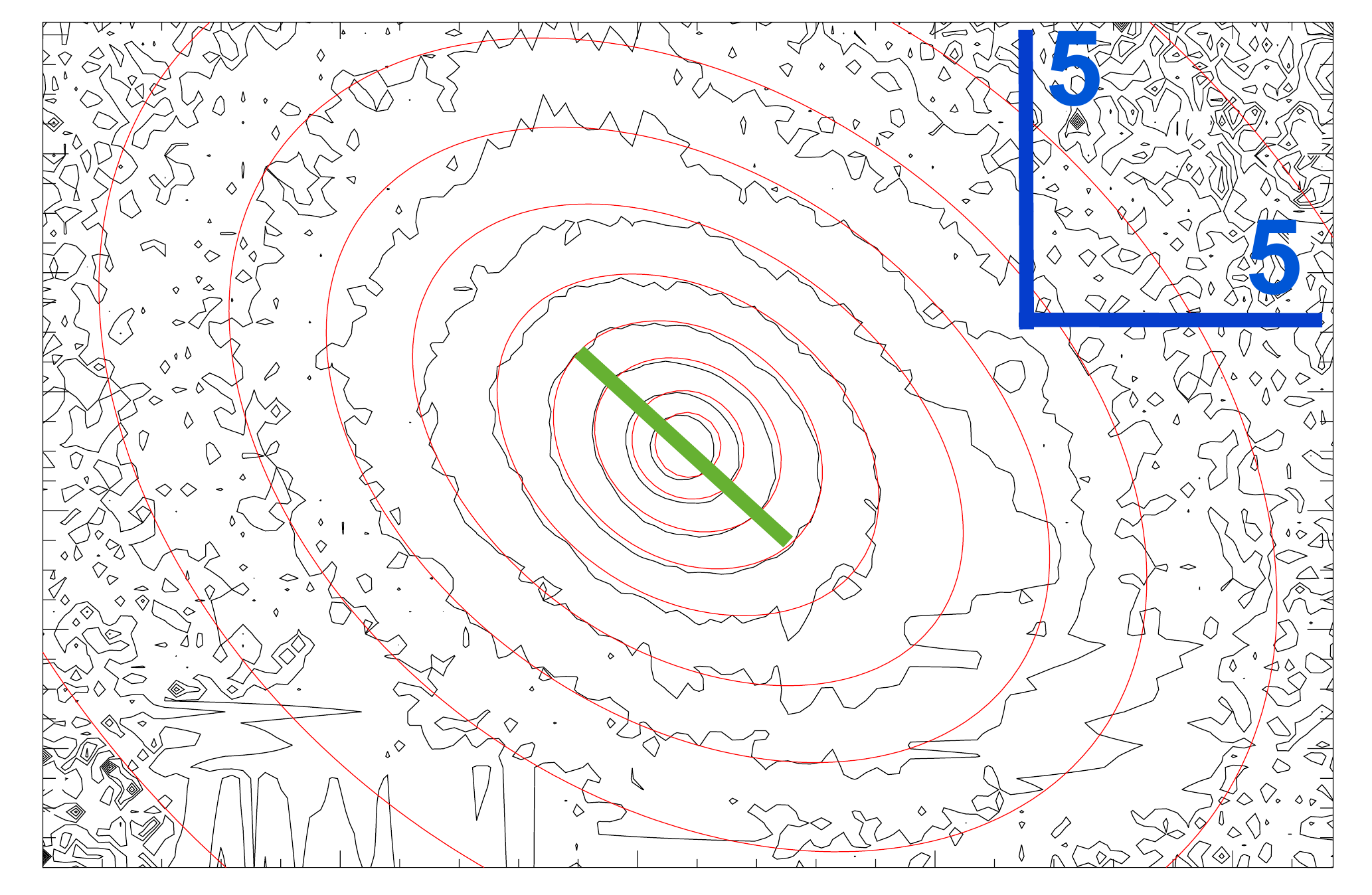}}
                              \subfloat[MS1455+22]{\includegraphics[height=2.9cm, width=3.3cm,  trim = 0mm 60mm 0mm 60mm, clip]{MS14_nuclear_slit3.pdf}}\\
                  \caption{MGE fits to the $r-$band images. The green line indicates the position of the slit, and the blue lines the scale in arcsec.}
\label{MGEFig}
\end{figure*}

Figure \ref{MGEFig} shows the MGE surface brightness modelling overplotted on the contours of the surface brightness in the $r$-band images. The green line indicates the position of the slit, and the blue lines the scale in arcsec. We only show the images of the nuclei where the slit was located and the kinematics measured. Our wide-field CFHT imaging, and the MGE modelling, extend far beyond the images of the nuclei shown in Figure \ref{MGEFig} and the spatial extent of the available kinematics. Our MGE fits extend beyond the effective radii ($R_{e}$) for all the BCGs modelled here (on average it extends to $\sim$3.2$R_{e}$), with the exception of Abell 68 for which it extends to $\sim$0.73$R_{e}$ (where $R_{e}$ is 41 kpc). Therefore in all cases, the MGE modelling extends well beyond the available kinematics (>15 kpc).

For more details on our photometric observations for this cluster sample see \citet{Hoekstra2007}, \citet{Bildfell2008}, \citet{Hoekstra2012}, \citet{Sand2012}, and \citet{Bildfell_thesis}. The unique combination of long integration times for the imaging and sub-arcsecond resolution allows us to simultaneously resolve the cores of BCGs and reliably trace their surface brightness profiles out to large radii, which makes this data set ideal for the purpose of our study. 

We measure the mean absolute deviation between the fitted MGE and the data (across the whole measured area extending well beyond the nuclei as described above) expressed as a fraction. Of the 25 BCGs we model with MGE, we find an average mean absolute deviation of 10.2 per cent, with Abell 646 and Abell 1689 showing the highest deviation (both 22.2 per cent). We emphasise that this is deviation as measured across the whole radial region (on average ~3.2Re), and mostly driven by objects in the line-of-sight in the outer parts of the galaxies. The deviation across the nuclei where we fit the kinematics is much smaller. To quantify the possible effect on our dynamical models, we increase the stellar mass input by 10 per cent (likely to be an overestimate) in the fits for Abell 68 and Abell 646 as a test (for the dynamical fits using just a stellar mass component). For Abell 68, we find that $\beta_{z}$ decreased by 0.05 and $\Upsilon_{\star \rm DYN}$ decreased by 0.30 (from the values given in Table \ref{AdditionalTable}). For Abell 646, we find that $\beta_{z}$ decreased by 0.02 and $\Upsilon_{\star \rm DYN}$ decreased by 0.51.

\section{Slit Position Angle (PA)}
\label{SlitPA}

The slit PA of the long-slit observations and major axis PA derived from the MGE fitting procedure is within 15 degrees of each other for all except two of the BCGs analysed here, with the average difference being six degrees. The two exceptions are the BCG in Abell 1689, where the difference is 84 degrees, and Abell 2055 where the difference is 62 degrees. For Abell 2055, we find a very low $\beta_{z}$, and as stated in Section \ref{sec:results} do not include the galaxy in further analysis since it is a known BL Lac. We have also checked than none of our conclusions change if we would exclude Abell 1689.

\section{Effect of the PSF on the solutions}
\label{PSF}

\begin{figure}
\centering
   \subfloat{\includegraphics[scale=0.3]{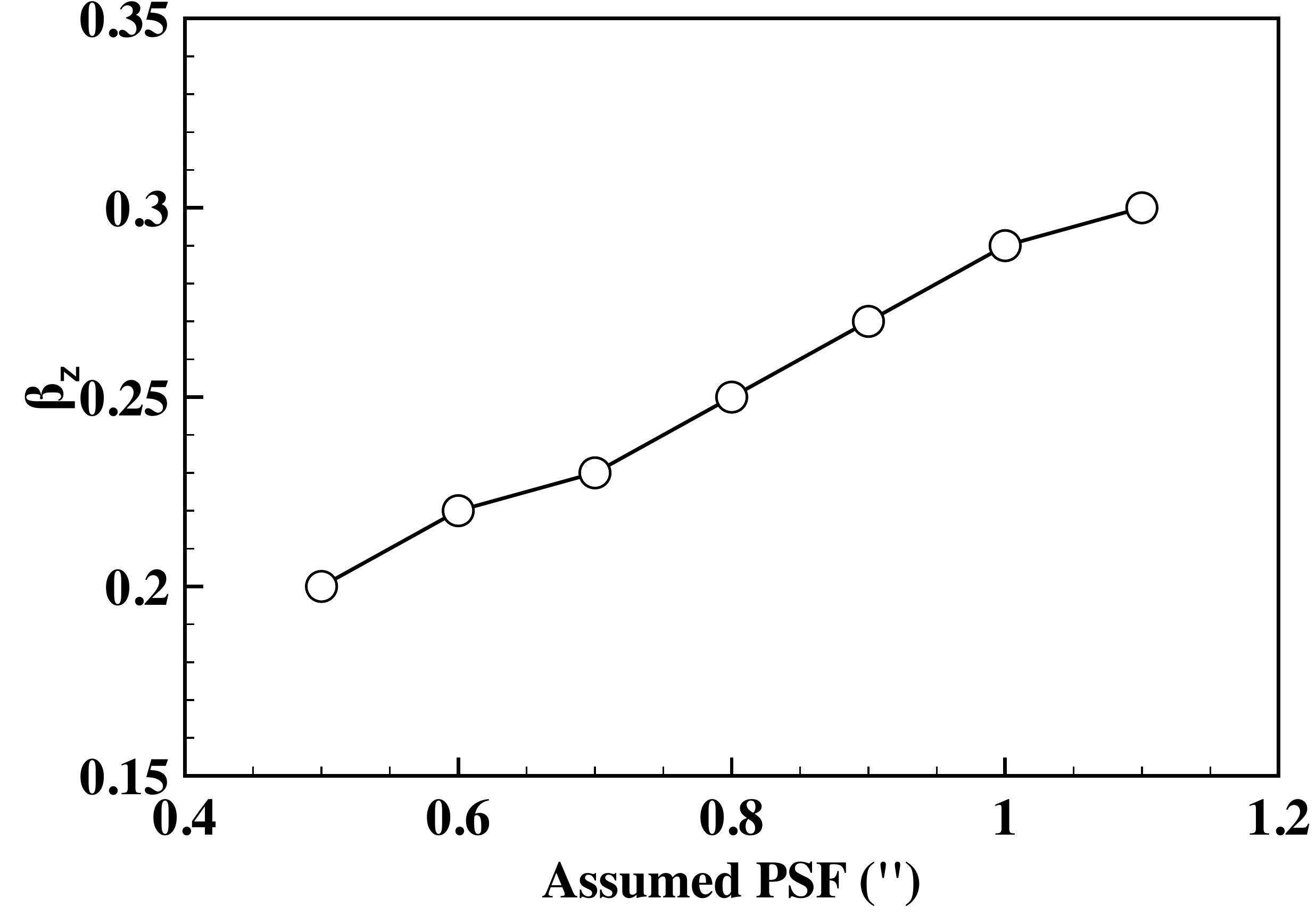}}
   \caption{Illustrating the sensitivity of our dynamical modelling is to variations in the PSF, using the BCG in Abell 68 (with a measured PSF of 0.8$\arcsec$) as an example. Here, we use just a stellar mass component, and illustrate the increase in $\beta_{z}$ with an increase in the PSF. As the PSF increases from 0.5$\arcsec$ to 1.1$\arcsec$, $\Upsilon_{\star \rm DYN}$ also decreases from 5.10 to 4.84. The uncertainties of the measured PSFs are small ($<$0.05$\arcsec$). }
\label{table_PSF}
\end{figure}

We test how sensitive our dynamical modelling is to variations in the PSF by using the BCG in Abell 68 as an example in Figure \ref{table_PSF}, and varying the PSF (used for convolution in the MGE and JAM analysis before comparison to the data) between 0.5$\arcsec$ and 1.1$\arcsec$. Here, we use just a stellar mass component. The $\beta_{z}$ parameter seems the most sensitive to the PSF and we plot it in Figure \ref{table_PSF}, and describe the other parameters in the figure caption. The uncertainties of the measured PSFs are small ($<$0.05$\arcsec$). 

\section{Effect of the radius and mass of the black hole}
\label{sec:BH}

In Table \ref{BH}, we illustrate how sensitive the best-fitting parameters are to changes in the mass or radius of the black hole mass component. We again use the BCG in Abell 68 as an example, where $M_{\rm CEN}$ = $M_{\rm BH}$ = 2.22$\times 10^{9}$ M$_{\sun}$ (0.2\arcsec) was used in the original model. The mass and radius assumed for the black hole do not have a significant effect on the best-fitting parameters.

\begin{table}
\caption{Influence of the central mass component on the best-fitting parameters in the dynamical modelling of the BCG in Abell 68.}
\label{BH}
\centering
\begin{tabular}{lccc}
\toprule
\multicolumn{4}{c}{\textbf{Abell 68}} \\
fit  & $\beta_{z}$  &  $\Upsilon_{\star \rm DYN}$ & $\chi^{2}/DOF$\\
$\star$ + CEN  &  & (M$_{\sun}$) & \\
\midrule
$M_{\rm BH}$ (0.2\arcsec)  & 0.26 & 4.97 & 2.10 \\
\hline
0.1$\times M_{\rm BH}$ (0.2\arcsec) & 0.25 & 4.97 & 2.10 \\
5.0$\times M_{\rm BH}$ (0.2\arcsec)     & 0.25 & 4.95 & 2.10 \\
10.0$\times M_{\rm BH}$ (0.2\arcsec)       & 0.24 & 4.94 & 2.09 \\
\hline
$M_{\rm BH}$ (0.05\arcsec)         & 0.25 & 4.96 & 2.10 \\
$M_{\rm BH}$ (0.5\arcsec)           & 0.25 & 4.97 & 2.09 \\
$M_{\rm BH}$ (1.0\arcsec)             & 0.25 & 4.97 & 2.09 \\
$M_{\rm BH}$ (2.0\arcsec)              & 0.25 & 4.97 & 2.09 \\  
\hline
10.0$\times M_{\rm BH}$ (2.0\arcsec)              & 0.25 & 4.97 & 2.09 \\ 
\bottomrule
\end{tabular}
\end{table}

\section{Effect of the uncertainties on the weak lensing masses and the concentration parameter}
\label{Delta}

In Table \ref{Test_Delta}, we illustrate how sensitive the best-fitting parameters are to changes in the weak lensing masses used as well as the value used for the concentration parameter by incorporating the 1$\sigma$ errors on the weak lensing masses, as well as an estimated error of $\pm$10 per cent on the calculated value for the concentration parameter. We again use the BCG in Abell 68 as an example.

\begin{table}
\caption{Influence of the uncertainties on the weak lensing mass and concentration parameter on the best-fitting parameters in the dynamical modelling of the BCG in Abell 68.}
\label{Test_Delta}
\centering
\begin{tabular}{lccc}
\toprule
\multicolumn{4}{c}{\textbf{Abell 68}} \\
fit  & $\beta_{z}$  &  $\Upsilon_{\star \rm DYN}$ & $\chi^{2}/DOF$\\
$\star$ + CEN + DM  &  & (M$_{\sun}$) & \\
\midrule
M$_{\rm WL}$ $\&$\ C  & 0.32 & 4.19 & 2.99 \\
\hline
(M$_{\rm WL} + \delta M_{\rm WL})\ \&$\ C & 0.33 & 4.03 & 3.18 \\
(M$_{\rm WL} - \delta M_{\rm WL})\ \&$\ C    & 0.30 & 4.36 & 2.79 \\
M$_{\rm WL}$ $\&$\ (C$+ \delta \rm C)$     & 0.32 & 4.11 & 3.07 \\
M$_{\rm WL}$ $\&$\ (C$- \delta \rm C)$     & 0.31 & 4.28 & 4.28 \\
\bottomrule
\end{tabular}
\end{table}

\section{Best-fit solutions for the `\textbf{$\star$}' and `\textbf{$\star$ + CEN}' scenarios}
\label{Additional}

We present the best-fitting solutions for the `\textbf{$\star$}' and `\textbf{$\star$ + CEN}' scenarios in Table \ref{AdditionalTable}.

\begin{table}
\centering
\begin{scriptsize}
\caption{Table with best-fit solutions for the `\textbf{$\star$}' and `\textbf{$\star$ + CEN}' scenarios. $\Upsilon_{\star \rm DYN}$ is constant with radius. See discussion in Section \ref{param_distrib} regarding the best-fitting solutions for Abell 963 and Abell 2055.}
\begin{tabular}{lrrr|rrr}
\toprule
 & \multicolumn{3}{c}{`\textbf{$\star$}'}  &  \multicolumn{3}{c}{`\textbf{$\star$ + CEN}'} \\
 & $\beta_{z}$  &  $\Upsilon_{\star \rm DYN}$  & $\chi^{2}/DOF$ & $\beta_{z}$  &  $\Upsilon_{\star \rm DYN}$  & $\chi^{2}/DOF$ \\  
 &  &   &   &  &   &  \\         
\midrule
 A68 & 0.25 & 4.97 & 2.10 & 0.25 & 4.96 & 2.10  \\
 A267 & 0.12 & 3.83 & 11.84 & 0.12 & 3.83 & 11.83 \\ 
 A383 & --0.08 & 2.17  & 7.22 &  --0.10 &  2.16 &  7.14 \\
 A611 & --0.28 & 3.15 & 6.70 & --0.28 & 3.14  & 6.69 \\
 A644 & --0.07 & 6.81 & 6.11 & --0.08 & 6.71 & 5.82 \\
 A646 & --0.27 & 5.22 & 0.32 & --0.28 & 5.19 & 0.33  \\
 A754 & 0.37 & 2.78 & 0.79 & 0.36 & 2.76  & 0.82 \\
 A780 & --0.37 & 3.20 & 4.25 & --0.40 & 3.17 & 4.29 \\
 A963 & --1.24 & 2.31 & 11.53 & --1.27 & 2.30  & 11.57 \\
 A1650 & 0.00 & 6.42 & 3.34 & --0.06 & 6.22  & 3.34 \\
 A1689 & --0.79 & 3.98 & 8.32 & --0.79 & 3.96  & 8.30 \\
 A1763 & --0.17 & 2.05 & 3.20 & --0.18 & 2.04  & 3.21 \\
 A1795 & --0.10 & 3.68 & 5.32 & --0.11 & 3.67 & 5.31 \\
 A1942 & --0.73 & 1.37 & 3.22 & --0.73 & 1.36  & 3.22 \\
 A1991 & 0.05 & 5.79 & 6.86 & 0.04 & 5.74 & 6.71 \\
 A2029 & 0.12 & 7.09 & 2.74 & 0.12 & 7.07 & 2.70 \\
 A2050 & --0.06 & 4.24 & 1.24 & --0.07 & 4.23 & 1.24 \\
 A2055 & --2.93 & 1.34 & 2.72 & --2.95 & 1.33  & 2.72  \\
 A2142 & 0.07 & 6.61 & 3.13 & 0.05 & 6.54  & 3.19  \\
 A2259 & 0.20 & 4.44 & 3.28 & 0.20 & 4.41 & 3.24 \\
 A2261 & --0.31 & 3.73 & 0.95 & --0.32 & 3.71  & 0.94 \\
 A2319 & --0.11 & 6.28 & 7.62 & --0.38 & 6.23 & 6.52 \\
 A2420 & 0.05 & 5.91 & 1.66 & 0.05 & 5.88 & 1.65 \\
 A2537 & --0.63 & 2.40 & 1.19 & --0.65 & 2.40 & 1.19 \\
 MS1455 & 0.08 & 2.03 & 0.70 & 0.08  & 2.02  & 0.67 \\
\bottomrule
\end{tabular}
\label{AdditionalTable}
\end{scriptsize}
\end{table}

\section{Dynamical modelling results}
\label{DynMods}

\begin{figure*}
\centering
         \subfloat[Abell 68]{\includegraphics[scale=0.27]{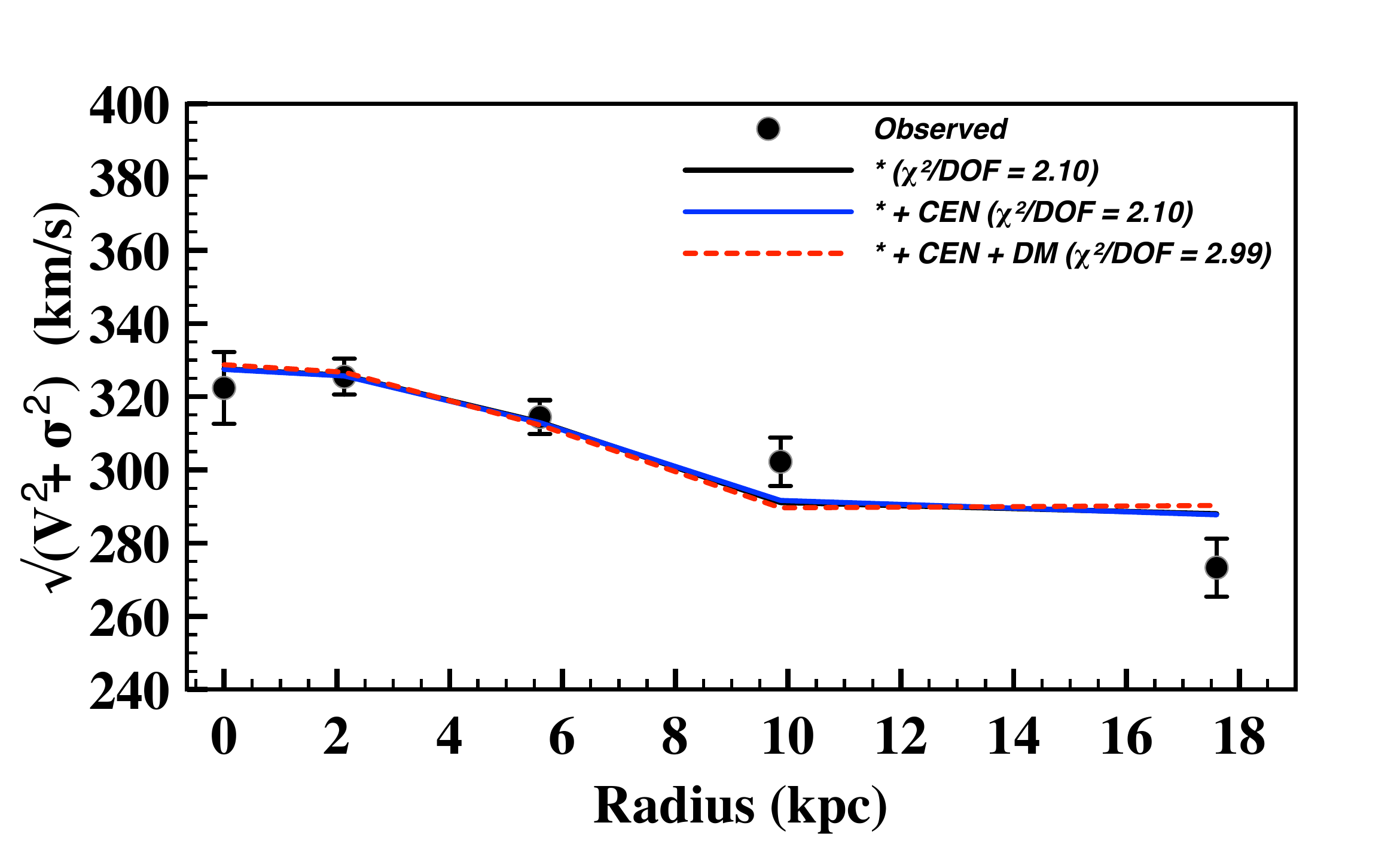}}
         \subfloat[Abell 267]{\includegraphics[scale=0.27]{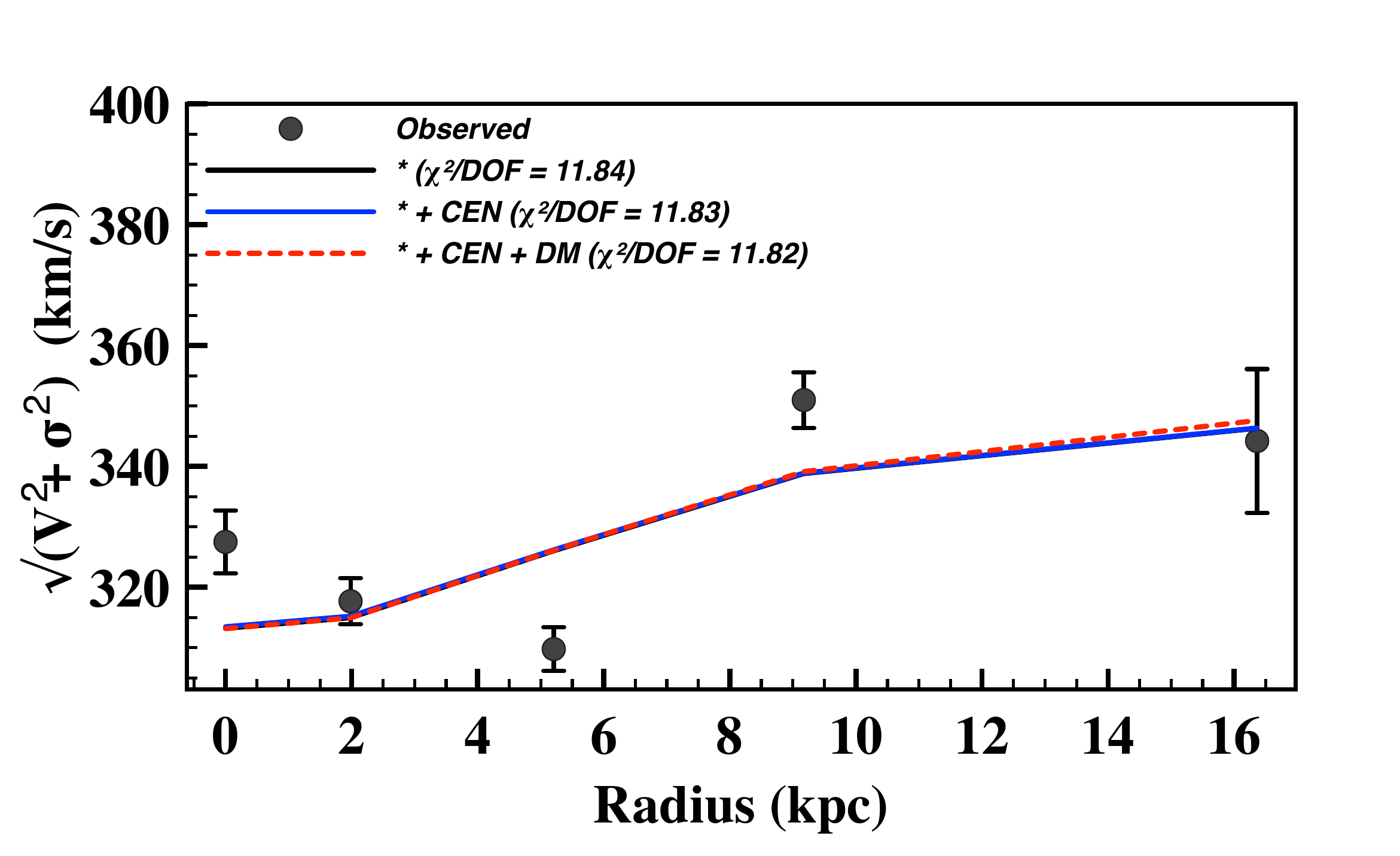}}
         \subfloat[Abell 383]{\includegraphics[scale=0.27]{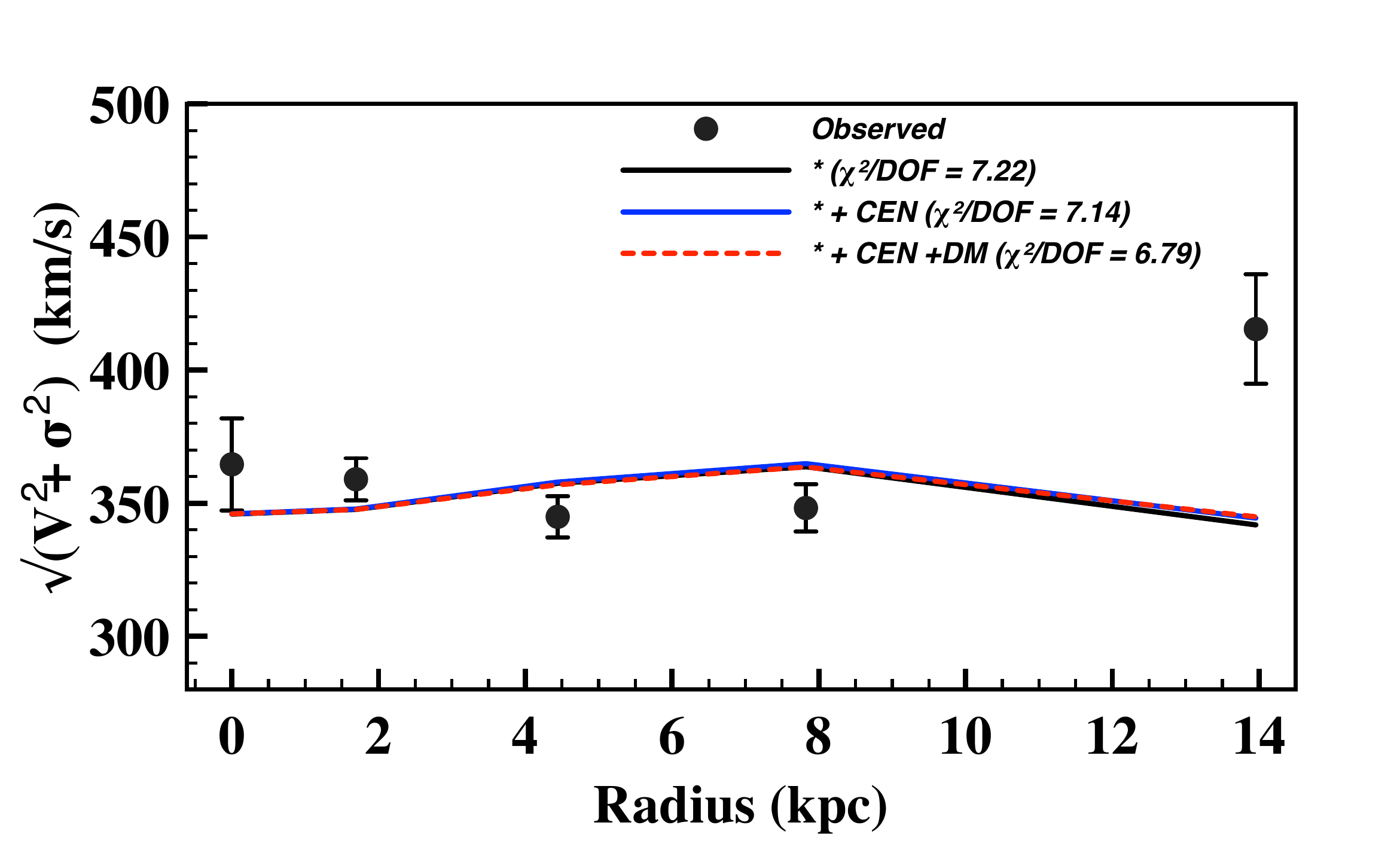}}\\
         \subfloat[Abell 611]{\includegraphics[scale=0.27]{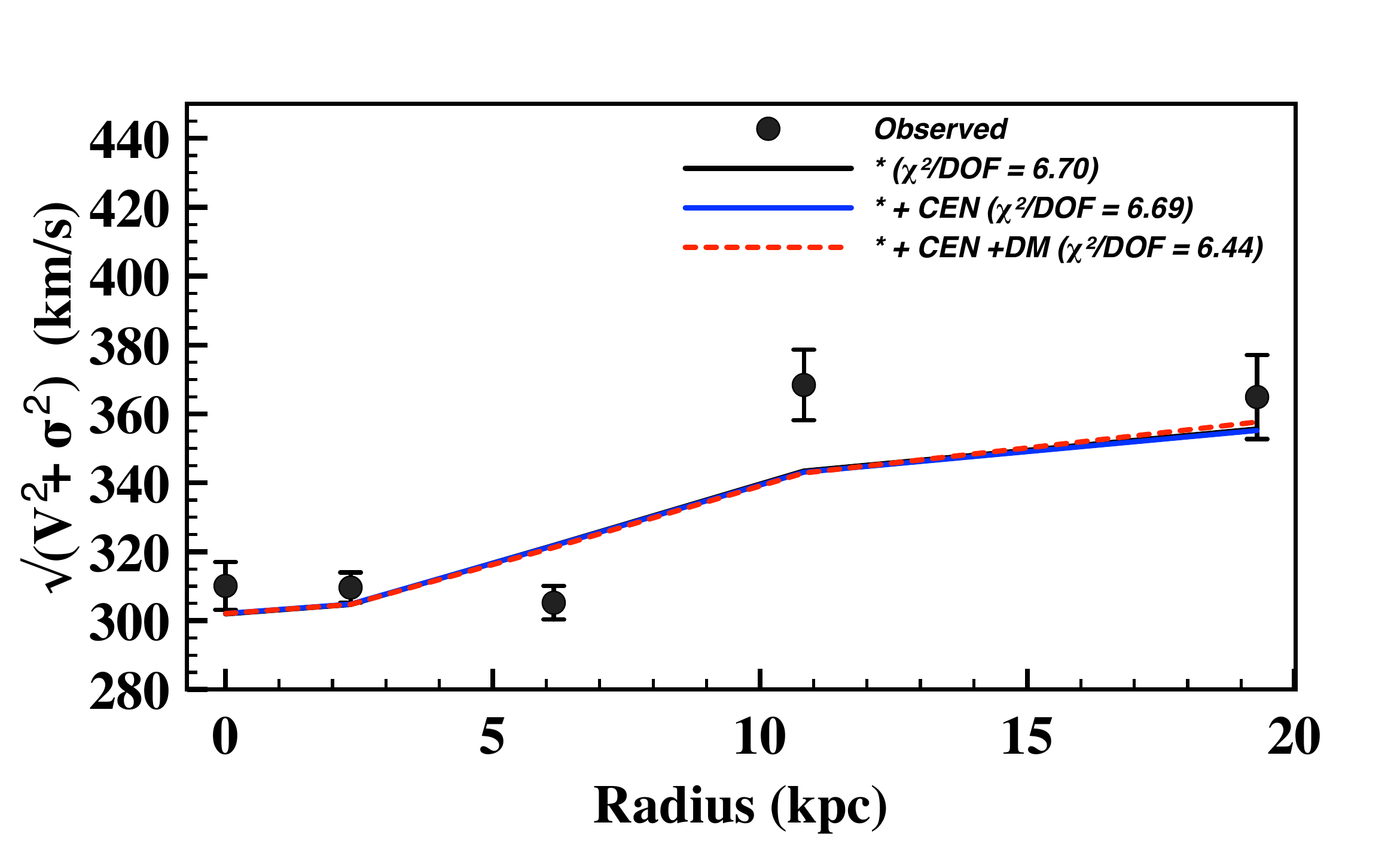}}
          \subfloat[Abell 644]{\includegraphics[scale=0.27]{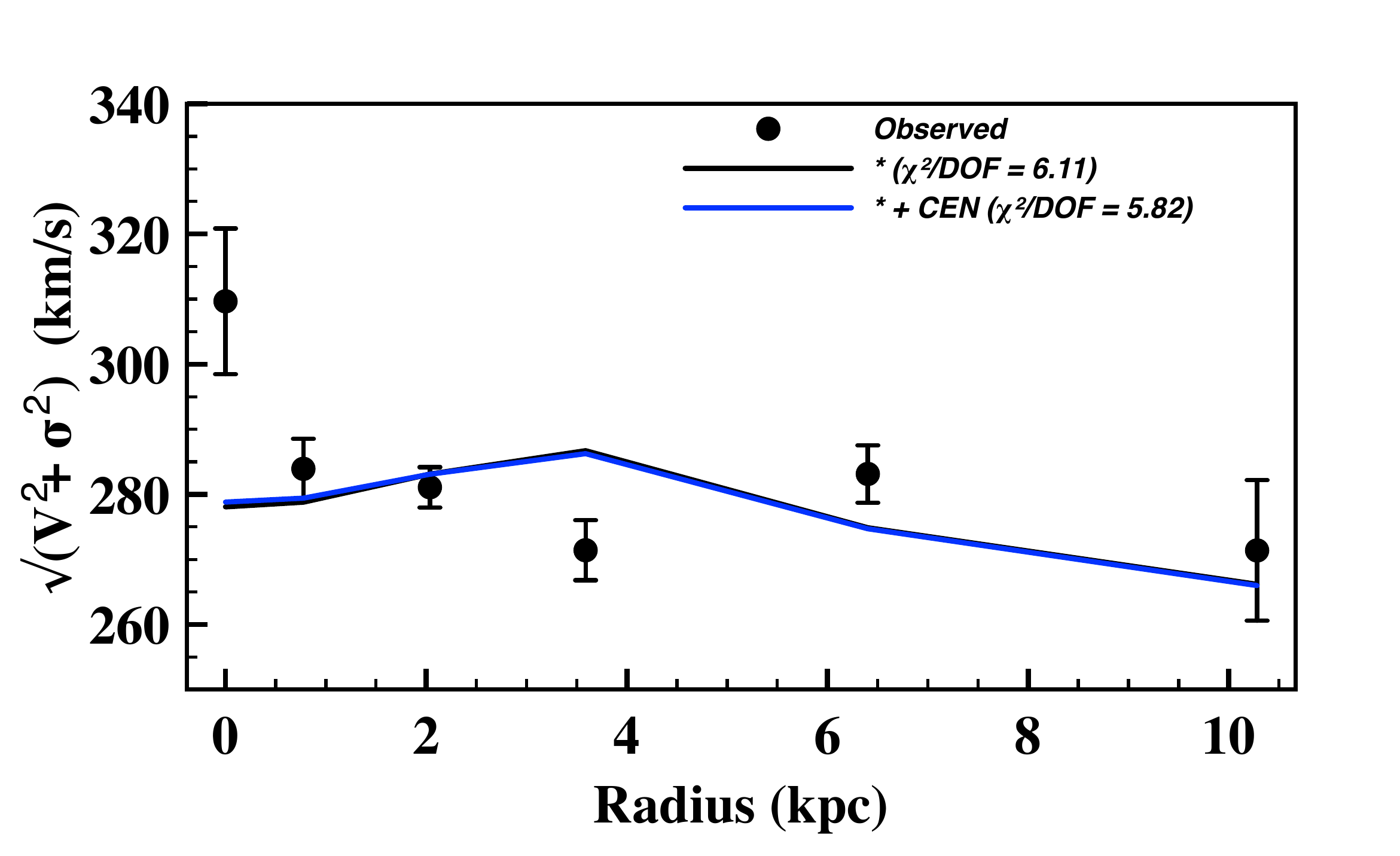}}
           \subfloat[Abell 646]{\includegraphics[scale=0.27]{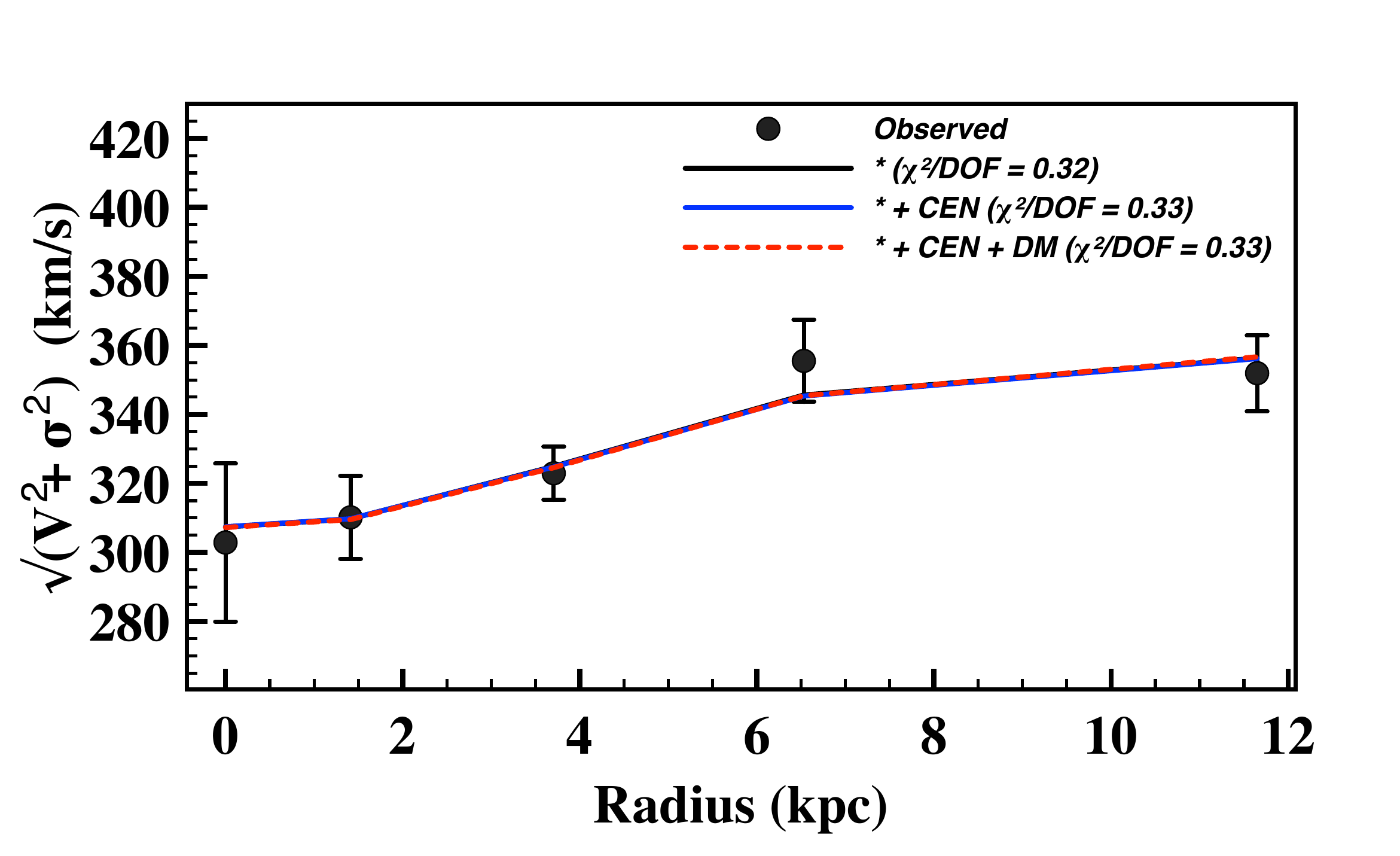}}\\
            \subfloat[Abell 754]{\includegraphics[scale=0.27]{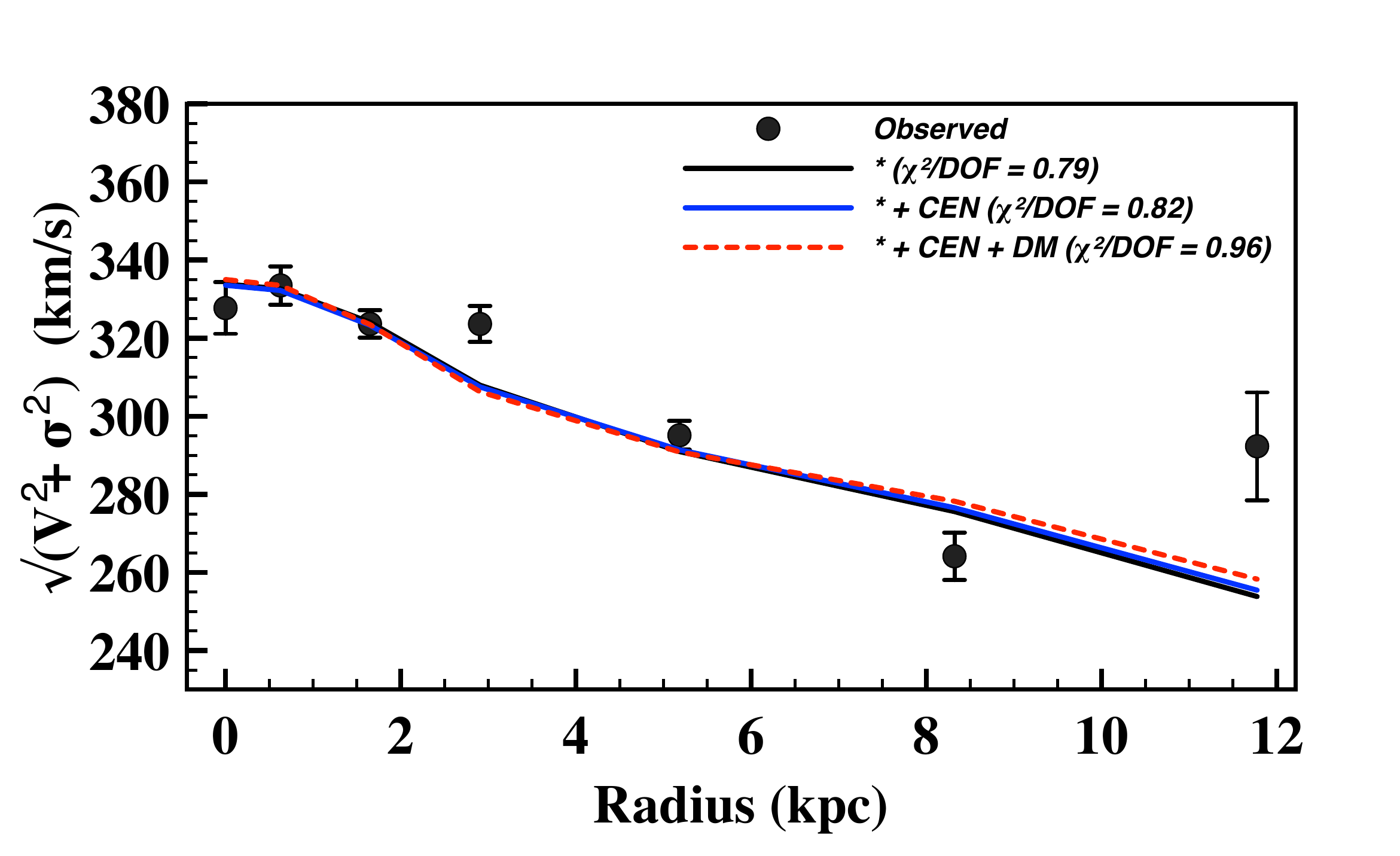}}
             \subfloat[Abell 780]{\includegraphics[scale=0.27]{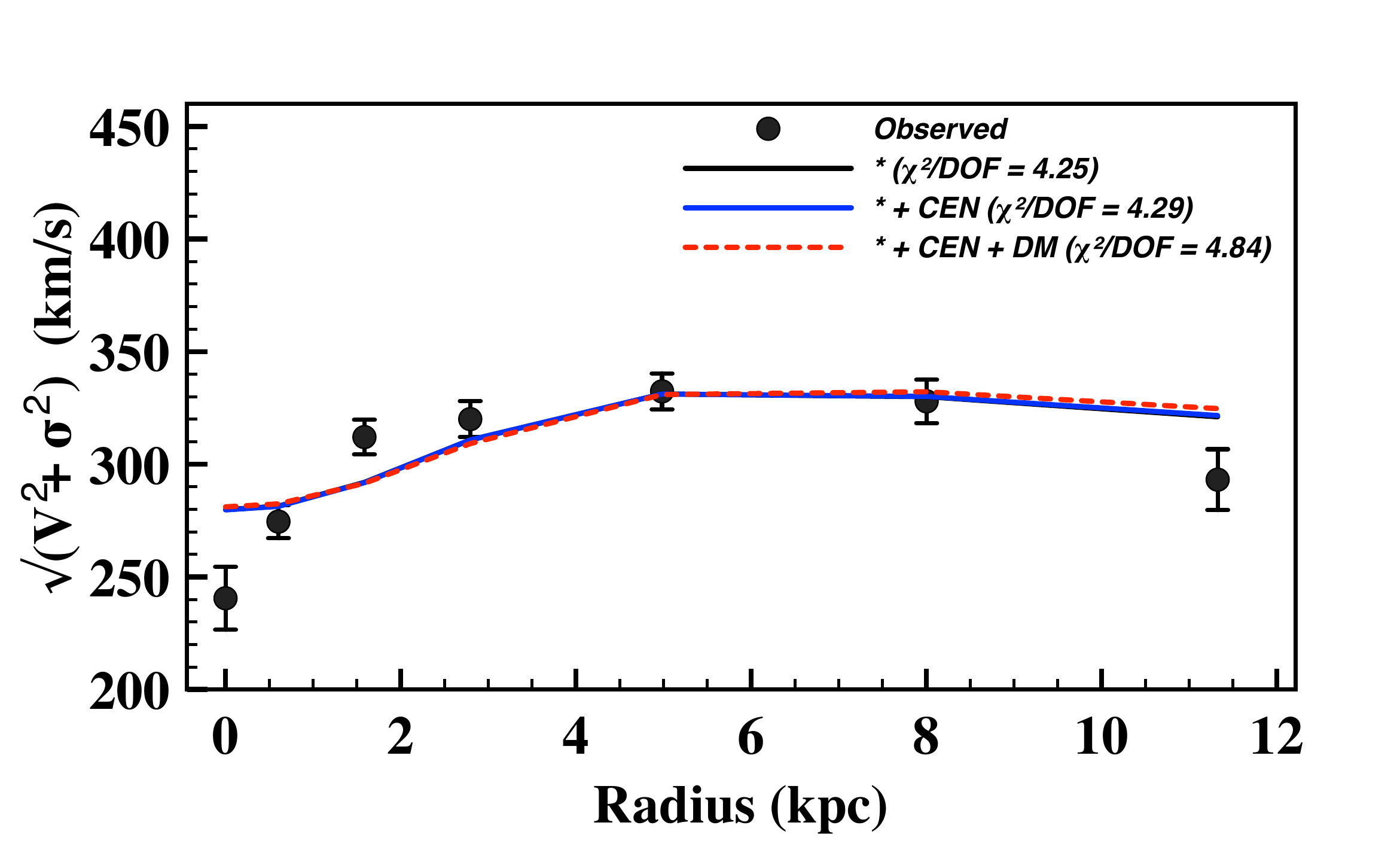}}
              \subfloat[Abell 963]{\includegraphics[scale=0.27]{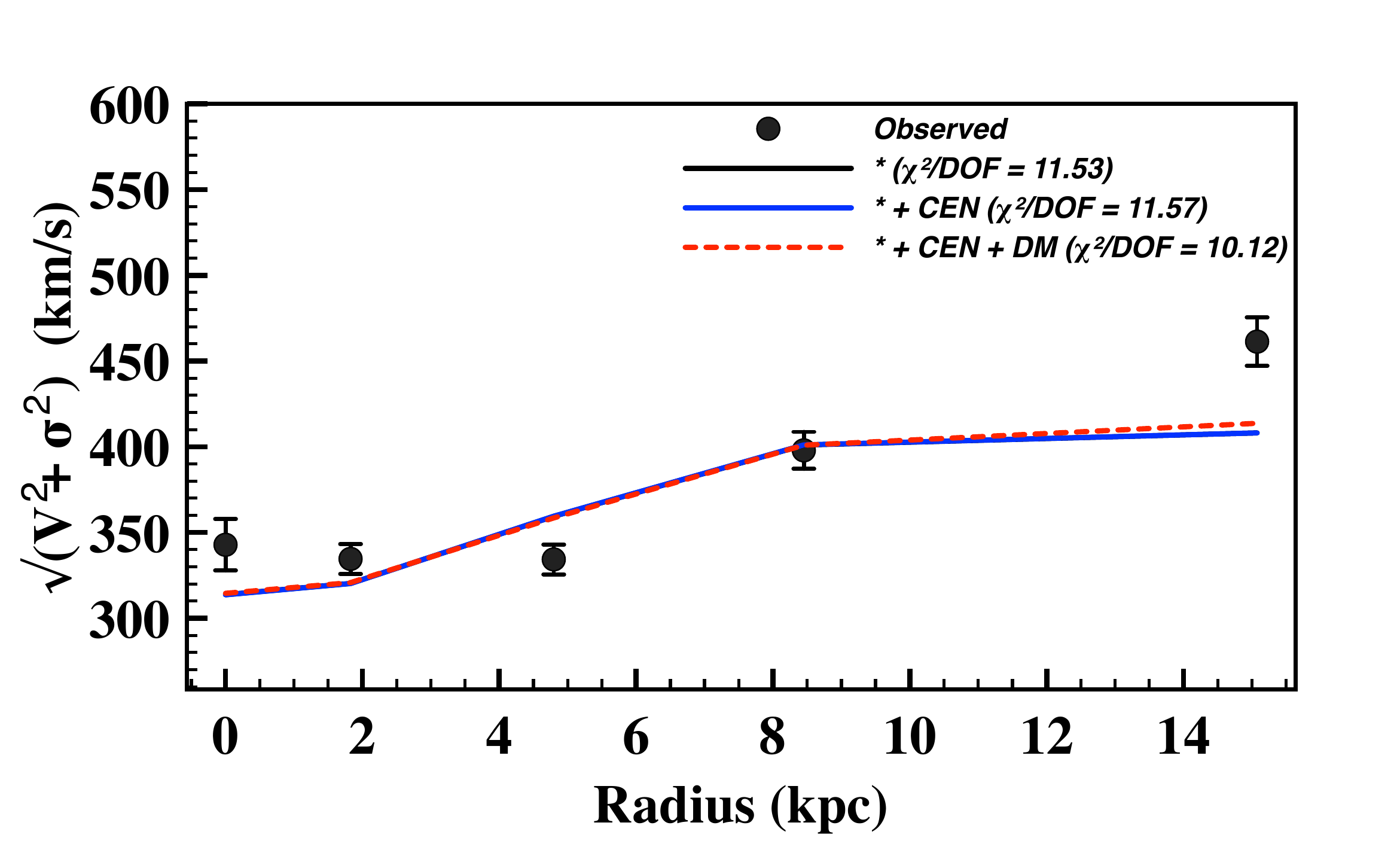}}\\
               \subfloat[Abell 1650]{\includegraphics[scale=0.27]{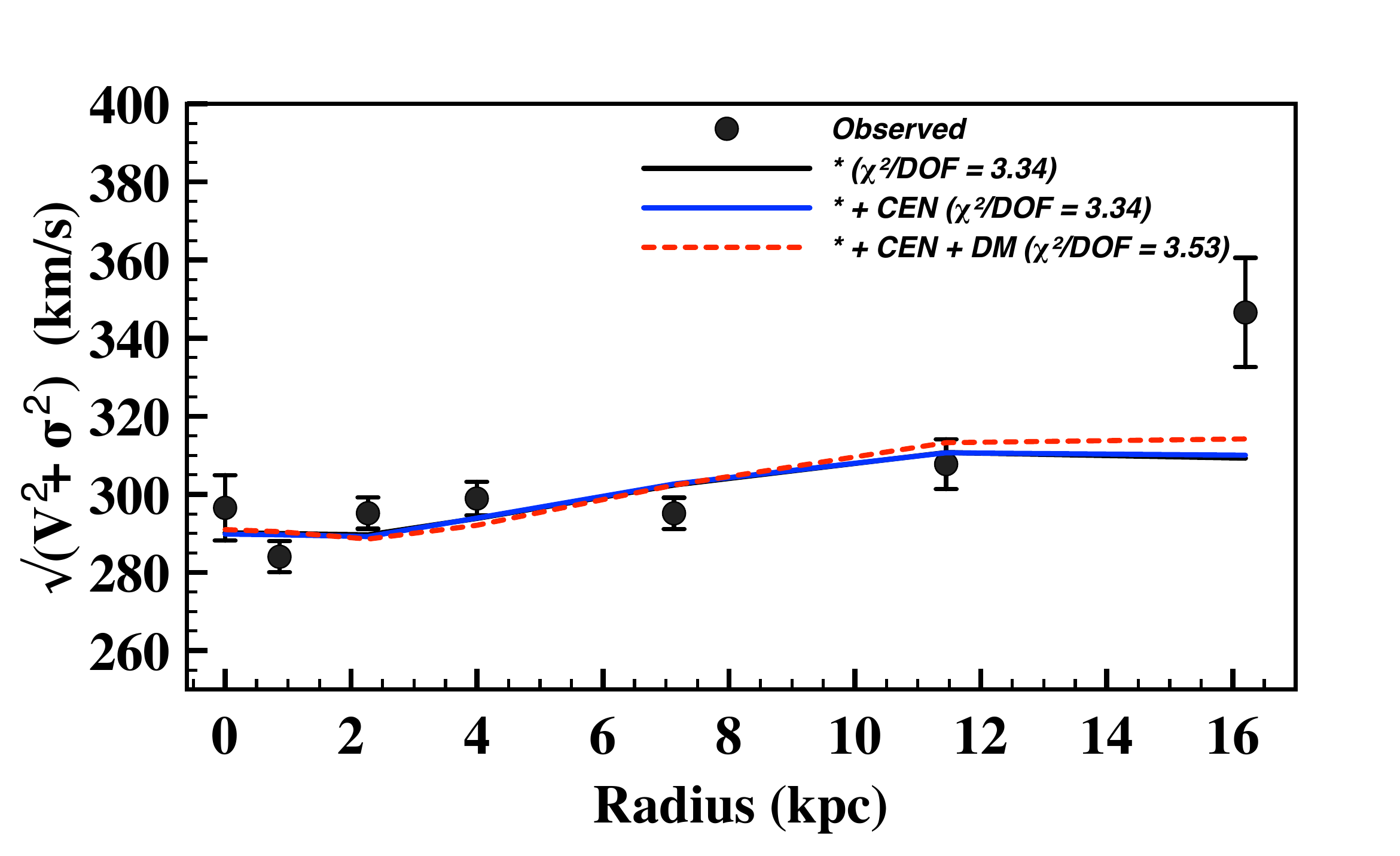}}
                \subfloat[Abell 1689]{\includegraphics[scale=0.27]{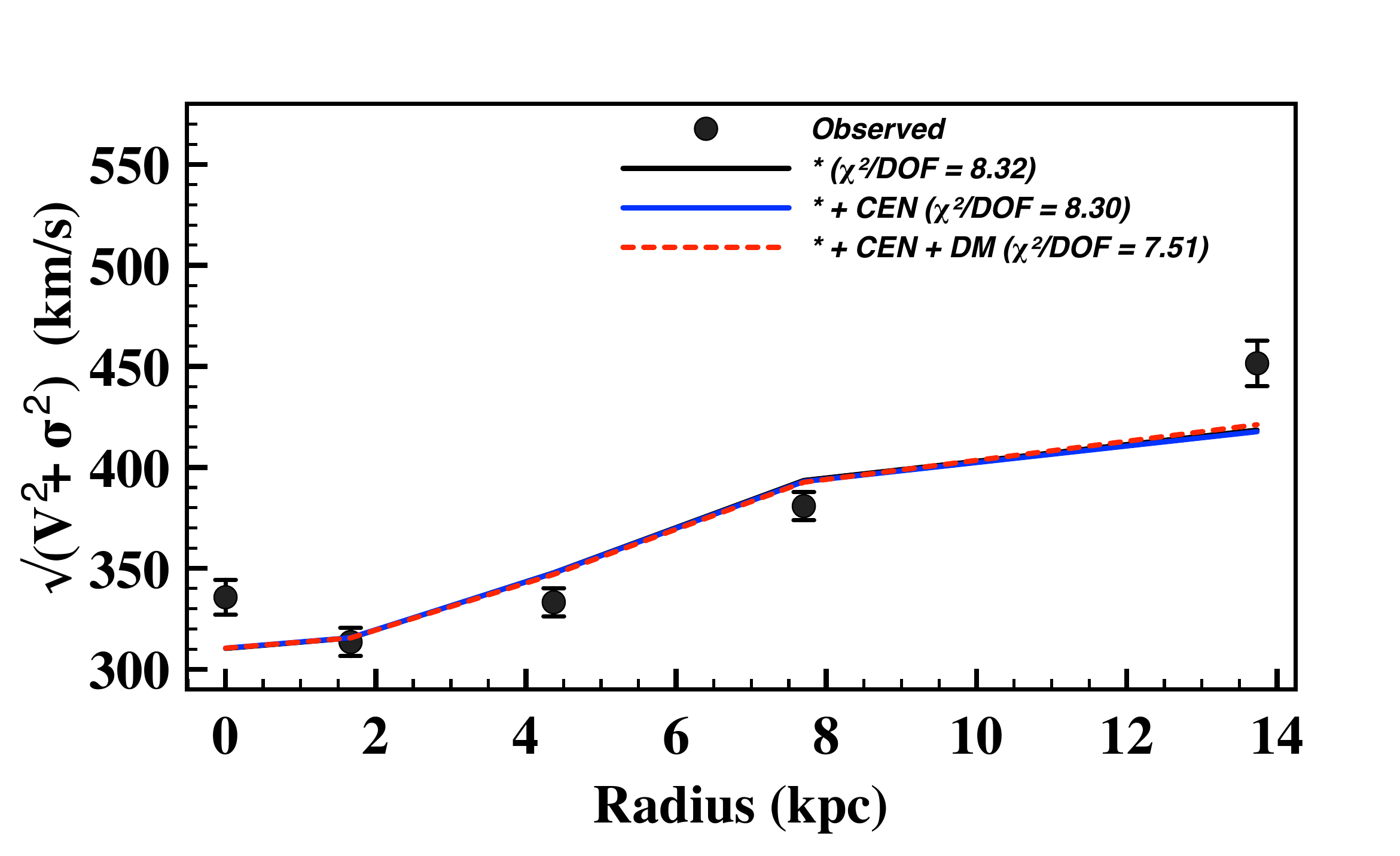}}
                 \subfloat[Abell 1763]{\includegraphics[scale=0.27]{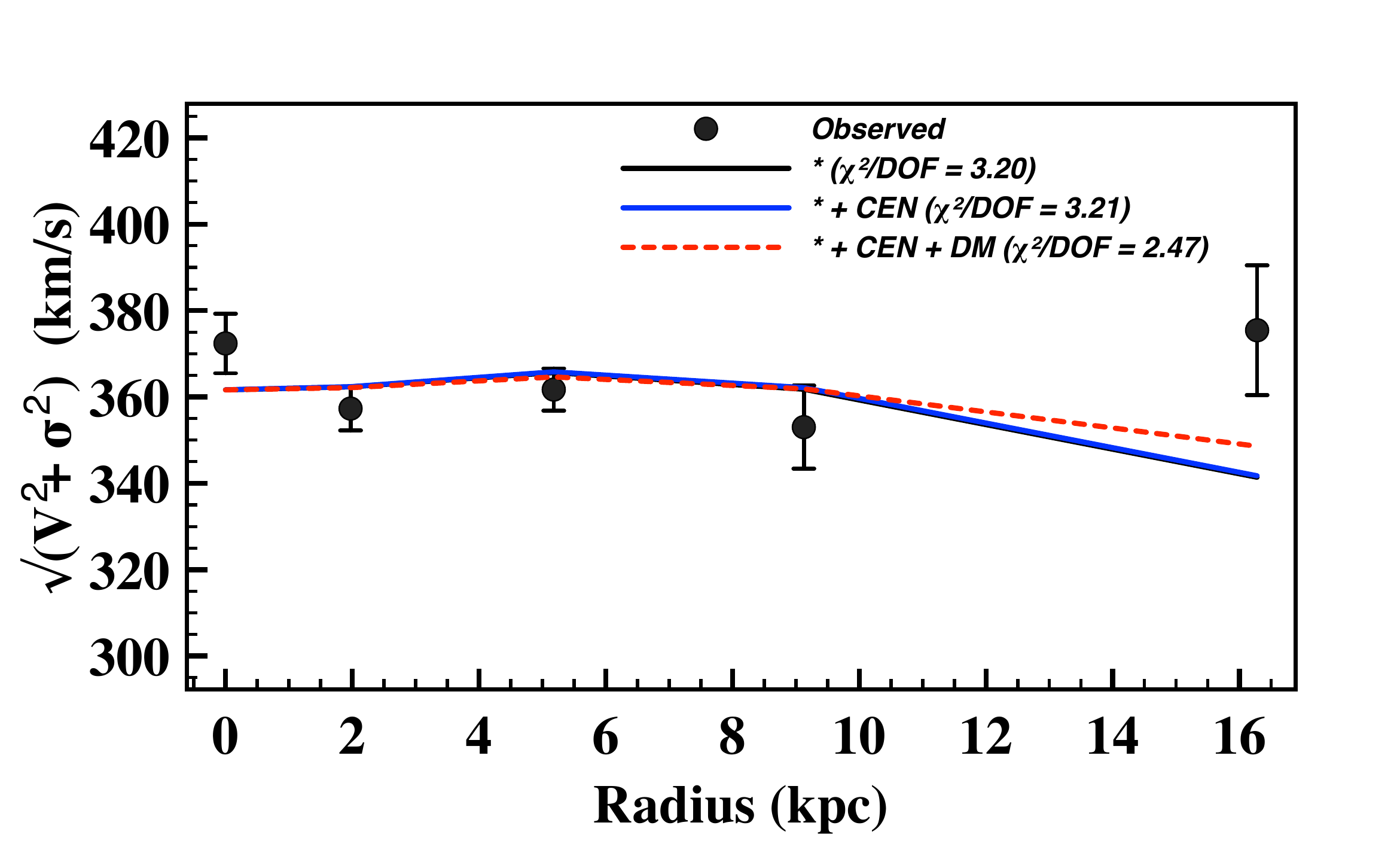}}\\
                  \subfloat[Abell 1795]{\includegraphics[scale=0.27]{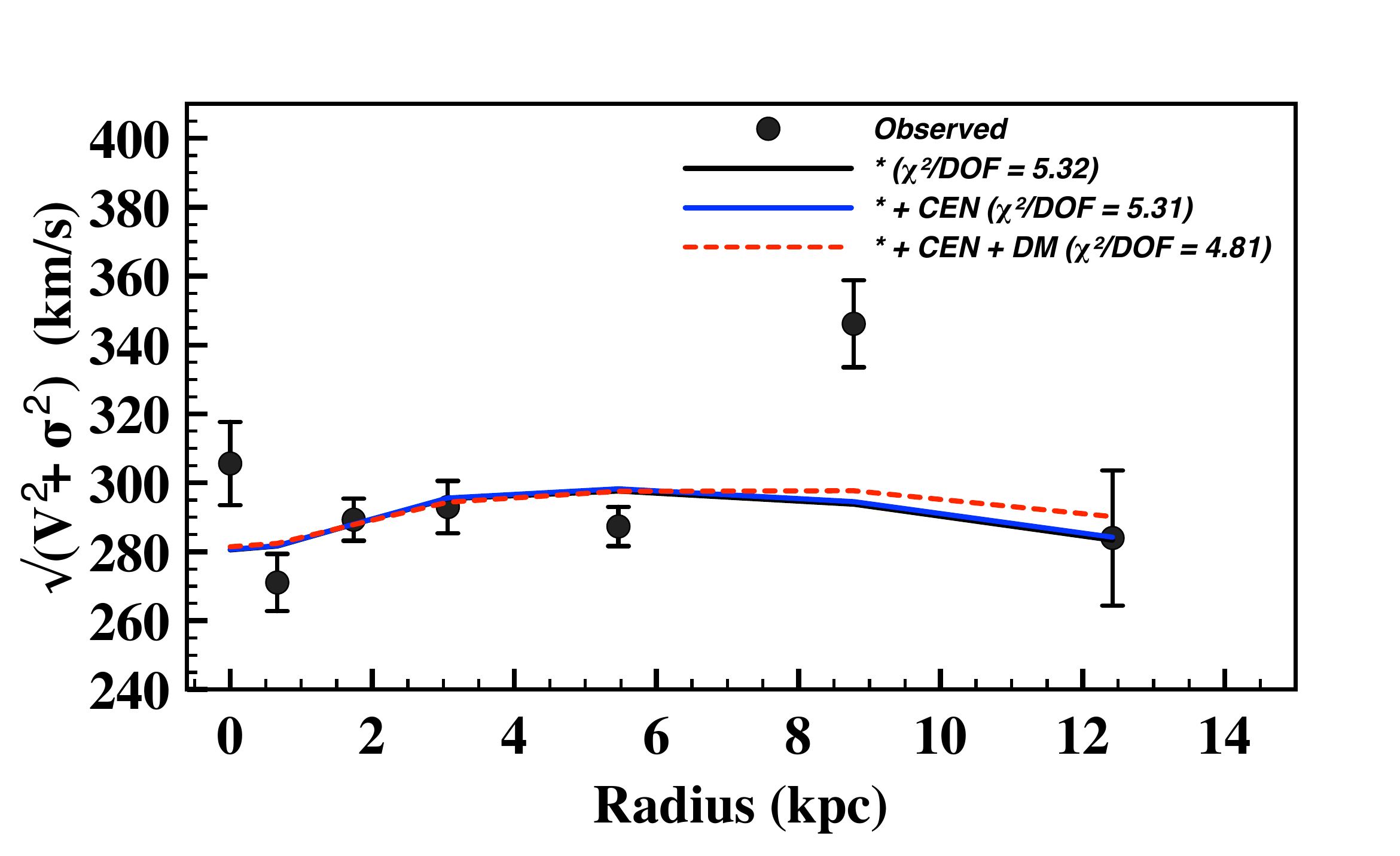}}
                   \subfloat[Abell 1942]{\includegraphics[scale=0.27]{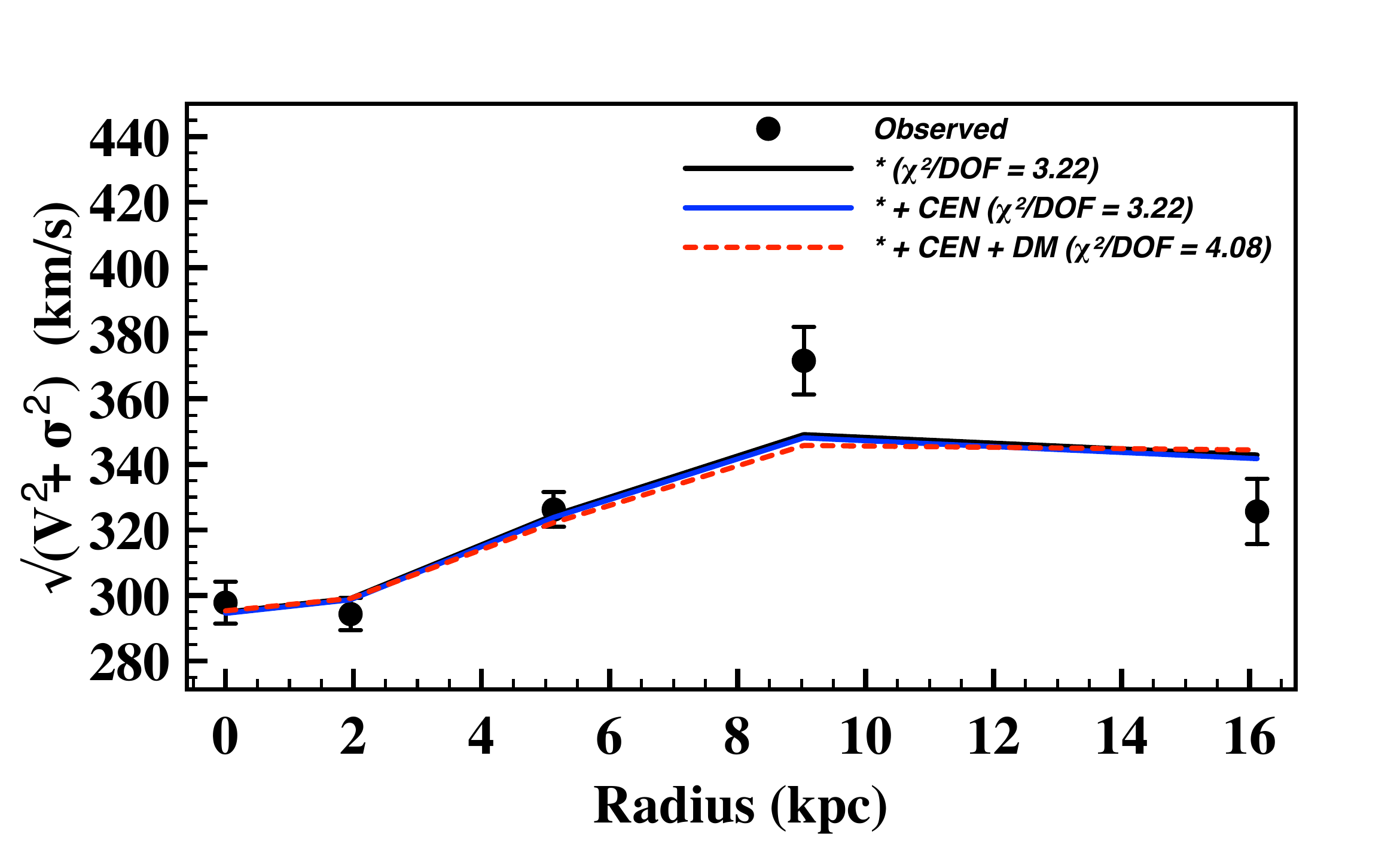}}
                    \subfloat[Abell 1991]{\includegraphics[scale=0.27]{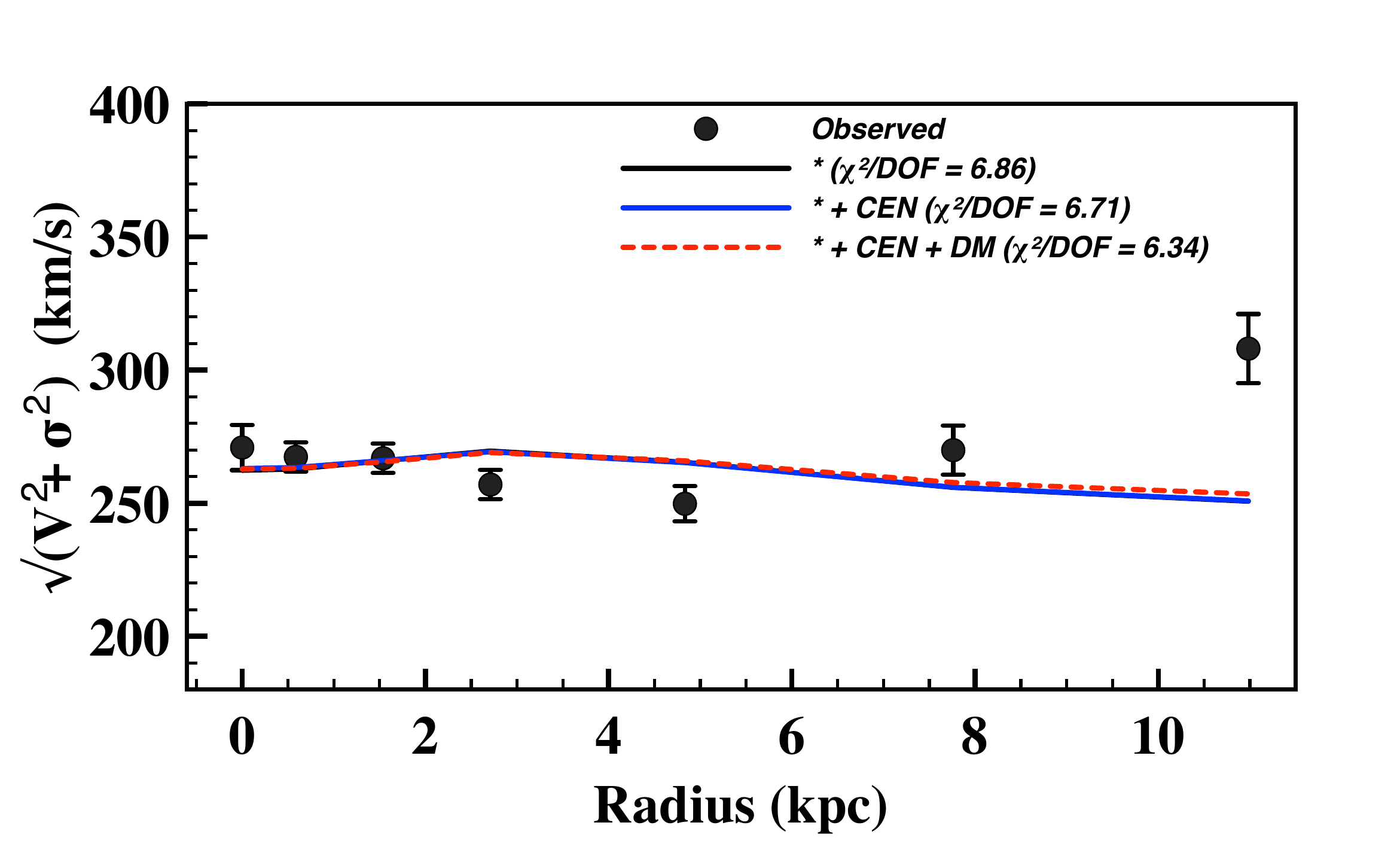}}\\
  \caption{The averaged second moment of velocity ($\sqrt{V^{2} + \sigma^{2}}$) profile.}
\label{DynModsFig}
\end{figure*}

\begin{figure*}
\centering                    
                     \subfloat[Abell 2029]{\includegraphics[scale=0.27]{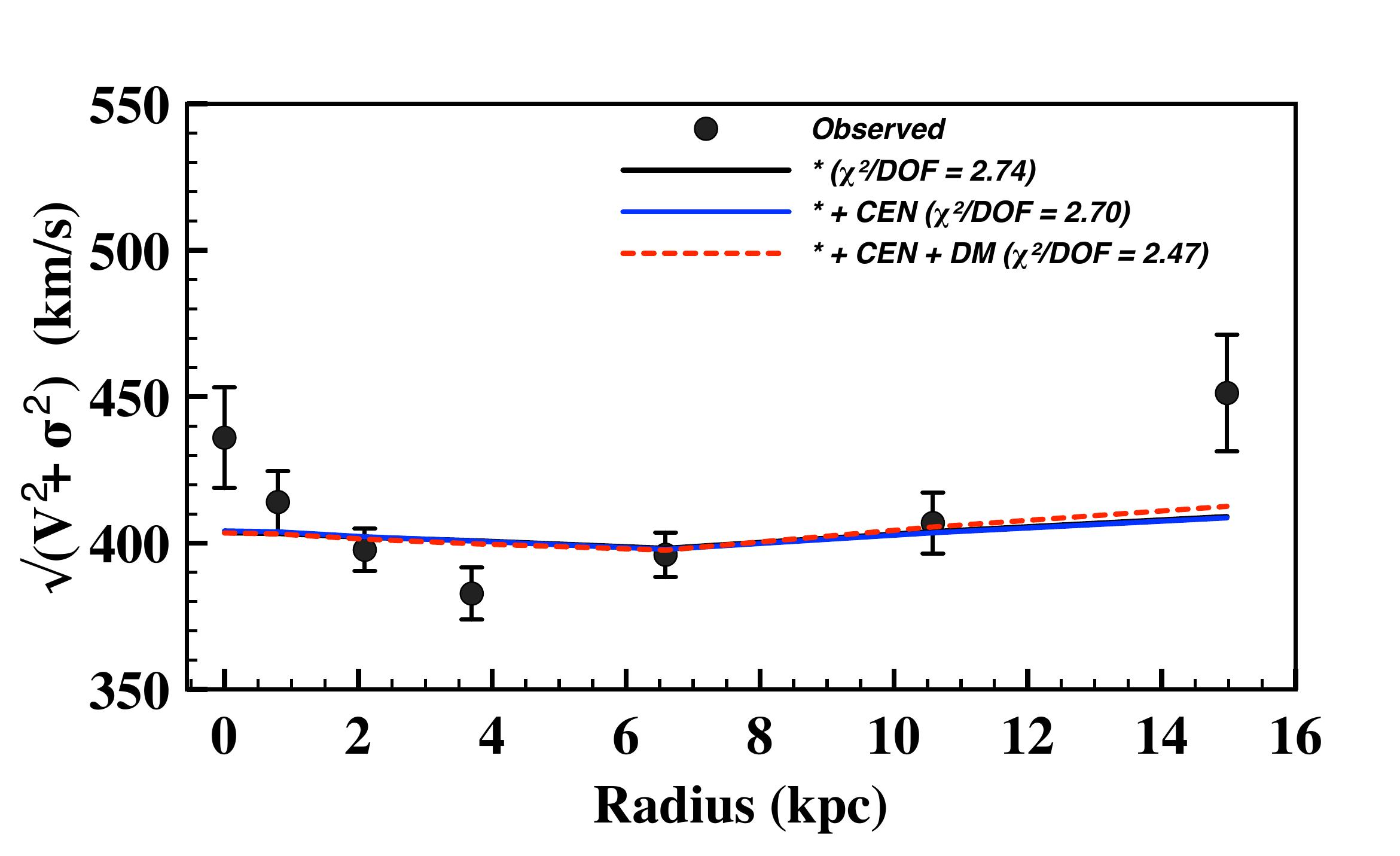}}
                      \subfloat[Abell 2050]{\includegraphics[scale=0.27]{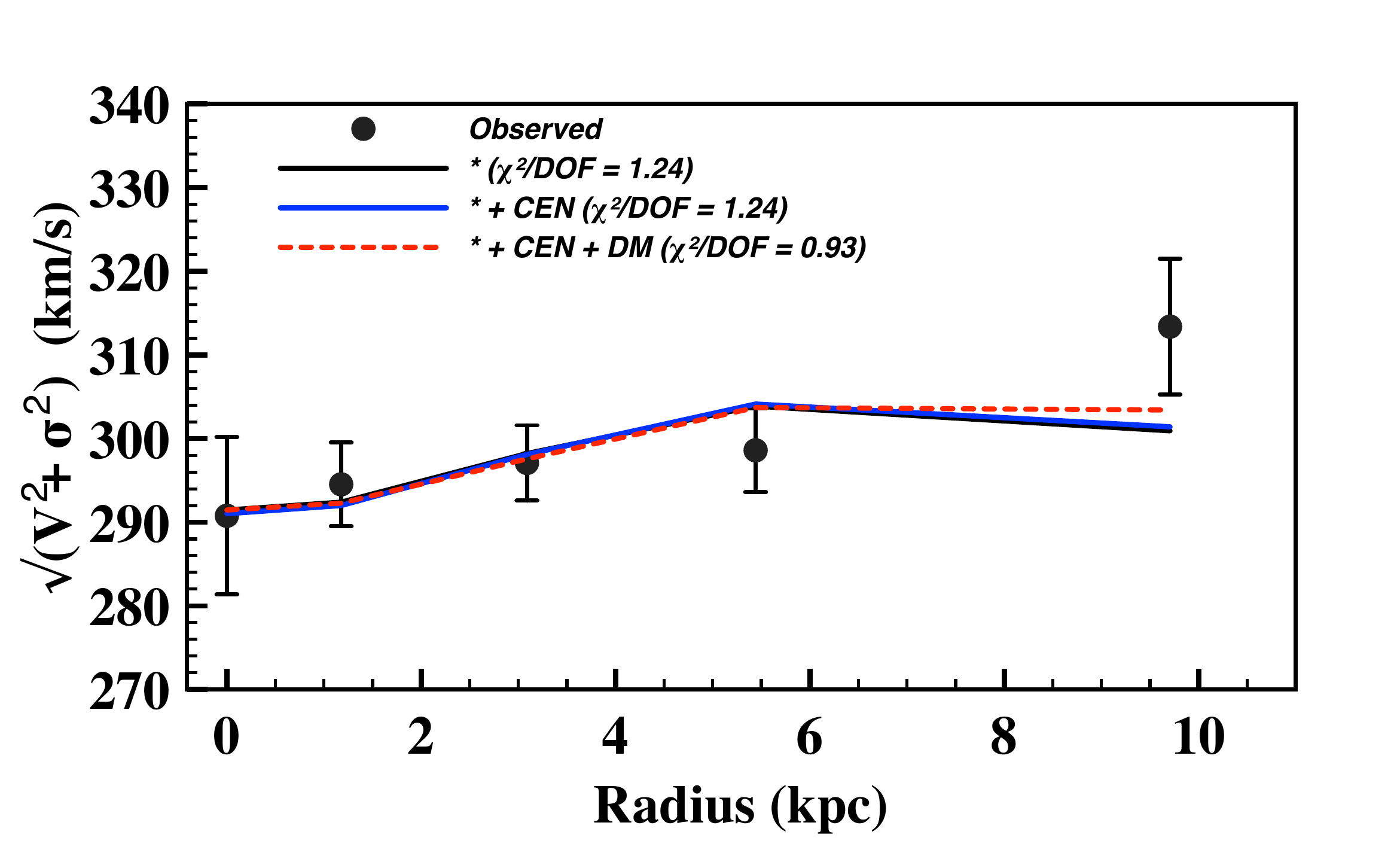}}
                       \subfloat[Abell 2055]{\includegraphics[scale=0.27]{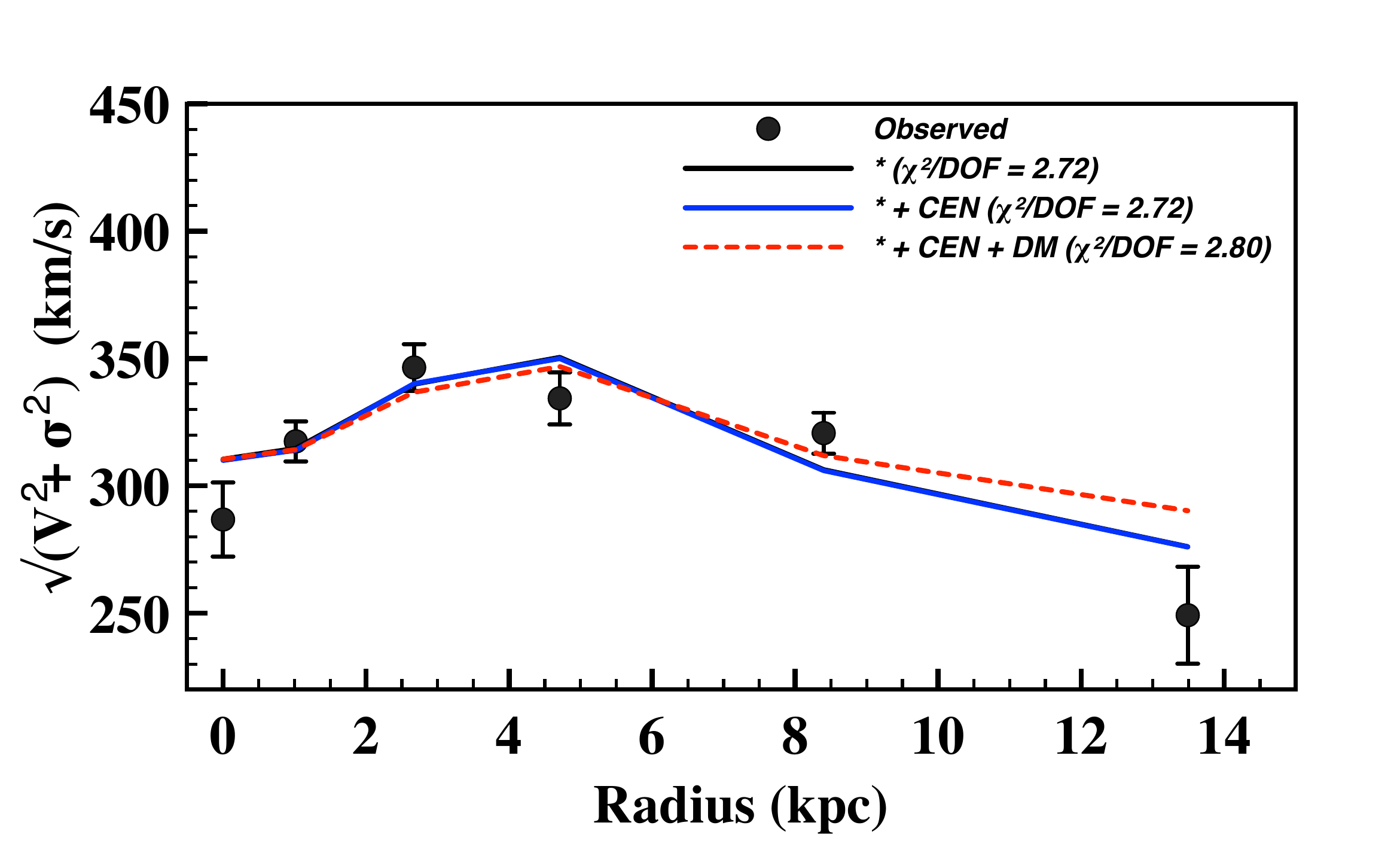}}\\
                        \subfloat[Abell 2142]{\includegraphics[scale=0.27]{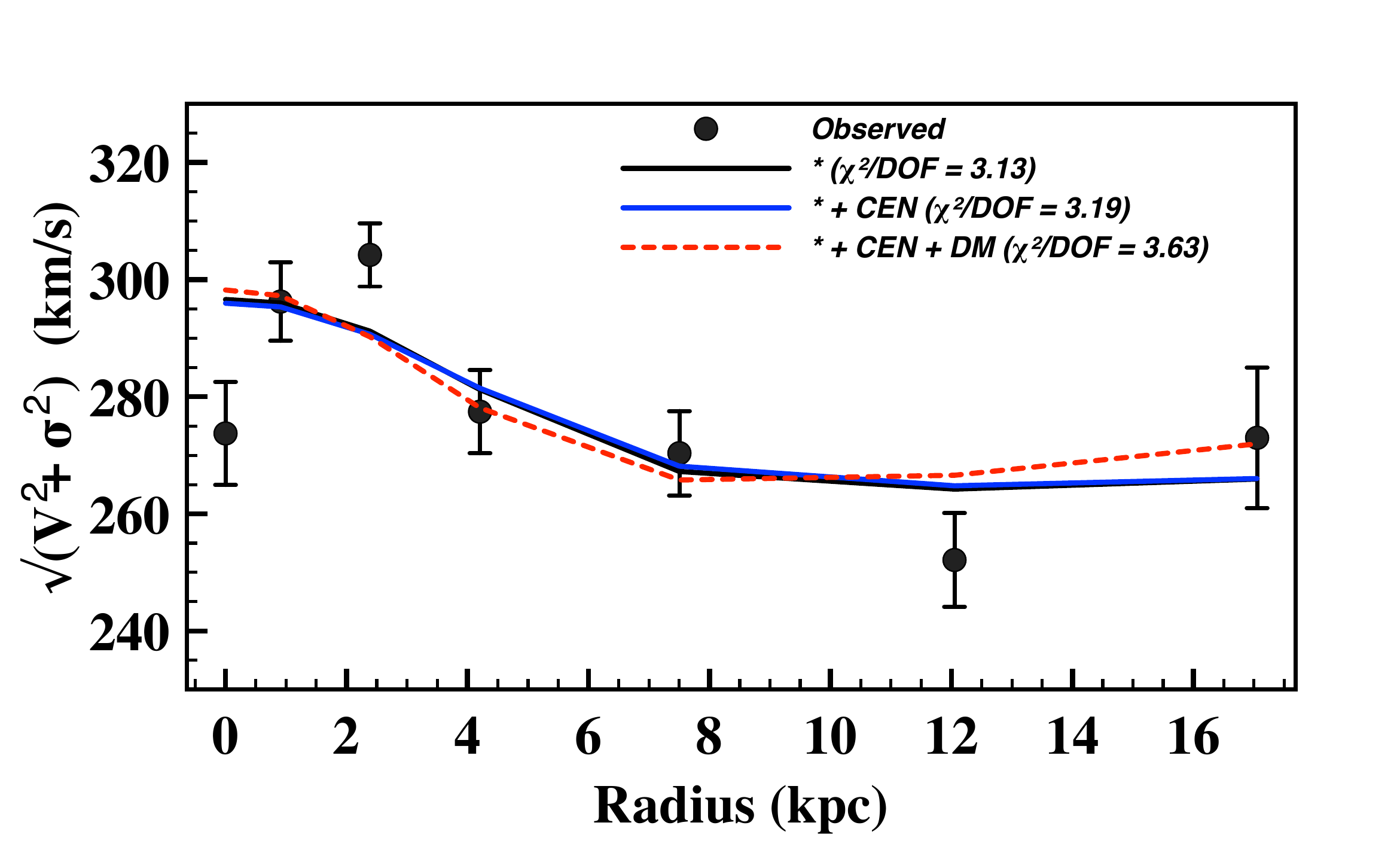}}
                         \subfloat[Abell 2259]{\includegraphics[scale=0.27]{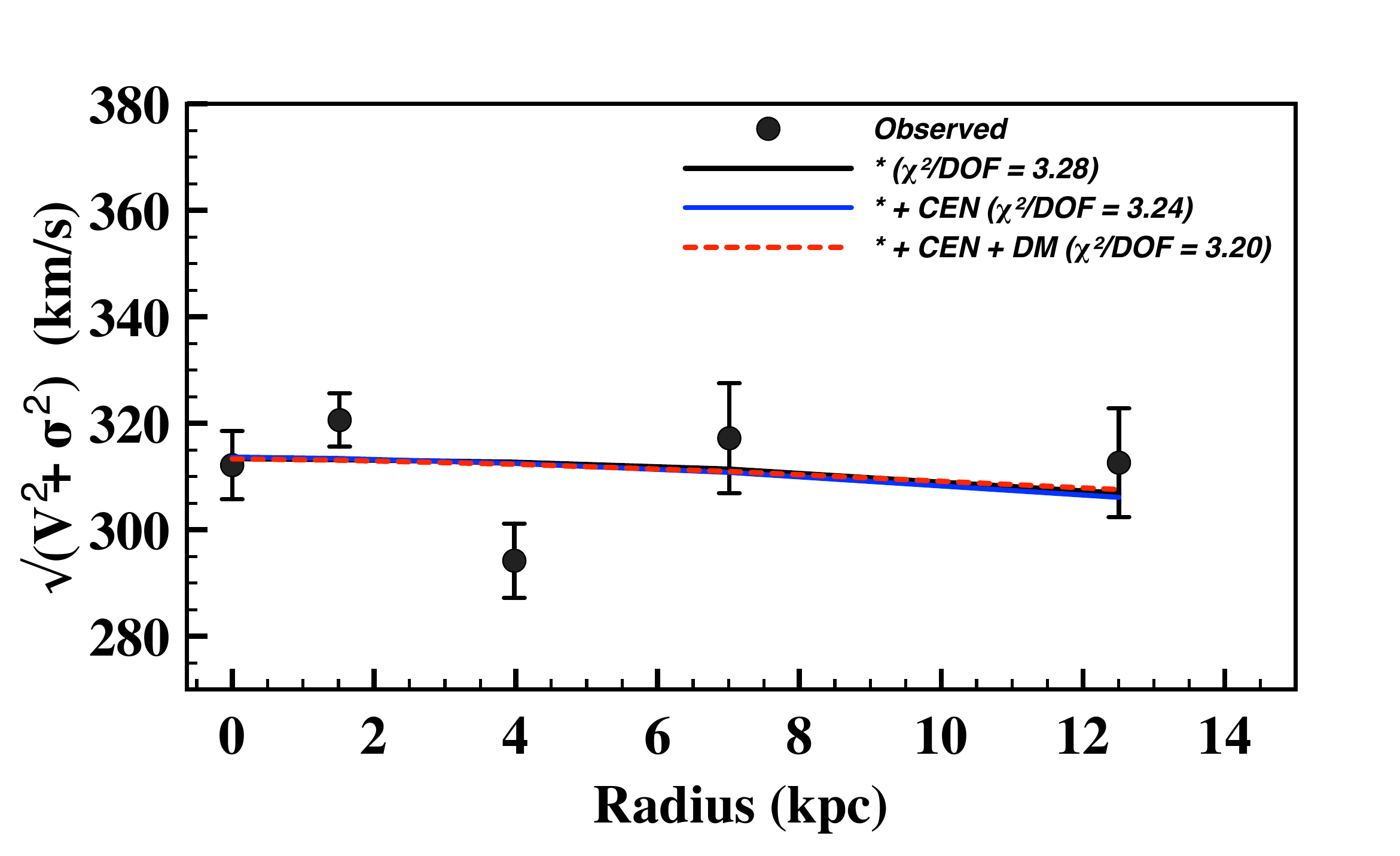}}
                          \subfloat[Abell 2261]{\includegraphics[scale=0.27]{A2261_V2Sig2.pdf}}\\
                           \subfloat[Abell 2319]{\includegraphics[scale=0.27]{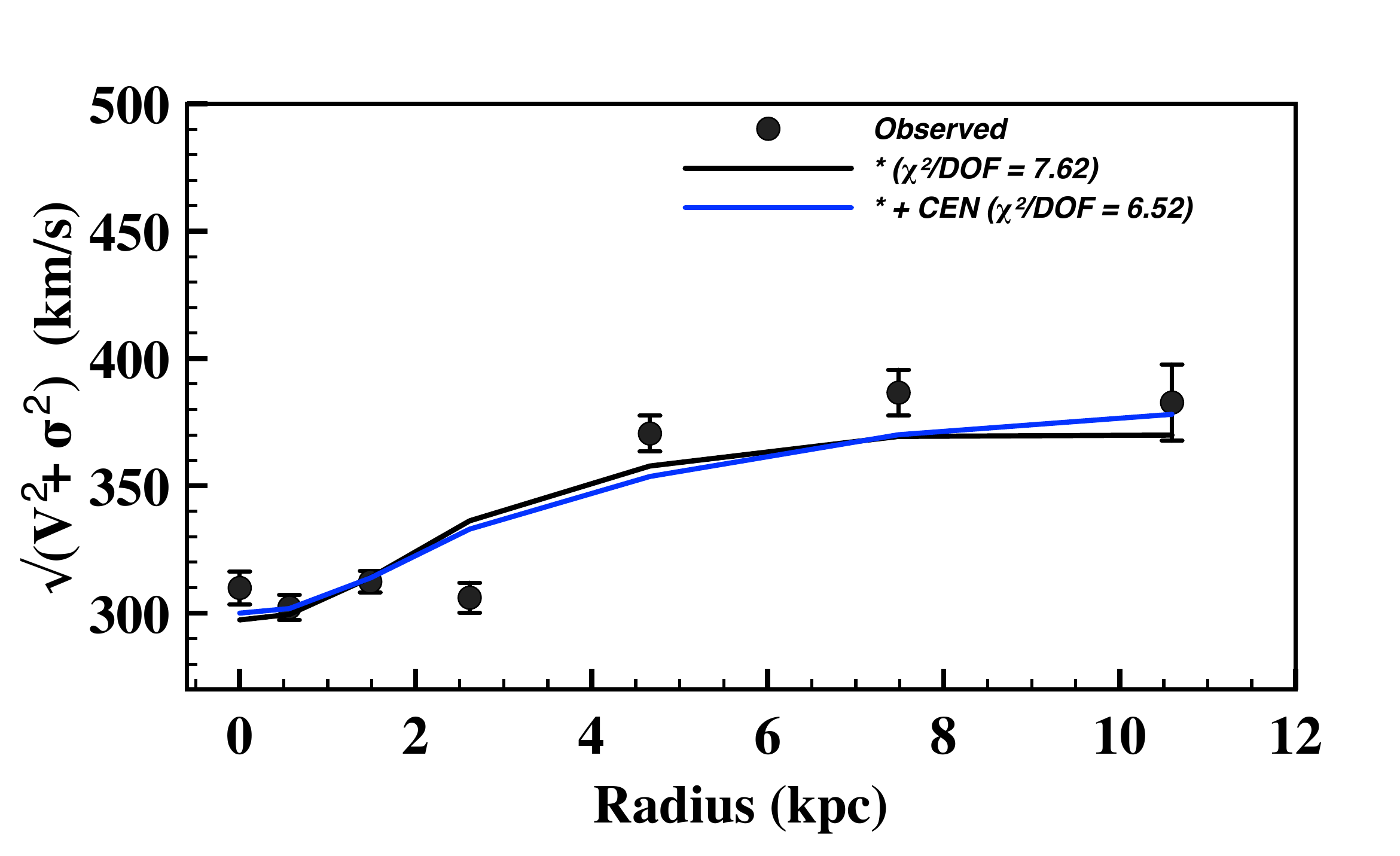}}
                            \subfloat[Abell 2420]{\includegraphics[scale=0.27]{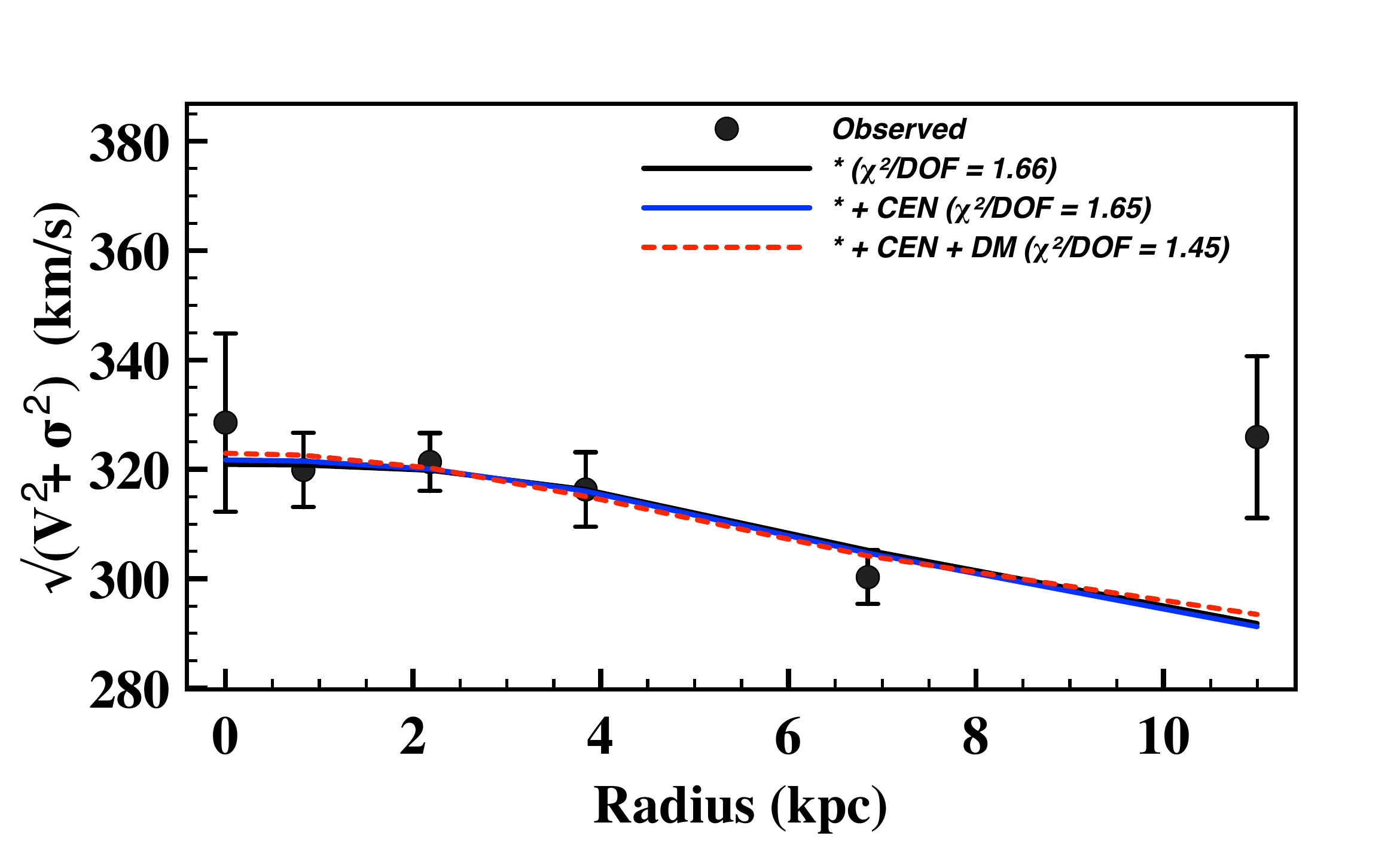}}
                             \subfloat[Abell 2537]{\includegraphics[scale=0.27]{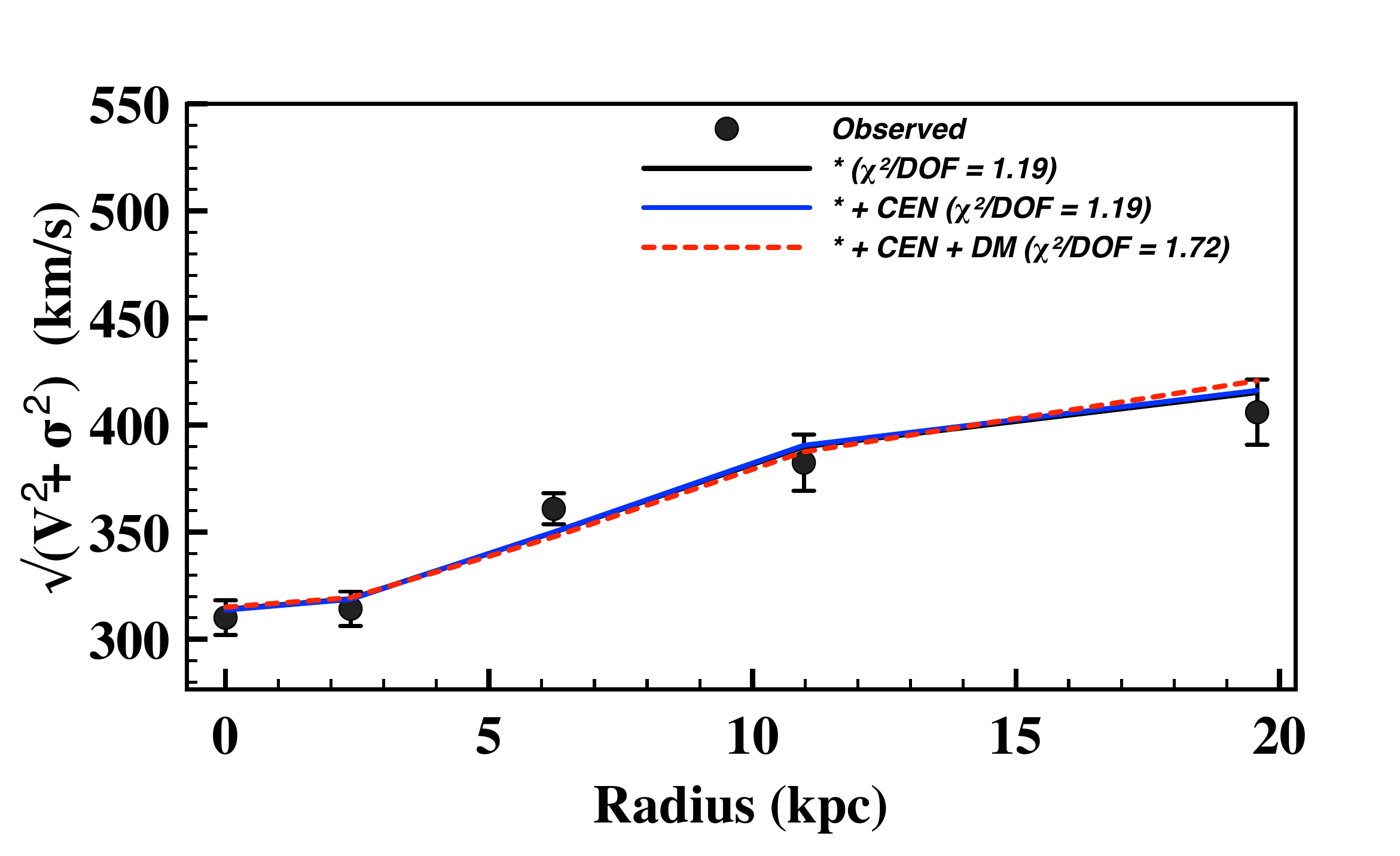}}\\
                              \subfloat[MS1455+22]{\includegraphics[scale=0.27]{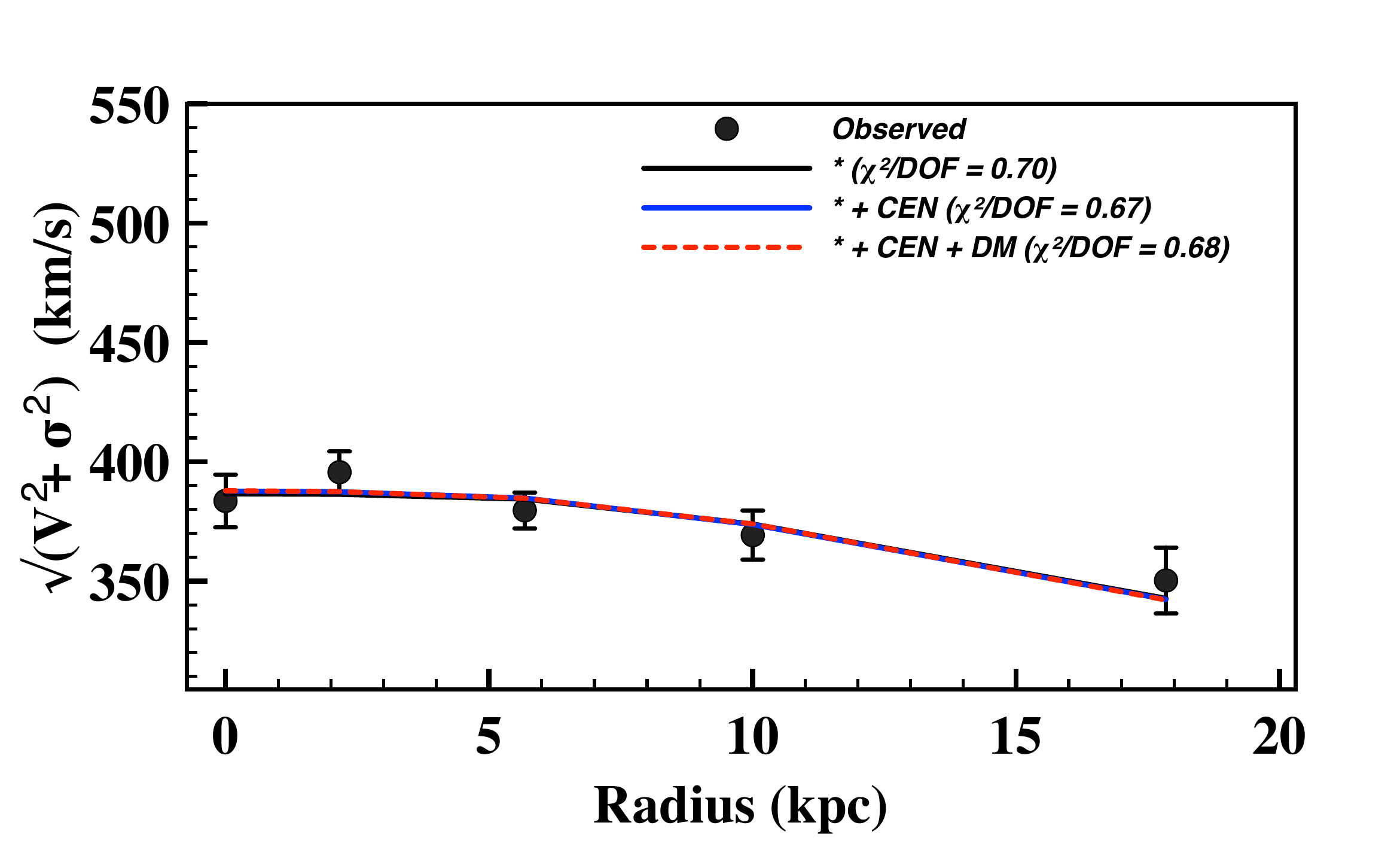}}\\
                  \caption{The averaged second moment of velocity ($\sqrt{V^{2} + \sigma^{2}}$) profile.}
\label{DynModsFig2}
\end{figure*}

Figures \ref{DynModsFig} and \ref{DynModsFig2} show the averaged second moment of velocity ($\sqrt{V^{2} + \sigma^{2}}$) profile. We assume symmetry and average the measurements on both sides of the galaxy centre (inversely weighted by the errors on $\sqrt{V^{2} + \sigma^{2}}$).

\bsp	
\label{lastpage}


\end{document}